\documentclass[aps,prd,preprintnumbers,superscriptaddress,nofootinbib]{revtex4}

\usepackage{hyperref}
\hypersetup{colorlinks=true,linkcolor=purple,anchorcolor=blue,citecolor=blue, filecolor=blue,urlcolor=red,bookmarksnumbered=true,
pdfview=FitB
}
\usepackage{color}
\usepackage{xcolor}
\colorlet{purple1}{blue!70!red}
\colorlet{darkred}{red!50!black}

\usepackage{graphicx}
\usepackage[bf,SL,BF]{subfigure}
\usepackage{psfrag}
\usepackage{color}
\usepackage{amssymb}
\usepackage{amsmath}
\usepackage{epstopdf}
\usepackage{bbold}
\usepackage[normalem]{ulem}

\newcommand{\be}{\begin{eqnarray}}
\newcommand{\ee}{\end{eqnarray}}

\newcommand{\nn}{\nonumber}

\newcommand{\bfb}{{\bf b}_{\perp}}

\newcommand{\bfp}{{\bf p}_{\perp}}

\newcommand{\bfd}{{\bf \Delta}_{\perp}}

\newcommand{\bfD}{{\bf D}_{\perp}}

\begin{document}


\title{Gluon Wigner Distributions in boost-invariant longitudinal position space}

\author{Sonia } 
\affiliation{Department of Physics, National Institute of Technology Kurukshetra, India 136119}

\author{Tanmay Maji}
\email{tanmayphy@nitkkr.ac.in} 
\affiliation{Department of Physics, National Institute of Technology Kurukshetra, India 136119}

\author{Hemant Kumar}
\affiliation{Department of Physics, National Institute of Technology Kurukshetra, India 136119}

\date{\today}

\begin{abstract}
The Wigner distributions (WDs) in boost-invariant longitudinal space for unpolarized, longitudinally polarized and linearly polarized gluons in a proton are presented in the framework of the light-front gluon spectator model inspired by AdS/QCD. The boost-invariant longitudinal space defined by the coordinate $\sigma=\frac{1}{2}b^- P^+$,  can be accessed through the Fourier transformation over skewness $\xi$ to the gluon-gluon correlator of generalized transverse momentum dependent distributions (GTMDs). 
The model results for gluon WDs in $\sigma$-space show an oscillatory behavior analogous to the diffraction scattering of a wave in optics. The diffraction pattern is more sensitive to the momentum fraction $x$ and passively varies with the total energy transfer to the electron-proton scattering $-t$, both of which are analogous to an effective slit-width. The leading-twist gluon WDs in the impact-parameter space and their skewness sensitivity are investigated extensively for unpolarized, longitudinally polarized, and transversely polarized protons. The gluon spin-OAM correlation is also reported and compared with existing models and lattice results.  

\end{abstract}

\maketitle

\section{Introduction\label{intro}}

Hadrons are bound states of quarks and gluons held together by the strong force governed by Quantum Chromodynamics (QCD), and understanding their internal partonic structure is crucial. The deep inelastic scattering (DIS) is one of the key processes for investigating hadrons structure measuring parton distribution functions. The parton distribution functions (PDFs)~\cite{Collins:1981uw, Martin:1998sq, Gluck:1994uf, Gluck:1998xa} provide one-dimensional momentum distribution with longitudinal momentum fraction $x$. The single spin asymmetry(SSA) measured in semi-inclusive DIS experiments \cite{Anselmino:2007fs,Anselmino:2008jk,Anselmino:2013vqa,Anselmino:2005nn, Anselmino:2005ea, Anselmino:2008sga} indicate a significant importance of the three-dimensional transverse momentum distribution(TMDs)~\cite{Mulders:2000sh,Meissner:2007rx} which provides transverse momentum $\bfp$ distribution of constituent parton along with the longitudinal momentum fraction $x$ and give access to the spin-transverse momentum correlation e.g, Collins asymmetry, Sivers asymmetry in a hadron \cite{COLLINS1993161,Sivers:1989cc}.  The TMDs can be probed in semi-inclusive deep inelastic scattering (SIDIS) and the Drell-Yan process~\cite{Mulders:1995dh,Bacchetta:2006tn,Barone:2001sp,Brodsky:2002cx,Bacchetta:2017gcc}. While, the three-dimensional phase-space distribution of constituent partons is given by generalized parton distributions (GPDs)~\cite{Muller:1994ses,Ji:1996nm, Radyushkin:1997ki, Goeke:2001tz, Diehl:2003ny,Ji:2004gf, Belitsky:2005qn, Boffi:2007yc}, which can be probed through hard exclusive reactions like deeply virtual Compton scattering (DVCS) or deeply virtual meson production(DVMP)~\cite{Ji:1996nm,Diehl:2003ny,Belitsky:2005qn,Goeke:2001tz}. In recent time, investigation of the spin and momentum structure of proton has drawn more attention aiming towards the Electron-ion-collider(EIC) experiment \cite{AbdulKhalek:2021gbh,AbdulKhalek:2022hcn}. 

A more general six-dimensional momentum space distribution is the generalized transverse momentum distribution (GTMDs), which encodes complete information of the three momenta $(x, \bfp)$ of constituents as well as the momentum transferred $(\xi, \bfd)$ in the scattering process.
At twist-2, there are sixteen GTMDs for quark and gluon in the nucleon respectively \cite{Meissner:2009ww, Lorce:2013pza}, with different polarization combinations of proton and its constituents. 
They are characterized by revealing different spin-orbit and spin-spin correlations between the nucleon and a parton inside the nucleon, for example, $F_{1,4}$ and $G_{1,1}$ give important information about the canonical orbital angular momentum (OAM)~\cite{Lorce:2011kd,Hatta:2011ku,Ji:2012sj,Lorce:2012ce} and the spin-orbit correlations~\cite{Lorce:2014mxa,Tan:2021osk} of partons, respectively. In certain kinematical limits, GTMDs reduce to TMDs and GPDs. The quark GTMDs can be probed through the exclusive double Drell-Yan process~\cite{Bhattacharya:2017bvs}.  

The Fourier transformation with respective parameters gives different Wigner Distributions (WDs), which are a quasi-probabilistic interpretation and reveal spin-spin, spin-orbital momentum correlations. 
Fourier transform of GTMDs with respect to the skewness variable $\xi = -\Delta^+/(2P^+)$ and the transverse momentum $\bfD=\bfd/(1-\xi^2)$ provides access to the longitudinal coordinate space, defined by $\sigma = \frac{1}{2} b^- P^+$, and the impact parameter space $(\mathbf{b}_\perp)$ respectively. The Wigner distributions are the quantum-mechanical analogues of classical phase-space distributions, providing a unified description of both momentum and spatial information of partons inside hadrons, which is first introduced in Ji~\cite{Ji:2003ak}. The quark sector have since been extensively studied to explore the multidimensional imaging of hadronic structure. The GTMDs and WDs for spin-1/2 composite particles are calculated using different theoretical models, such as the light-cone constituent quark model~\cite{Lorce:2011ni,Lorce:2011kd,Lorce:2011dv}, the light-front dressed quark model~\cite{Mukherjee:2014nya,Mukherjee:2015aja,More:2017zqq}, the chiral soliton model~\cite{Lorce:2011ni,Lorce:2011kd}, light-cone spectator model~\cite{Liu:2015eqa}, the light-front quark-diquark model~\cite{Chakrabarti:2016yuw,Chakrabarti:2017teq,Chakrabarti:2019wjx,Gutsche:2016gcd,Kaur:2019lox,Kumar:2017xcm}, quark target model~\cite{Kanazawa:2014nha}.
While for gluon sector GTMDs and Wigner distribution study is remain limited by impact parameter space in few models calculation.  Ref.\cite{More:2017zqp} investigates gluon WDs in impact parameter space for unpolarized, longitudinally polarized and linearly polarized gluon in a target state consist of a quark dressed with a gluon at one loop~\cite{Mukherjee:2015aja,More:2017zqp,Jana:2023btd}.
Considering a two-body
composite system of a spin-$1$ gluon $g$ and a spin-$1/2$ three quark spectator, the impact parameter space gluon WDs are presented in the the light-front gluon triquark  model ~\cite{Tan:2024dmz} and spectator model \cite{Tan:2023vvi}.  The $\bfb$-space gluon WDs are also studied in the light-front gluon spectator model~\cite{Chakrabarti:2025qba, Sonia2025GTMD} for vanishing skewness ($\xi=0$). 
The model prediction to non vanishing skewness dependency is very important as experimental measurement includes skewness. 
 However, the WDs in the boost-invariant longitudinal space ($\sigma$-space) has drown special attention in recent time as it completes the three-dimension position space information. For the quark sector,  $\sigma$-space WDs shows an oscillatory behavior analogs to the diffraction pattern of single-slit diffraction in optics. The total energy transfer $-t$ and the longitudinal momentum fraction carried by the constituents $x$ behave like an effective slit-with, that shows an inverse behavior to the width of the central maxima~\cite{Maji:2022tog}. The analogy between optics and quantum field on the light cone was first explored in \cite{Sudarshan:1982gv, Mukunda:1982gu} where, a correspondence between paraxial wave optics and the light-cone dynamics of scalars are demonstrated. 
 As longitudinal momentum and light-front position are Fourier-conjugate variables, a variation in the longitudinal momentum of the gluon leads to a variation in its light-front position. The change in the longitudinal light-front coordinate of the gluon is described by the variable $\sigma=\frac{1}{2}b^- P^+$. 

In this work, the gluon sector of the $\sigma$-space WDs is presented for the first time. We use the framework of light-front gluon spectator model~\cite{Chakrabarti:2023djs}, and calculate the Wigner distribution in both the boost-invariant longitudinal space $\sigma$ and impact parameter space $\bfb$ for non-zero longitudinal momentum transferred ($\xi \neq 0$) for unpolarized, longitudinally polarized and linearly polarized (R and L) gluon for the three configuration of proton: unpolarized, longitudinally polarized and transversely polarized. The gluon spin-OAM correlation is also discussed in brief. 

 The experiments probe nonzero skewness and it is important to dive deeper into Wigner distribution at nonzero skewness~\cite{Maji:2022tog,Tan:2024dmz,Ojha:2022fls,Jana:2023btd}.
 The gluon GTMDs are measurable in diffractive dijet production in deep-inelastic lepton-nucleon and lepton-nucleus scattering~\cite{Hatta:2016dxp,Ji:2016jgn,Hatta:2016aoc,Bhattacharya:2022vvo} and ultraperipheral proton-nucleus collisions~\cite{Hagiwara:2017fye}, as well as in virtual photon-nucleus quasielastic scattering~\cite{Zhou:2016rnt}. Recently, the leading-twist gluon and sea-quark generalized transverse momentum distributions (GTMDs) in the small-$x$ is presented in the framework of eikonal, approximation at vanishing skewness $\xi =0$ \cite{Benic:2026idy}.

The paper is organized as follows: in Sec.-\ref{model} we briefly discuss the gluon spectator model, the Sec.-\ref{kin} includes the model results of GTMDs with analytical expressions, Sec.-\ref{sigma} covers the Wigner distribution in boost-invariant longitudinal space with detailed model results for all possible polarizations of the proton and their numerical plots. The Sec-\ref{b-space} contains the discussion on the variation of $\bfb$-space Wigner distribution with skewness in transverse momentum and impact parameter plane, and we conclude in Sec.-\ref{con}.

\section{Light-front gluon spectator model \label{model}}
The minimal Fock-state description of the proton consists solely of valence quarks. Contributions from gluons and sea quarks arise when higher Fock sectors are taken into account. In the Light-front gluon spectator model, the proton is described as a composite system comprising one active gluon and a spin-$\frac{1}{2}$ spectator \cite{Lu:2016vqu, Chakrabarti:2023djs} of effective mass $M_X$.
The kinematics of the proton with longitudinal four-momentum $P$ can be written as $ P=(P^+,\frac{M^2}{P^+},\textbf{0}_\perp)$, where transverse momentum of proton vanishes. Whereas, the active gluon has momentum $ p =(x P^+, \frac{p^2+\textbf{p}_\perp^2}{x P^+}, \textbf{p}_\perp)$ with longitudinal momentum fraction $x=p^{+}/P^{+}$ carried by the active gluon and the spectator system has momentum  $ P_X=((1-x) P^+, \frac{M^2_X + \bfp^2}{(1-x)P^+}, -\textbf{p}_\perp)$.
The proton state can be represented by spin projections as a two-particle Fock-state expansion $J_{z}=\pm\frac{1}{2}$~\cite{Brodsky:2000ii} as,
	\begin{eqnarray}\label{state}\nonumber
		|P;\uparrow(\downarrow)\rangle
		= \int \frac{\mathrm{d}^2 \bfp \mathrm{d} x}{16 \pi^3 \sqrt{x(1-x)}}\times \Bigg[\psi_{+1+\frac{1}{2}}^{\uparrow(\downarrow)}\left(x, \bfp\right)\left|+1,+\frac{1}{2} ; x P^{+}, \bfp\right\rangle+\psi_{+1-\frac{1}{2}}^{\uparrow(\downarrow)}\left(x, \bfp \right)\left|+1,-\frac{1}{2} ; x P^{+}, \bfp \right\rangle\\ 
		+\psi_{-1+\frac{1}{2}}^{\uparrow(\downarrow)}\left(x, \bfp \right)\left|-1,+\frac{1}{2} ; x P^{+}, \bfp \right\rangle+\psi_{-1-\frac{1}{2}}^{\uparrow(\downarrow)}\left(x, \bfp\right)\left|-1,-\frac{1}{2} ; x P^{+}, \bfp\right\rangle\bigg],
        \label{proton state}
	\end{eqnarray}
where $|\lambda_{g},\lambda_{X};xP^{+},\bfp \rangle$ denotes the two-particle state having light-front wave function $\psi_{\lambda_{g}\lambda_{X}}^{\lambda_N}(x,\bfp)$ with
the helicity components of the active gluon $\lambda_{g}$, spectator $\lambda_{X}$, and  proton helicity $\lambda_N=\uparrow(\downarrow)$. 
The light-front wave functions in Eq.~(\ref{proton state}) is inspired by the well-known structure of the physical electron wave function~\cite{Brodsky:2000ii}, which consists of a spin-1 photon and a spin-$\tfrac{1}{2}$ electron. The light-front wave functions for the Fock-state expansion of a proton with spin projection $J_z = +\tfrac{1}{2}$ can be written in the following form:
	\begin{eqnarray} \label{LFWFsuparrow}   \nonumber
		\psi_{+1+\frac{1}{2}}^{\uparrow}\left(x,\bfp\right)&=&-\sqrt{2}\frac{(-p^{1}_{\perp}+ip^{2}_{\perp})}{x(1-x)}\varphi(x,\bfp^2), \\ \nonumber
		\psi_{+1-\frac{1}{2}}^{\uparrow}\left(x, \bfp\right)&=&-\sqrt{2}\bigg( M-\frac{M_{X}}{(1-x)} \bigg) \varphi(x,\bfp^2), \\ \nonumber
		\psi_{-1+\frac{1}{2}}^{\uparrow}\left(x, \bfp\right)&=&-\sqrt{2}\frac{(p^{1}_{\perp}+ip^{2}_{\perp})}{x}\varphi(x,\bfp^2), \\
		\psi_{-1-\frac{1}{2}}^{\uparrow}\left(x, \bfp\right)&=&0,
	\end{eqnarray}\\
likewise, the proton with $J_{z}=-1/2$ has light-front wave functions\\
	\begin{eqnarray} \label{LFWFsdownarrow}   \nonumber
		\psi_{+1+\frac{1}{2}}^{\downarrow}\left(x, \bfp\right)&=& 0, \\ \nonumber
		\psi_{+1-\frac{1}{2}}^{\downarrow}\left(x,\bfp\right)&=&-\sqrt{2}\frac{(-p^{1}_{\perp}+ip^{2}_{\perp})}{x}\varphi(x,\bfp^2), \\ \nonumber
		\psi_{-1+\frac{1}{2}}^{\downarrow}\left(x, \bfp\right)&=&-\sqrt{2}\bigg( M-\frac{M_{X}}{(1-x)} \bigg) \varphi(x,\bfp^2),  \\
		\psi_{-1-\frac{1}{2}}^{\downarrow}\left(x, \bfp \right)&=& -\sqrt{2}\frac{(p^{1}_{\perp}+ip^{2}_{\perp})}{x(1-x)}\varphi(x,\bfp^2),
	\end{eqnarray}
where $M$ is the proton mass and $M_X$ is the spectator mass.  $\varphi(x,\bfp^2)$ represents the modified form of the soft-wall AdS/QCD wave function~\cite{Gutsche:2013zia} shaped by introducing the parameters $a$ and $b$ of Eq.(\ref{AdSphi}).
The asymptotes of the parton distribution functions for the small $x$ region are based on the Regge trajectory predicted from high-energy scattering processes. For the large $x$ region, the constraints are imposed from the power-counting rules for hard scattering ~\cite{Brodsky:1994kg,Brodsky:1989db,Brodsky:1994kg}. The soft-wall AdS/QCD wave function $\phi(x,\boldsymbol{p}_\perp^{\,2})$ is given by
    \be\label{AdSphi}
	\varphi(x,\bfp^2)=N_{g}\frac{4\pi}{\kappa}\sqrt{\frac{\log[1/(1-x)]}{x}}x^{b}(1-x)^{a}\exp{\bigg[-\frac{\log[1/(1-x)]}{2\kappa^{2}x^2}\bfp^{2}\bigg]},
	\ee	
where $a$ and $b$ represents our model parameters. The model parameters $a$ and $b$, along with the normalization constant $N_g$, are determined by fitting the gluon unpolarized PDF at the scale $Q_0 = 2~\mathrm{GeV}$ with NNPDF3.0 data \cite{Chakrabarti:2023djs}. The stable bound state property of proton restricts the lower-bound of the spectator mass $M_{X}$, i.e., $M_X>M$ and in this model $M_{X}=0.985$ GeV is considered. The sensitivity of spectator mass on the gluon distributions is discussed in \cite{Chakrabarti:2023djs}.

\section{GTMDs with non-zero skewness in LFGSM\label{kin}}
 We consider the DIS process where an electron is scattered by target proton of initial momentum $P'$ and final momentum $P''$. The Energy transfer to the process is denoted by $-t$, via the exchange of a virtual photon carrying a four-momentum $q$, with virtuality $q^2=-Q^2$. The average momentum of proton is $P= \frac{1}{2} (P^{\prime\prime}+P^\prime)$ with total momentum transfer $\Delta=(P^{\prime\prime}-P^\prime)$ and $t= \Delta^2$. 
 In this section we present a detailed calculation of the GTMDs within the Light Front Gluon Spectator Model (LFGSM)~\cite{Chakrabarti:2023djs}. 
At fixed light-cone time $z^+=0$, the gluon-gluon correlator for GTMDs is defined as \cite{Lorce:2013pza}
\be
&W^{[\Gamma^{ij}]}_{\lambda^{\prime}\lambda^{\prime\prime}}(x,\xi,\bfd, \bfp)=\int \frac{dz^-d^2z_\perp}{xP^+(2\pi)^3} e^{ip.z}
&\times\langle P^{\prime\prime}; \lambda^{\prime\prime} |\Gamma^{ij} G^{+i}(-z/2) \mathcal{W}_{[-z/2,z/2]} G^{+j}(z/2) |P^\prime;\lambda^{\prime}\rangle \bigg|_{z^+=0},
\label{Wdef}
 \ee
 where, $|P^\prime;\lambda^{\prime}\rangle$ and $|P^{\prime\prime};\lambda^{\prime\prime}\rangle$ denote the initial and final proton states with helicities $\lambda'$ and $\lambda''$, respectively, and $G^{+i}$ ($G^{+j}$) represents the gluon fields. The matrix $\Gamma^{ij}$ corresponds to the leading-twist Dirac structures,
 
\begin{equation}
\Gamma^{ij} = \left\{ \delta_\perp^{ij}, -i\epsilon_\perp^{ij}  \right\},
\end{equation}
which describe unpolarized and longitudinally polarized gluons, respectively. Under light-front gauge $A^+=0$, Wilson line $\mathcal W_{[-z/2,z/2]}$ becomes unity. Throughout this work, we adopt the convention $x^\pm=(x^0 \pm x^3)$ , and the relevant kinematics are given by
\be 
P &\equiv& \bigg(P^+,\frac{M^2+\bfd^2/4}{(1-\xi^2)P^+},\textbf{0}_\perp\bigg)\,,\\
p &\equiv& \bigg(xP^+, p^-,\bfp \bigg)\,,\\
\Delta &\equiv& \left(-2\xi P^+, \frac{t + \bfd^2}{-2 \xi P^+},\bfd \right)\,.
\ee
We choose the symmetric frame. In symmetric frame, the initial and final four momenta of the proton are given by
\be
P^{\prime} &\equiv& \bigg((1+\xi)P^+,\frac{M^2+\bfd^2/4}{(1+\xi)P^+},-\bfd/2\bigg)\, ,\label{Pp}\\
P^{\prime\prime} &\equiv& \bigg((1-\xi)P^+,\frac{M^2+\bfd^2/4}{(1-\xi)P^+},\bfd/2\bigg), \label{Ppp}
\ee
where, the skewness $\xi$ is defined as $\xi=- \Delta^+/2P^+$.
The transverse momentum of active gluon in the initial and final state is given by
\be
\bfp' = \bfp - (1-x')\frac{\bfd}{2}, \hspace{0.4cm}
\bfp'' = \bfp + (1-x'')\frac{\bfd}{2},
\ee
with
$ x'=\frac{x+\xi}{1+\xi}, $ and $
x''=\frac{x-\xi}{1-\xi}$.
Using $\Delta^-=( P^{\prime \prime -} - P^{\prime  -})$, relation between total energy transferred and the momentum transferred in longitudinal and transverse direction can be derived as
 \be
- t= \frac{4 \xi^2 M^2 + \bfd^2}{(1-\xi^2)}. \label{mt_def}
\ee 
The bilinear decomposition of the fully unintegrated gluon–gluon correlator, Eq. (\ref{Wdef}), is expressed in terms of leading-twist GTMDs \cite{Stephan} as 
\begin{align}
	W^{[\delta_\perp^{ij}]}_{\lambda^\prime\lambda''}
	=&\frac{1}{2M}\bar{u}(\textbf{P}'',\lambda'')\bigg[F_{1,1}^g+\frac{i\sigma^{i+}\bfp^i}{P^+}F^g_{1,2} +\frac{i\sigma^{i+}\bfd^i}{P^+}F^g_{1,3}+\frac{i\sigma^{ij}\bfp^i\Delta^j_\perp}{M^2}F^g_{1,4}\bigg]u(\textbf{P}^\prime,\lambda^\prime),
	\label{eq:F-GTMDsDef}\\
 W^{[-i\epsilon_\perp^{ij}]}_{\lambda^\prime\lambda''}
	=&\frac{1}{2M}\bar{u}(\textbf{P}'',\lambda'')\bigg[-\frac{i\epsilon_\perp^{ij}\bfp^i\bfd^j}{M^2} G_{1,1}^g+\frac{i\sigma^{i+}\gamma_5\bfp^i}{P^+}G^g_{1,2} +\frac{i\sigma^{i+}\gamma_5 \bfd^i}{P^+}G^g_{1,3}+i\sigma^{+-}\gamma_5 G^g_{1,4}\bigg]u(\textbf{P}^\prime,\lambda^\prime).
    \label{eq:G-GTMDsDef}
\end{align}
The explicit form of the spinors $u(k,\lambda) $ with the momentum $k$ and the helicity $\lambda\,(=\pm )$ are given in Appendix-\ref{AppA}. 
In this model, the correlator $W^{[\Gamma^{ij}]}_{\lambda^{\prime}\lambda^{\prime\prime}}$, can be written in terms of overlaps of the light-front wave functions (LFWFs), given in Eqs.(\ref{LFWFsuparrow}), and (\ref{LFWFsdownarrow}) and reads for the unpolarized and longitudinally polarized gluon as
\begin{align}\label{eq:correlator1}
    W^{[\delta_\perp^{ij}]}_{\lambda^{\prime} \lambda^{\prime\prime}}(x,\xi,\bfd,\bfp)=&\frac{1}{16\pi^{3}}\sum_{\lambda_{g},\lambda_{X}}\psi^{\lambda^{\prime\prime\dagger}}_{\lambda_{g}\lambda_{X}}(x^{\prime\prime},\bfp^{\prime\prime})\psi^{\lambda^{\prime}}_{\lambda_{g}\lambda_{X}}(x^{\prime},\bfp^{\prime})(\varepsilon^{1}_{\lambda_{g}}\varepsilon^{1\ast}_{\lambda_{g}}+\varepsilon^{2}_{\lambda_{g}}\varepsilon^{2\ast}_{\lambda_{g}}),\\
    {W}^{[-i\epsilon_\perp^{ij}]}_{\lambda^{\prime}\lambda^{\prime\prime}}(x,\xi,\bfd,\bfp)=&-\frac{i}{16\pi^{3}}\sum_{\lambda_{g},\lambda_{X}}\psi^{\lambda^{\prime\prime\dagger}}_{\lambda_{g}\lambda_{X}}(x^{\prime\prime},\bfp^{\prime\prime})\psi^{\lambda^{\prime}}_{\lambda_{g}\lambda_{X}}(x^{\prime},\bfp^{\prime})(\varepsilon^{1}_{\lambda_{g}}\varepsilon^{2\ast}_{\lambda_{g}}-\varepsilon^{2}_{\lambda_{g}}\varepsilon^{1\ast}_{\lambda_{g}}),
    \label{eq:correlator2}
\end{align}
 respectively. While, the right-handed ($R$) and left-handed ($L$) linearly polarized gluon GTMDs correlators read as
\begin{align}
 W^{R}_{\lambda^{\prime}\lambda''}(x,\xi,\bfd,\bfp)=&\frac{1}{16\pi^{3}}\sum_{\lambda_{g},\lambda_{X}}\psi^{\lambda^{\prime\prime\dagger}}_{\lambda_{g}\lambda_{X}}(x^{\prime\prime},\bfp^{\prime\prime})\psi^{\lambda^{\prime}}_{\lambda_{g}\lambda_{X}}(x^{\prime},\bfp^{\prime})\varepsilon^{R}_{\lambda_{g}}\varepsilon^{R\ast}_{\lambda_{g}},
 \label{eq:correlator3}\\
 W^{L}_{\lambda^{\prime}\lambda^{\prime\prime}}(x,\xi,\bfd,\bfp)=&\frac{1}{16\pi^{3}}\sum_{\lambda_{g},\lambda_{X}}\psi^{\lambda^{\prime\prime\dagger}}_{\lambda_{g}\lambda_{X}}(x^{\prime\prime},\bfp^{\prime\prime})\psi^{\lambda^{\prime}}_{\lambda_{g}\lambda_{X}}(x^{\prime},\bfp^{\prime})\varepsilon^{L}_{\lambda_{g}}\varepsilon^{L\ast}_{\lambda_{g}},
\label{eq:correlator4}
\end{align}
with $\varepsilon^{R(L)}_{\lambda_g}=\varepsilon^{1}_{\lambda_g}\pm i\varepsilon^{2}_{\lambda_g}$ are the right and left-handed gluon polarization vectors, respectively.
 Using the LFWFs in Eq.(\ref{eq:correlator1})$-$(\ref{eq:correlator4}) and the relations given in Eqs.(\ref{eq:F-GTMDs}), (\ref{eq:G-GTMDs}), we  parametrize the correlators and obtain leading-twist GTMDs for gluon. The explicit analytical form of the gluon GTMDs in this model, are listed corresponding to the different polarization as\\
$(i)$ for unpolarized gluons with Dirac matrix structure $\Gamma^{ij}=\delta_{\perp}^{ij}$
 \begin{align}
F^g_{1,1}=& \frac{2}{16\pi^3}\sqrt{1-\xi^2}\Bigg[N_1(x')N_2(x'')+B(x',x'')\left(\bfp^2-\frac{1}{2}((1-x')-(1-x'')) \bfp .\bfd-\frac{1}{4}(1-x')(1-x'') \bfd^2\right)\Bigg] \nonumber \\
& \hspace{12cm} \times\phi(x',\bfp'^2)\phi(x'',\bfp''^2), \label{eq:F11} \\
 F^g_{1,2}=& \frac{2M}{16\pi^3}\frac{1}{\sqrt{1-\xi^2}}\left(\frac{1}{x'}N_2(x'')\frac{1}{x''}N_1(x')\right)\phi(x',\bfp'^2)\phi(x'',\bfp''^2) 
  -\frac{\Delta^2_\perp}{2M^2}\frac{\xi}{(1-\xi^2)} F^g_{1,4},\\
F^g_{1,3}=& \frac{-M}{16\pi^3}\frac{1}{\sqrt{1-\xi^2}}\Bigg[\frac{1-x'}{x'}N_2(x'')+\frac{1-x''}{x''}N_1(x')\Bigg]\phi(x',\bfp'^2)\phi(x'',\bfp''^2)+\frac{1}{2(1-\xi^2)}F^g_{1,1} 
+\frac{\xi}{2(1-\xi^2)}\frac{\bfp.\bfd}{M^2}F^g_{1,4},\\
F^g_{1,4}=&\frac{M^2}{16\pi^3}\sqrt{1-\xi^2}B(x',x'')((1-x')+(1-x''))\phi(x',\bfp'^2)\phi(x'',\bfp''^2). \label{eq:F14} 
\end{align}
$(ii)$ for longitudinally polarized gluons with Dirac matrix structure $\Gamma^{ij}=-i\epsilon_{\perp}^{ij}$
\begin{align}
    G^g_{1,1}=&-\frac{M^2}{16\pi^3}\sqrt{1-\xi^2}B(x',x'')((1-x')+(1-x''))\phi(x',\bfp'^2)\phi(x'',\bfp''^2), \label{eq:G11}  \\
G^g_{1,2}=& \frac{-2M}{16\pi^3}\frac{1}{\sqrt{1-\xi^2}}\Bigg[\frac{1}{x'}N_2(x'')+\frac{1}{x''}N_1(x')\Bigg]\phi(x',\bfp'^2)\phi(x'',\bfp''^2)
  +\frac{\Delta^2_\perp}{2M^2}\frac{\xi}{(1-\xi^2)} G^g_{1,1}, \\ 
G^g_{1,3}=&\frac{M}{16\pi^3}\frac{1}{\sqrt{1-\xi^2}}\Bigg[\frac{(1-x')}{x'}N_2(x'')-\frac{(1-x'')}{x''}N_1(x')\Bigg]\phi(x',\bfp'^2)\phi(x'',\bfp''^2)
-\frac{\bfp.\bfd}{2M^2(1-\xi^2)}G^g_{1,1}+\frac{\xi}{2(1-\xi^2)}G^g_{1,4},   \\
G^g_{1,4}=& \frac{2}{16\pi^3}\sqrt{1-\xi^2}\Bigg[N_1(x')N_2(x'')+C(x',x'')\left(\bfp^2-\frac{1}{2}((1-x')-(1-x'')) \bfp .\bfd-\frac{1}{4}(1-x')(1-x'')\bfd^2\right)\Bigg] \nonumber \\
& \hspace{12cm} \times\phi(x',\bfp'^2)\phi(x'',\bfp''^2), \label{eq:G14} 
\end{align}
where,

\begin{eqnarray}
&\phi(x',\bfp'^2)\phi(x'',\bfp''^2)=& \hspace{-1.5cm} A(x^{\prime})A(x^{\prime\prime})\exp\Big(-a(x^{\prime})\bfp^{\prime 2}-a(x^{\prime\prime})\bfp^{\prime\prime 2}\Big) ,\\  
& a(x^\prime)=\frac{1}{2\kappa^2(x^\prime)^2}log\Big(\frac{1}{1-x^\prime}\Big),\hspace{0.2cm}
 & a(x^{\prime\prime})=\frac{1}{2\kappa^2(x^{\prime\prime})^2}log\Big(\frac{1}{1-x^{\prime\prime}}\Big), \\
& A(x^{\prime})=N_g\frac{4\pi}{\kappa}\sqrt{\frac{log\Big(\frac{1}{1-x^{\prime}}\Big)}{x^{\prime}}}(x^{\prime})^b(1-x^{\prime})^a,\hspace{0.2cm}
& A(x^{\prime\prime})=N_g\frac{4\pi}{\kappa}\sqrt{\frac{log\Big(\frac{1}{1-x^{\prime\prime}}\Big)}{x^{\prime\prime}}}(x^{\prime\prime})^b(1-x^{\prime\prime})^a,\\
 & B(x',x'')= \frac{1+(1-x')(1-x'')}{x'x''(1-x')(1-x'')},\hspace{0.2cm}
 & C(x',x'')=\frac{1-(1-x')(1-x'')}{x'x''(1-x')(1-x'')},\\
& N_1(x')=\left(M-\frac{M_X}{1-x'}\right),\hspace{0.2cm}
 &N_2(x'')=\left(M-\frac{M_X}{1-x''}\right).
\end{eqnarray}
All the GTMDs are function of $(x,\xi,\bfd^2,\bfp^2,\bfd.\bfp)$ and Fourier transform over $\xi$ and $\bfD$ provides Wigner distribution in the boost-invariant longitudinal position space $\sigma$ and  transverse impact parameter $\bfb$. These analytical expressions of GTMDs of Eqs.(\ref{eq:F11}-\ref{eq:G14}) are used for further numerical investigations of WDs. 
\section{Wigner Distributions in $\sigma$-space \label{sigma}}
The boost invariant longitudinal space $\sigma$, defined as $\sigma=\frac{1}{2}b^-P^+$, is Fourier conjugate to the skewness $\xi$ and Fourier transform of correlator $W^{[\Gamma^{ij}]}_{\lambda'\lambda''}(x,\xi,\bfd,\bfp)$ with respect to $\xi$ provides WDs in $\sigma$-space. 
Here, we concentrate on the gluon sector, and the Wigner distribution for gluons in boost invariant longitudinal ($\sigma$) space is defined as 
\begin{align}
 \rho^{[\Gamma^{ij}]}(x,\sigma,\bfd,\bfp)&=\int_{0}^{\xi_{max}}\frac{d\xi}{2\pi}e^{i\sigma\cdot\xi}W^{[\Gamma^{ij}]}_{\lambda'\lambda''}(x,\xi,\bfd,\bfp),
 \label{rs_def}
\end{align}
 where the  upper limit of the integration is restricted by the energy transfer $t$ to the system as  
         \begin{align}
             \xi_{max}=\frac{-t}{2M^2}\Big(\sqrt{1+\frac{4M^2}{-t}}-1\Big),
          \label{zmax_def}
          \end{align}
         \text{and}
         \begin{align}
             -t=\frac{4\xi^2M^2+\Delta^2_\perp}{1-\xi^2} .
             \label{rel_t_Del}
        \end{align} 
Depending on the different polarization states of the proton and gluon, we define the phase-space distribution $\rho_{XY}$, with $X$ indicating the proton's polarization and $Y$ indicating the gluon's polarization. 
\begin{figure}[h]
    \centering
    \subfigure[]{\includegraphics[width=0.35\linewidth, trim=80 240 80 240, clip]{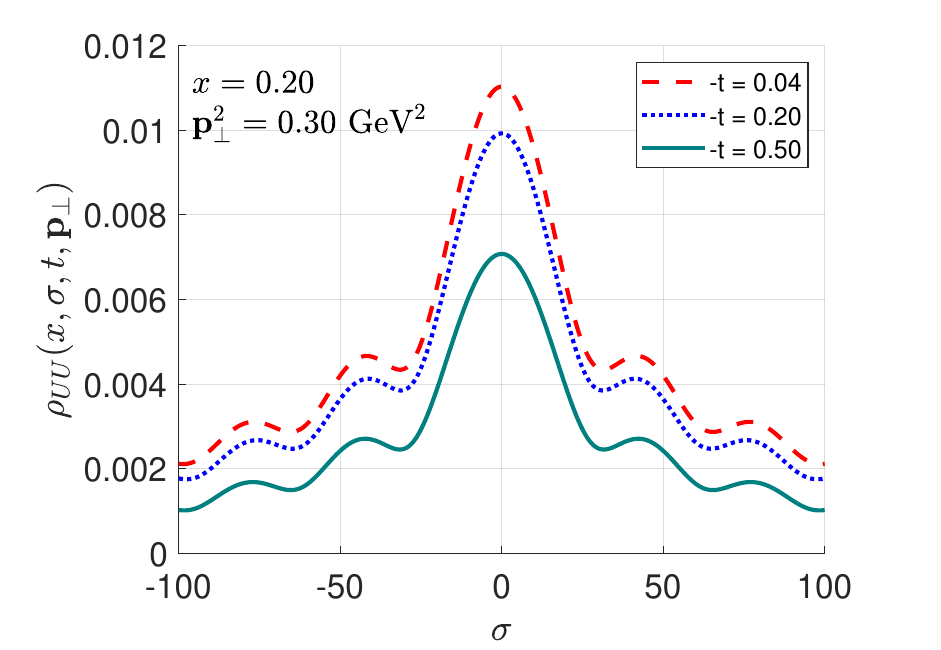}}
    \subfigure[]{\includegraphics[width=0.35\linewidth, trim=80 240 80 240, clip]{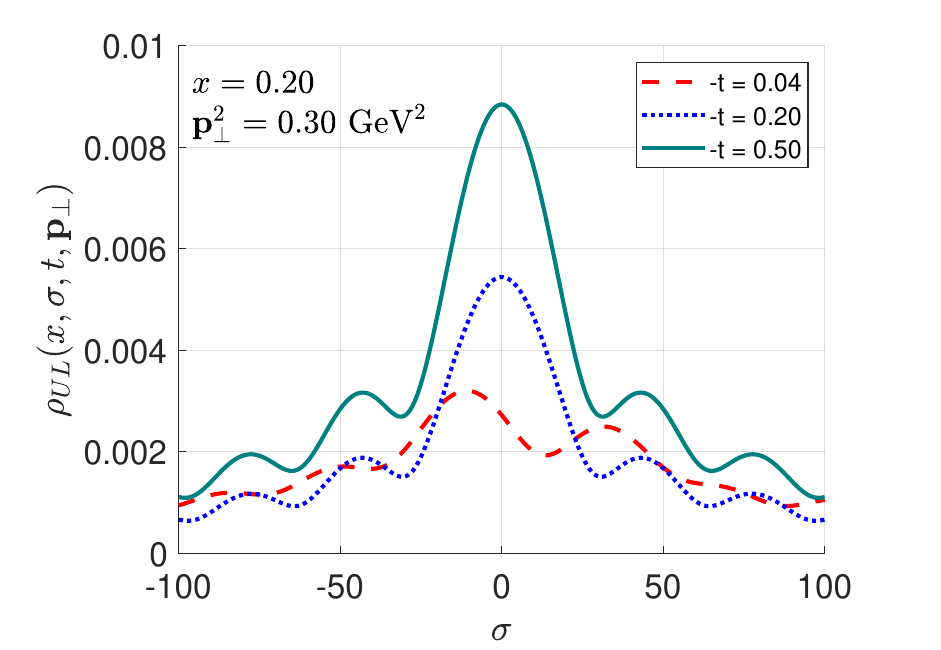}}
    \subfigure[]{\includegraphics[width=0.35\linewidth, trim=80 240 80 240, clip]{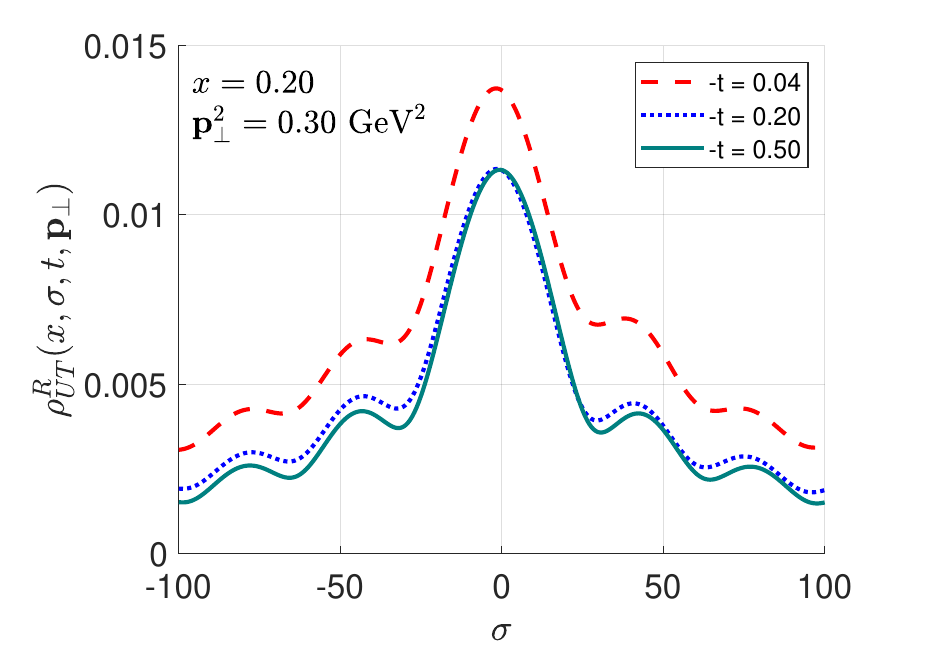}}
    \subfigure[]{\includegraphics[width=0.35\linewidth, trim=80 240 80 240, clip]{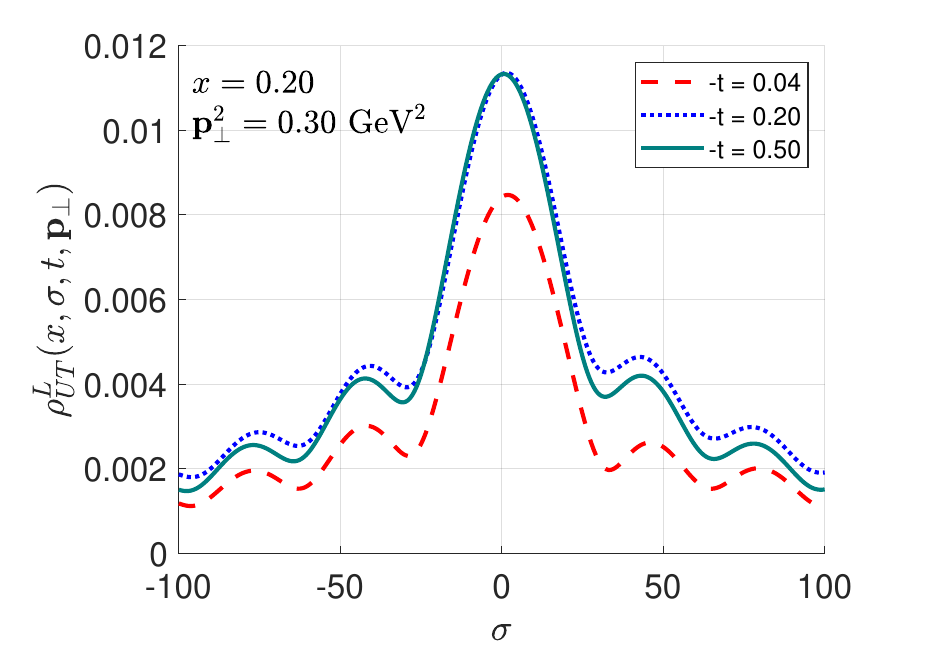}}
    \caption{The Wigner distribution $\rho^Z_{UY}$, (Y = U, L, T; Z = R, L) for unpolarized proton in the boost invariant longitudinal position space $\sigma$ for different values of $-t$ in GeV$^2$ at fixed $x=0.2$, $\bfp^2=0.3$ GeV$^2$ and $\bfd\perp\bfp$.}
    \label{fig:rho_UU}
\end{figure}
\subsection{Unpolarized proton}
The Wigner distribution for an unpolarized and longitudinally polarized gluon in an unpolarized proton can be expressed as a linear combination of the gluon correlators $W^{[\Gamma^{ij}]}_{++}$ and $W^{[\Gamma^{ij}]}_{--}$. These distributions are defined as
\begin{align}
 \rho_{UU}(x,\sigma,\bfd,\bfp)&=\int_{0}^{\xi_{s}}\frac{d\xi}{2\pi}e^{i\sigma\cdot\xi}\frac{1}{2}\Big[W^{1}_{++}(x,\xi,\bfd,\bfp)+W^{1}_{--}(x,\xi,\bfd,\bfp)\Big],
 \label{def_UU}\\
 \rho_{UL}(x,\sigma,\bfd,\bfp)&=\int_{0}^{\xi_{s}}\frac{d\xi}{2\pi}e^{i\sigma\cdot\xi}\frac{1}{2}\Big[W^{2}_{++}(x,\xi,\bfd,\bfp)+W^{2}_{--}(x,\xi,\bfd,\bfp)\Big],
 \label{def_UL}
 \end{align}
 where, super script $1$ and $2$ corresponds to Dirac structures $\delta^{ij}_{\perp}$ and $-i\epsilon^{ij}_{\perp}$.
 The Wigner distributions corresponding to linearly polarized gluons (R and L) in an unpolarized proton can be constructed from linear combinations of the unpolarized and longitudinally polarized gluon distributions as \cite{Lorce:2013pza} 
 \begin{eqnarray}
    \rho^{R}_{UT}(x,\sigma,\bfd,\bfp)=&\rho_{UU}(x,\sigma,\bfd,\bfp)-\rho_{UL}(x,\sigma,\bfd,\bfp),
     \label{def_UTR} \\
    \rho^{L}_{UT}(x,\sigma,\bfd,\bfp)=&\rho_{UU}(x,\sigma,\bfd,\bfp)+\rho_{UL}(x,\sigma,\bfd,\bfp).
    \label{def_UTL}
 \end{eqnarray}
Furthermore, the Wigner distributions $\rho_{UU}$ and $\rho_{UL}$ can be expressed in terms of the leading-twist GTMDs through a one-to-one correspondence using Eq.~(\ref{eq:F-GTMDs}) as
\begin{align}
               \rho_{UU}(x,\sigma, \bfd,\bfp)&=\int_{0}^{\xi_{s}}\frac{d\xi}{2\pi}e^{i\sigma\cdot\xi}\frac{1}{\sqrt{1-\xi^2}}F^g_{1,1},
               \label{rs_UU}\\
 \rho_{UL}(x,\sigma,\bfd,\bfp)&=\int_{0}^{\xi_{s}}\frac{d\xi}{2\pi}e^{i\sigma\cdot\xi}\frac{-i}{M^2\sqrt{1-\xi^2}}\epsilon^{ij}_\perp \bfp^i\bfd^j G^g_{1,1}.
 \label{rs_UL}
\end{align}
The analytical results in Eqs.~(\ref{def_UU})-(\ref{rs_UL}) are used for further numerical computation. All the WDs are functions of ${\rho}_{XY}(x,\sigma,\bfd,\bfp)$, however, using the relation between the transverse momentum transfer $\bfd$ and the invariant momentum transfer square $-t$, given in Eq.~(\ref{rel_t_Del}), these distributions can equivalently be expressed as ${\rho}_{XY}(x,\sigma, t,\bfp)$. 
\begin{figure}[h]
    \centering
    \subfigure[]{\includegraphics[width=0.35\linewidth, trim=80 240 80 240, clip]{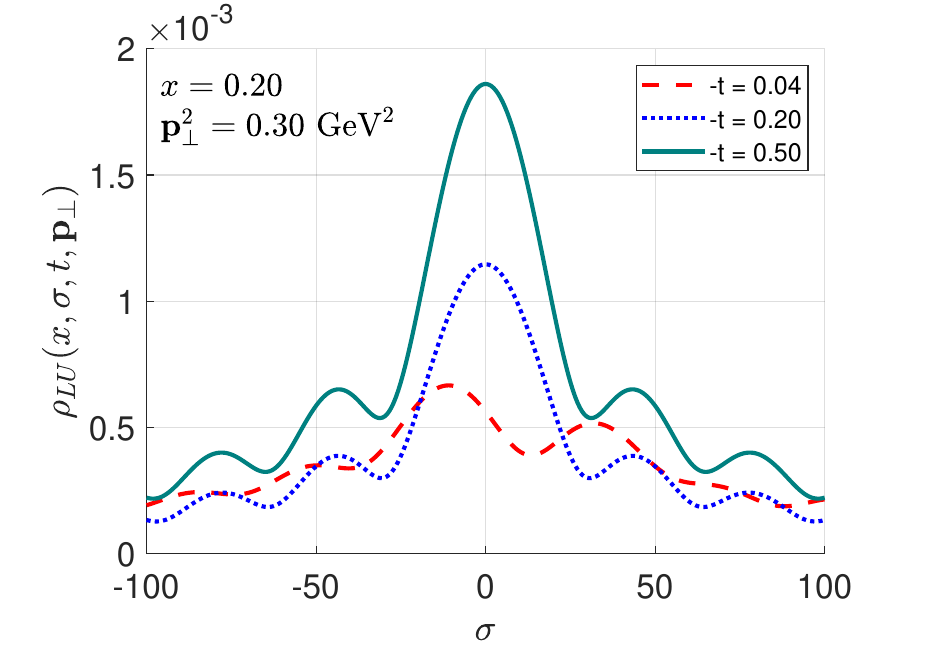}}
    \subfigure[]{\includegraphics[width=0.35\linewidth, trim=80 240 80 240, clip]{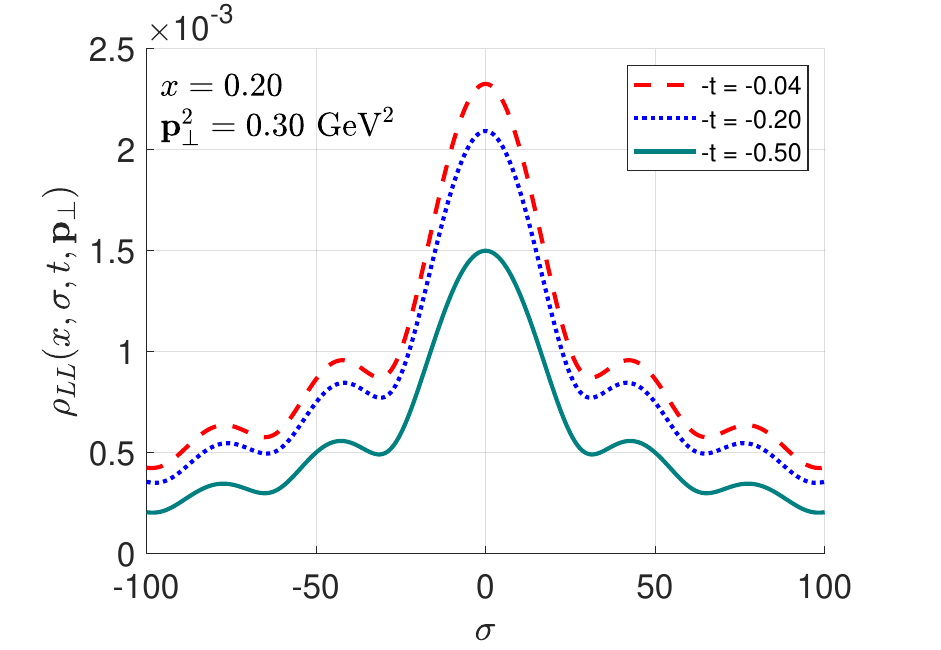}}
    \subfigure[]{\includegraphics[width=0.35\linewidth, trim=80 240 80 240, clip]{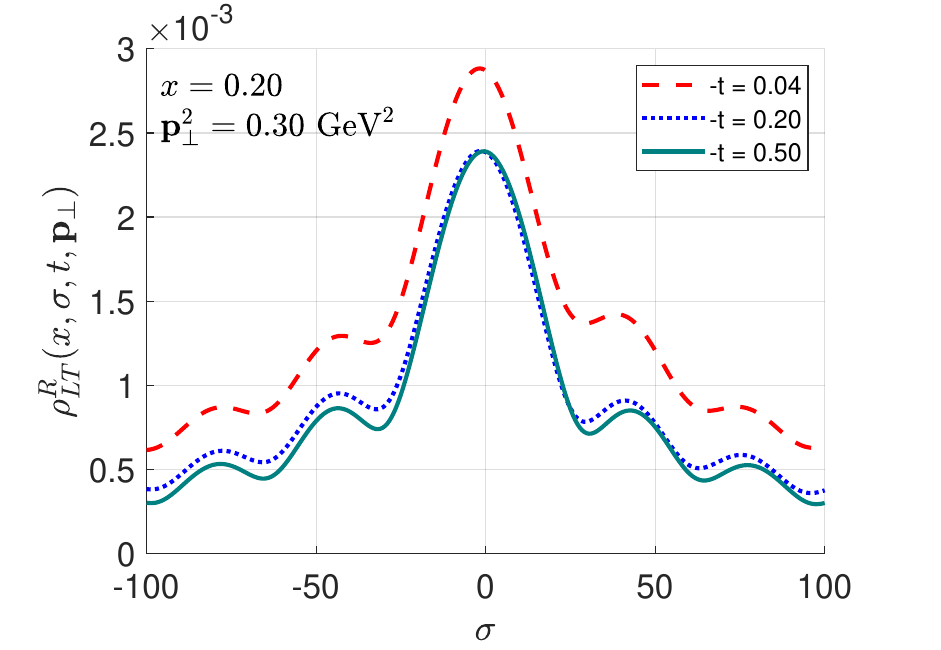}}
    \subfigure[]{\includegraphics[width=0.35\linewidth, trim=80 240 80 240, clip]{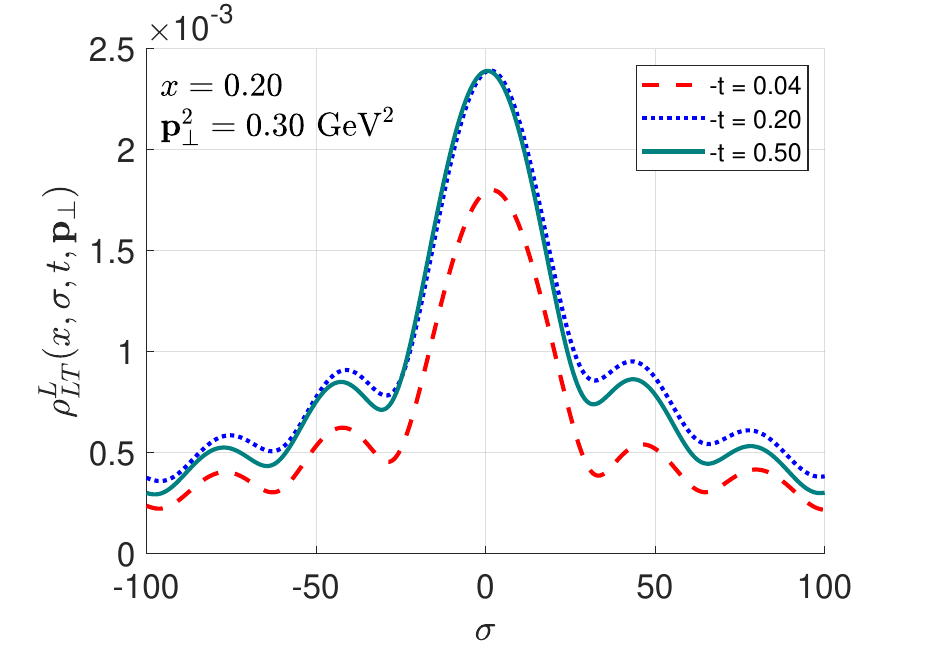}}
    \caption{The Wigner distribution $\rho^Z_{LY}$, (Y = U, L, T; Z = R, L) for longitudinally polarized proton in the boost invariant longitudinal position space $\sigma$ for different values of $-t$ in GeV$^2$ at fixed $x=0.2$, $\bfp^2=0.3$ GeV$^2$ and $\bfd\perp\bfp$.}
    \label{fig:rho_LL}
\end{figure}

For the unpolarized proton, numerical results of gluon Wigner distributions $\rho_{UY}$, $(Y = U, L, T)$ in $\sigma$-space are shown in the Fig.~\ref{fig:rho_UU} at fixed values of $x=0.2$, and $\bfp^2=0.3~\text{GeV}^2$.  The Wigner distribution is, in general, a complex-valued function; the vertical axis in Fig.~\ref{fig:rho_UU} denotes its absolute value. The three plots in each figure are for different values of $-t=\{0.04, 0.20, 0.50\}$ GeV$^2$, represented by red, blue, and teal colors, respectively, and the corresponding $\xi_{max}= \{0.1913, 0.3759, 0.5208\}$. For nonzero skewness, the DGLAP region $\xi < x < 1$ restricts the longitudinal coordinate to the interval $0 \le \xi \le \xi_s$, with
\begin{equation}
\xi_s =
\begin{cases}
\xi_{\rm max}, & \text{if } \xi_{\rm max} < x, \\
x, & \text{if } \xi_{\rm max} > x.
\end{cases}
\end{equation}
Where, $\xi_{\rm max}$ is determined by the momentum transfer $-t$ through Eq.~(\ref{zmax_def}). Accordingly, the upper limit of $\xi$ integration in Eq.~(\ref{rs_def}) is $\xi_s = 0.1913$ for $-t = 0.04~\text{GeV}^2$, while for $-t = 0.20~\text{GeV}^2$ and $-t = 0.50~\text{GeV}^2$, the condition $\xi_{\rm max} > x$ leads to $\xi_s = x = 0.2$. All these four distributions involve GTMDs $F_{1,1}$, $G_{1,1}$. We chose $\bfd $ and $ \bfp$ to be mutually perpendicular, which reduces the contribution from the terms containing $\bfd.\bfp$. 
We observe that the distribution $\rho_{UU}$ shows an oscillatory and symmetric behavior about $\sigma=0$ that can be interpreted as a diffraction-like pattern, analogous to single-slit diffraction in optics. This optical analogue is based on the behavior of the Fourier transform of the GTMDs in the longitudinal light-front coordinate $\sigma$. 
In optical diffraction, the finite width of a slit is a necessary condition for the formation of a diffraction pattern. Analogously, in Eq.~(\ref{rs_def}), the integration on $\xi$ is performed over the finite range ($0 \le \xi \le \xi_{max}$), that limited domain governed by $-t$  effectively acts as a slit of finite width. This provides the essential condition for the emergence of a diffraction pattern in the Fourier transform of GTMDs.
Similar diffraction patterns in longitudinal position space have been reported in DVCS amplitudes~\cite{Brodsky:2006in,Brodsky:2006ku}, GPDs~\cite{Chakrabarti:2008mw,Manohar:2010zm,Kumar:2015fta,Mondal:2015uha,Chakrabarti:2015ama,Mondal:2017wbf,Kaur:2018ewq}, and coordinate-space parton densities~\cite{Miller:2019ysh}, and thus the emergence of such structures of WDs in $\sigma$-space is not unexpected. In Fig.~\ref{fig:rho_UU}(a), we also observe that the height of the central maximum decreases with increasing values of $(-t)$, accompanied by a gradual shift of the distributions toward lower values along the $y-$axes. A similar behavior is observed in light-front dressed quark model \cite{Jana_2025}. 
Fig.~\ref{fig:rho_UU}(b) shows the Wigner distribution of longitudinally polarized gluons in an unpolarized proton, denoted by $\rho_{UL}$. The same color coding for the total momentum transfer $-t$ and the conventions used for $\rho_{UU}$ are retained here for consistency. In contrast to $\rho_{UU}$, the distribution exhibits a complementary behavior as the value of $-t$ increases, the height of the central maximum also increases. Furthermore, the distribution corresponding to the smallest value of $-t=0.04$ GeV$^2$ shows a noticeable asymmetry, with the central peak slightly shifted toward negative values of $\sigma$, while the distribution observed in \cite{Jana_2025} is symmetric for all $-t$ values.
Fig.~\ref{fig:rho_UU}(c) and (d) display the Wigner distributions of linearly polarized gluons in an unpolarized proton, corresponding to right-handed ($R$) and left-handed ($L$) polarizations, denoted by $\rho^{R}_{UT}$ and $\rho^{L}_{UT}$, respectively. For both cases, the distributions exhibit a slight asymmetry about $\sigma=0$ for all considered values of $-t$. In the case of $\rho^{R}_{UT}$, the distribution shows a left-shift toward negative values of $\sigma$. As $-t$ increases from $0.04$ GeV$^2$ to $0.20$ GeV$^2$, the height of the central peak decreases. However, when $-t$ is further increased from $0.20$ GeV$^2$ to $0.50$ GeV$^2$, the position of the peak remains unchanged, although a small shift in the secondary maxima along the $y-$axis is observed.
\begin{figure}[b]
    \centering
    \subfigure[]{\includegraphics[width=0.35\linewidth, trim=80 240 80 240, clip]{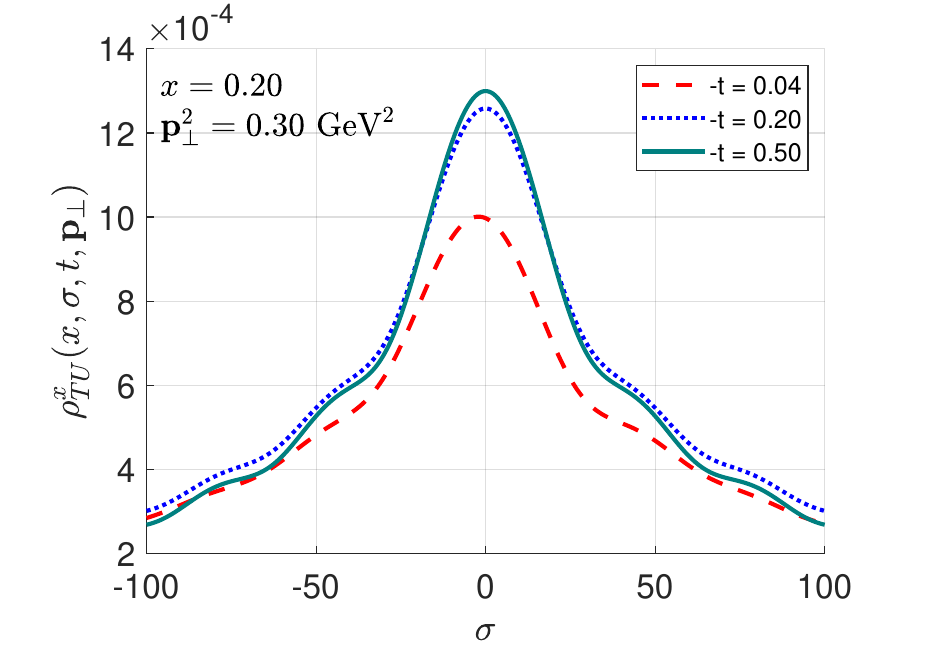}}
    \subfigure[]{\includegraphics[width=0.35\linewidth, trim=80 240 80 240, clip]{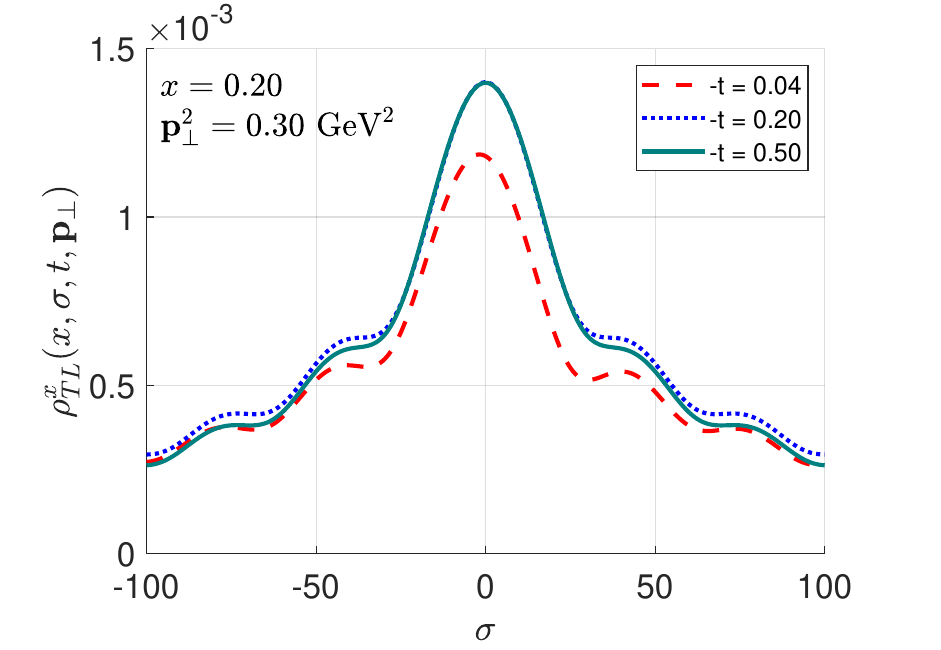}}
    \subfigure[]{\includegraphics[width=0.35\linewidth, trim=80 240 80 240, clip]{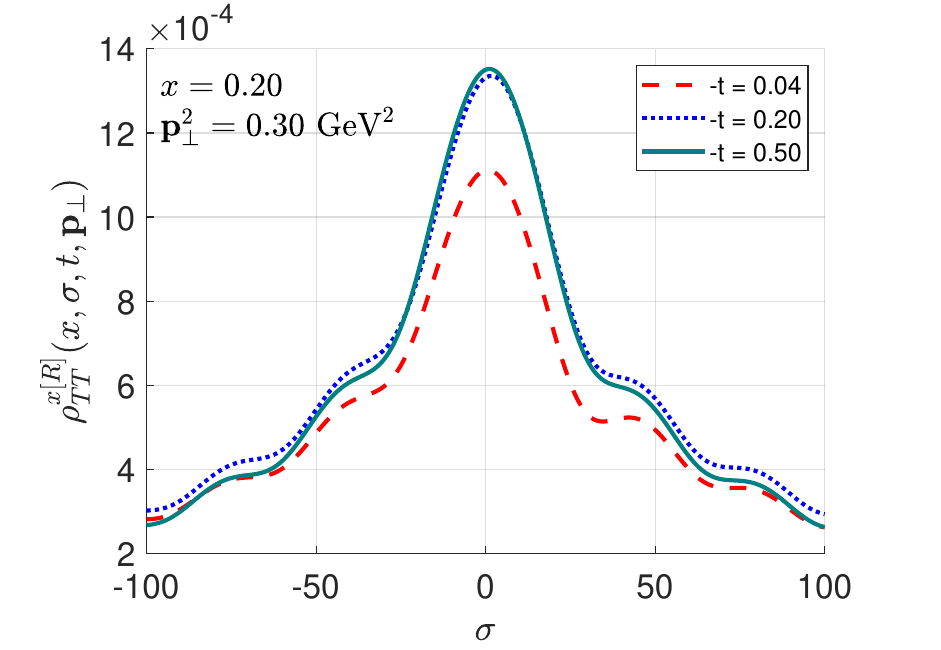}}
    \subfigure[]{\includegraphics[width=0.35\linewidth, trim=80 240 80 240, clip]{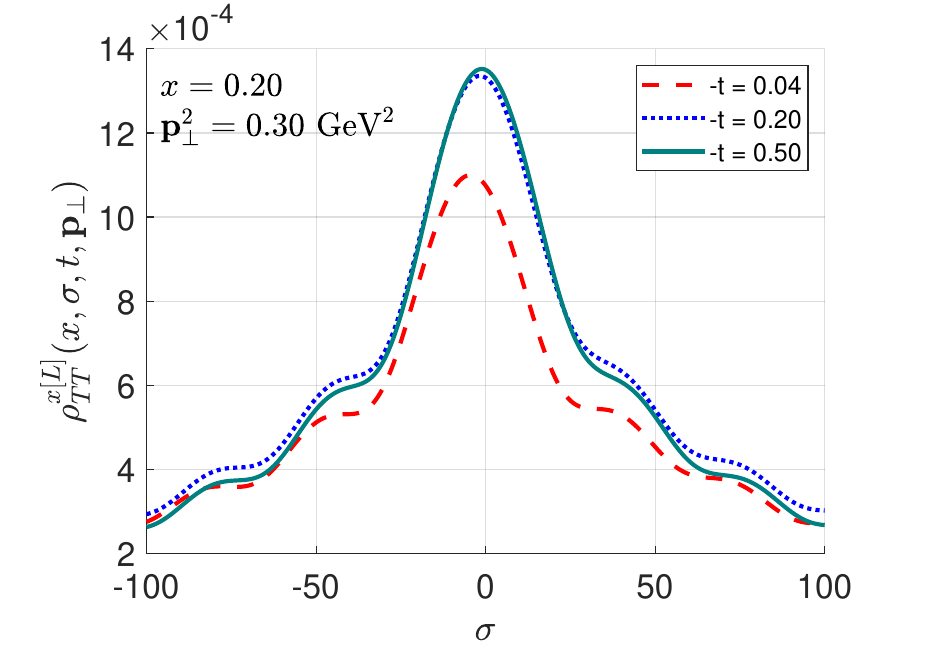}}
    \caption{The Wigner distribution $\rho^Z_{TY}$, (Y = U, L, T; Z = R, L) for transversely polarized proton in the boost invariant longitudinal position space $(\sigma)$ for different values of $-t$ in GeV$^2$ at fixed $x=0.2$, $\bfp^2=0.3$ GeV$^2$ and $\bfd\perp\bfp$.}
    \label{fig:rho_TU}
\end{figure}
For $\rho^{L}_{UT}$, the distribution exhibits right-shift  slightly toward positive values of $\sigma$. The height of the central maximum increases as $-t$ varies from $0.04$ GeV$^2$ to $0.20$ GeV$^2$, while it remains nearly unchanged for $-t=0.50$ GeV$^2$, accompanied by a slight displacement of secondary maxima along the $y$-axis. It is also evident that, for both $\rho^{R}_{UT}$ and $\rho^{L}_{UT}$, the peak positions are at same amplitude corresponding to $-t=0.20$ GeV$^2$ and $-t=0.50$ GeV$^2$.

\begin{figure}[b]
    \centering
    \subfigure[]{\includegraphics[width=0.35\linewidth, trim=80 240 80 240, clip]{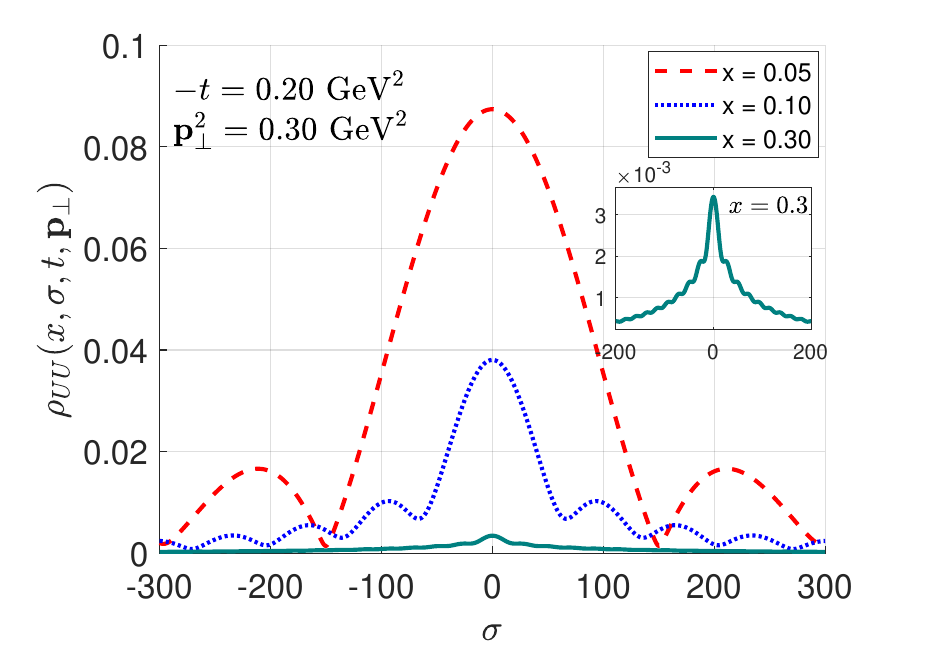}}\hspace{-.5cm}
    \subfigure[]{\includegraphics[width=0.35\linewidth, trim=80 240 80 240, clip]{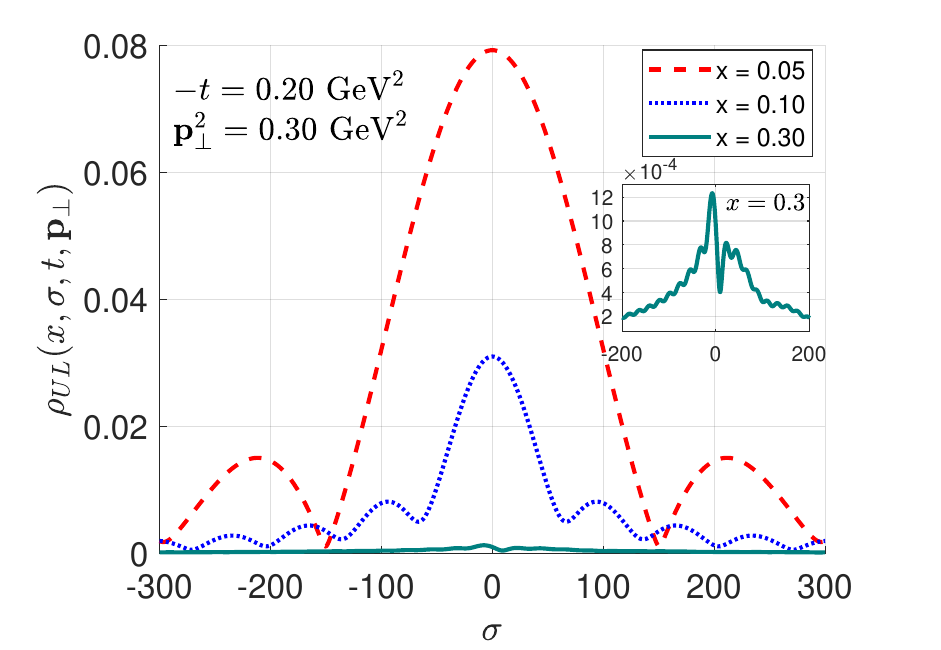}}\hspace{-.5cm}
    \subfigure[]{\includegraphics[width=0.35\linewidth, trim=80 240 80 240, clip]{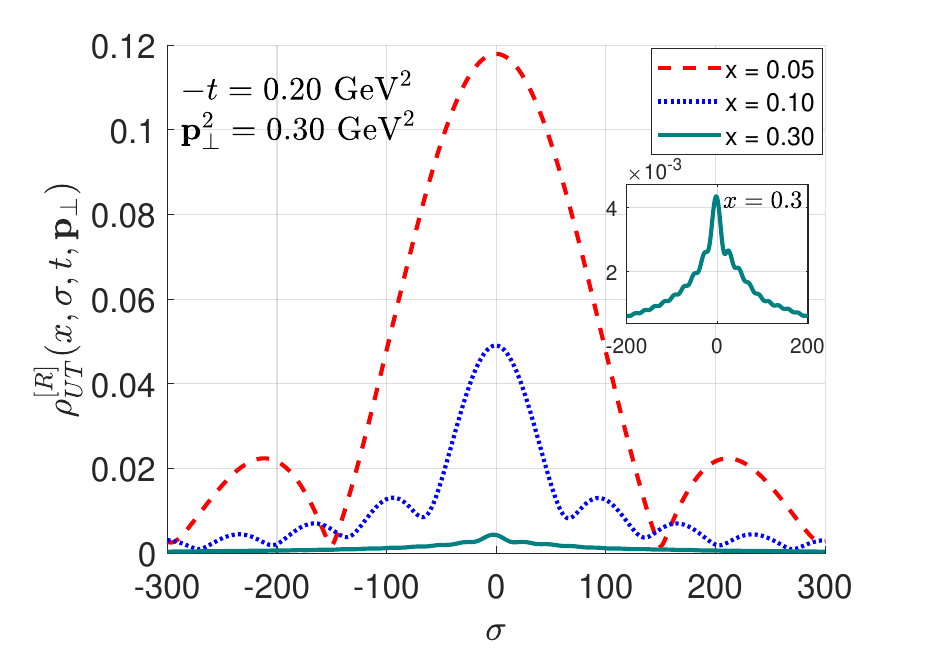}}\hspace{-.5cm}
    \subfigure[]{\includegraphics[width=0.35\linewidth, trim=80 240 80 240, clip]{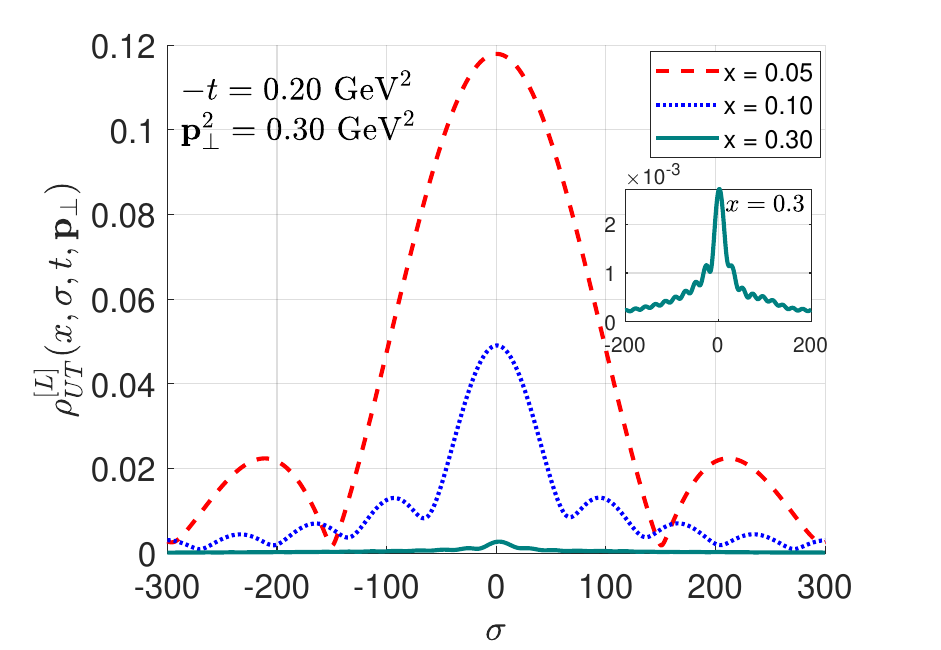}}
    \caption{The longitudinal momentum fraction $(x)$ sensitivity to Wigner distribution $\rho^Z_{UY}$, (Y = U, L, T; Z = R, L) for unpolarized proton in the boost invariant longitudinal position space $\sigma$ at fixed $-t=0.20$ GeV$^2$, $\bfp^2=0.30$ GeV$^2$ and $\bfd\perp\bfp$.}
    \label{fig:Uxvari}
\end{figure}
\subsection{Longitudinally proton}
Similarly, for a longitudinally polarized proton, a linear combination of the gluon correlators $W^{[\Gamma^{ij}]}_{++}$ and $W^{[\Gamma^{ij}]}_{--}$ can be used to express the Wigner distribution for an unpolarized and longitudinally polarized gluon as 
\begin{align}
 \rho_{LU}(x,\sigma,\bfd,\bfp)&=\int_{0}^{\xi_{s}}\frac{d\xi}{2\pi}e^{i\sigma\cdot\xi}\frac{1}{2}\Big[W^{1}_{++}(x,\xi,\bfd,\bfp)-W^{1}_{--}(x,\xi,\bfd,\bfp)\Big],
 \label{def_LU}\\
 \rho_{LL}(x,\sigma,\bfd,\bfp)&=\int_{0}^{\xi_{s}}\frac{d\xi}{2\pi}e^{i\sigma\cdot\xi}\frac{1}{2}\Big[W^{2}_{++}(x,\xi,\bfd,\bfp)-W^{2}_{--}(x,\xi,\bfd,\bfp)\Big],
 \label{def_LL}
 \end{align}
  and the Wigner distributions corresponding to linearly polarized gluons (R and L) in a longitudinally polarized proton can be constructed from linear combinations of the unpolarized and longitudinally polarized gluon distributions as 
  \begin{align}
    \rho^{R}_{LT}(x,\sigma,\bfd,\bfp)=-\rho_{LL}(x,\sigma,\bfd,\bfp)+\rho_{LU}(x,\sigma,\bfd,\bfp),
    \label{rs_LTR}\\
    \rho^{L}_{LT}(x,\sigma,\bfd,\bfp)=\rho_{LL}(x,\sigma,\bfd,\bfp)+\rho_{LU}(x,\sigma,\bfd,\bfp).
    \label{rs_LTL}
\end{align}
Additionally, the leading-twist GTMDs and the Wigner distributions $\rho_{LU}$ and $\rho_{LL}$ have a one-to-one correspondence from Eq.(\ref{eq:G-GTMDs}) as follows:
\begin{align}
\rho_{LU}(x,\sigma,\bfd,\bfp)&=\int_{0}^{\xi_{s}}\frac{d\xi}{2\pi}e^{i\sigma\cdot\xi}\frac{i}{M^2\sqrt{1-\xi^2}}\epsilon^{ij}_\perp \bfp^i\bfd^j F^g_{1,4},
\label{rs_LU}\\
 \rho_{LL}(x,\sigma,\bfd,\bfp)&=\int_{0}^{\xi_{s}}\frac{d\xi}{2\pi}e^{i\sigma\cdot\xi}\frac{1}{\sqrt{1-\xi^2}}G^g_{1,4}.
 \label{rs_LL}
\end{align}
Fig.~\ref{fig:rho_LL}(a),(b) present the Wigner distributions of unpolarized and longitudinally polarized gluons inside a longitudinally polarized proton, denoted by $\rho_{LU}$ and $\rho_{LL}$, respectively. We observe that the overall behavior of the distributions $\rho_{LU}$ and $\rho_{LL}$ is similar to that of $\rho_{UL}$ and $\rho_{UU}$, respectively. This similarity is expected due to the analogous structure of Eqs.~(\ref{rs_UL}), (\ref{rs_LU}) and Eqs.~(\ref{rs_UU}), (\ref{rs_LL}). The primary difference lies in the magnitude of the distributions: the height of the central peak for $\rho_{LU}$ and $\rho_{LL}$ is smaller than that of $\rho_{UL}$ and $\rho_{UU}$ for the considered values of $-t$. Ref.~\cite{Jana_2025} reports a more pronounced distribution of $\rho_{LU}$ then $\rho_{UL}$, while $\rho_{LL}$ is suppressed relative to $\rho_{UU}$.
Fig.~\ref{fig:rho_LL}(c),(d) show the Wigner distributions of linearly polarized gluons ($ R $ and $L$) in a longitudinally polarized proton, denoted by $\rho^{R}_{LT}$ and $\rho^{L}_{LT}$, respectively. The behavior of these distributions is also similar to that of $\rho^{R}_{UT}$ and $\rho^{L}_{UT}$, respectively, which is expected from the similarity in the mathematical structure of the expressions for $\rho_{UU}$, $\rho_{UL}$, $\rho_{LU}$ and $\rho_{LL}$.

\subsection{Transversely proton}
For transversely polarized proton, the Wigner distribution for an unpolarized, longitudinally polarized and linearly polarized (R and L) gluon is given as
\begin{align}
\rho^x_{TU}(x,\sigma,\bfd,\bfp)&=\int_{0}^{\xi_{s}}\frac{d\xi}{2\pi}e^{i\sigma\cdot\xi}\frac{1}{2}\Big[W^{1}(x,\xi,\bfd,\bfp,{+\hat{S_x}})-W^{1}(x,\xi,\bfd,\bfp,{-\hat{S_x}})\Big],
\label{def_TU}\\
\rho^x_{TL}(x,\sigma,\bfd,\bfp)&=\int_{0}^{\xi_{s}}\frac{d\xi}{2\pi}e^{i\sigma\cdot\xi}\frac{1}{2}\Big[W^{2}(x,\xi,\bfd,\bfp,{+\hat{S_x}})-W^{2}(x,\xi,\bfd,\bfp,{-\hat{S_x}})\Big], \label{def_TL}\\
\rho^{x[R]}_{TT}(x,\sigma,\bfd,\bfp)&=\rho_{TU}(x,\sigma,\bfd,\bfp)-\rho_{TL}(x,\sigma,\bfd,\bfp), \label{rs_TTR}\\
\rho^{x[L]}_{TT}(x,\sigma,\bfd,\bfp)&=\rho_{TU}(x,\sigma,\bfd,\bfp)+\rho_{TL}(x,\sigma,\bfd,\bfp), \label{rs_TTL}\\
\rho_{TU}^x(x,\sigma,\bfd,\bfp)&=\int_{0}^{\xi_{s}}\frac{d\xi}{2\pi}e^{i\sigma\cdot\xi}\frac{(2i)}{16\pi^3}\Bigg[ \bfp^{(2)} \Bigg(\frac{1}{x'}N_2(x'')-\frac{1}{x''}N_1(x')\Bigg)-\frac{\bfd^{(2)}}{2} \Bigg(\frac{1-x'}{x'}N_2(x'')+\frac{1-x''}{x''}N_1(x')\Bigg)\Bigg] \nn \\ 
&\times\phi(x',\bfp'^2)\phi(x'',\bfp''^2), \label{rs_TU} \\
\rho_{TL}^x(x,\sigma,\bfd,\bfp)&=\int_{0}^{\xi_{s}}\frac{d\xi}{2\pi}e^{i\sigma\cdot\xi}\frac{(-2)}{16\pi^3}\Bigg[ \bfp^{(1)} \Bigg(\frac{1}{x'}N_2(x'')+\frac{1}{x''}N_1(x')\Bigg)+\frac{\bfd^{(1)}}{2} \Bigg(-\frac{1-x'}{x'}N_2(x'')+\frac{1-x''}{x''}N_1(x')\Bigg)\Bigg] \nn \\
&\times\phi(x',\bfp'^2)\phi(x'',\bfp''^2), \label{rs_TL}
\end{align}
where $\hat{S_x}$ and $\hat{-S_x}$ denotes the proton's transverse polarization along the $x$-axis, represented as a linear combination of helicity states, $|\pm\hat{S_x}\rangle=\frac{1}{\sqrt{2}}(|\frac{1}{2}\rangle\pm|-\frac{1}{2}\rangle)$.

In Fig.~\ref{fig:rho_TU}, we present the Wigner distributions of unpolarized, longitudinally polarized, and linearly polarized gluons ($R$ and $L$) in a transversely polarized proton which are represented by $\rho^x_{TU}$, $\rho^x_{TL}$, $\rho^{x[R]}_{TT}$, and $\rho^{x[L]}_{TT}$, respectively. In Fig.~\ref{fig:rho_TU}(a), we show the distribution $\rho^x_{TU}$. While plotting $\rho^x_{TU}$, we make a choice of $\bfp$ and $\bfd$ as: $\bfp\equiv(0,|\bfp|)$, $\bfd\equiv(0,|\bfd|)$ to get a non-zero contribution from both the terms involved in Eq.(\ref{rs_TU}). We observe that the distribution $\rho^x_{TU}$ is symmetric for higher $-t$ values but becomes asymmetric for low $-t=0.04$ GeV$^2$, and is shifted towards the left of $\sigma=0$. For $\rho^x_{TL}$, shown in Fig.~\ref{fig:rho_TU}(b), we choose  $\bfp\equiv(|\bfp|,0)$, $\bfd\equiv(|\bfd|,0)$, to ensure that all the terms involved in Eq. (\ref{rs_TL}) contribute non-zero values. The distribution shows a pattern similar to that of $\rho^x_{TU}$, but its magnitude is higher than that of $\rho^x_{TU}$ for low $-t=0.04$ GeV$^2$ and the peaks merge for higher values of $-t$. One can further observe that the shift in the distribution is relatively large for $\rho^x_{TL}$ as $ -t $ increases from $-t=0.04$ GeV$^2$  to $-t=0.20$ GeV$^2$ as compared to that of $\rho^x_{TU}$. In Fig.~\ref{fig:rho_TU}(c), (d) we present the distribution $\rho^{x[R]}_{TT}$ and $\rho^{x[L]}_{TT}$, respectively. In this case, the preferable choice is $\bfp\equiv(|\bfp|/\sqrt{2},|\bfp|/\sqrt{2})$, $\bfd\equiv(|\bfd|/\sqrt{2},|\bfd|/\sqrt{2})$ to get the contribution of all the terms involved in Eqs.(\ref{rs_TU}) and (\ref{rs_TL}). We observe that the distributions are asymmetric about $\sigma=0$ for all values of $-t$ and magnitude of peak increases with increase in the values of $-t$. The only difference is that the distribution $\rho^{x[R]}_{TT}$ is shifted towards the right, while the distribution $\rho^{x[L]}_{TT}$ is shifted towards the left of $\sigma=0$.

\subsection{Longitudinal momentum sensitivity of WDs in $\sigma$-space}
\begin{figure}[h]
    \centering
    \subfigure[]{\includegraphics[width=0.35\linewidth, trim=80 240 80 240, clip]{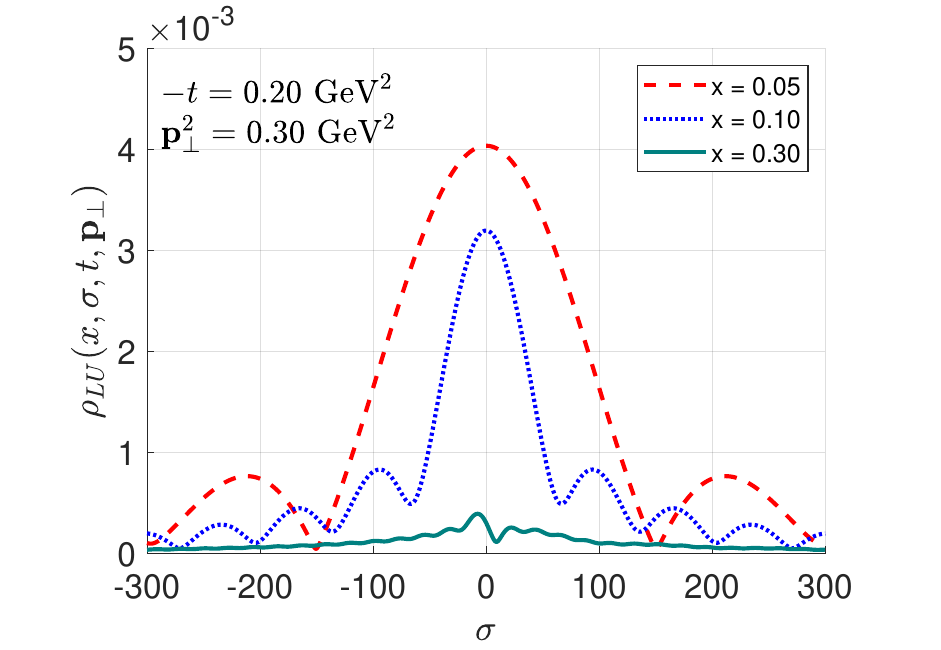}}\hspace{-.5cm}
    \subfigure[]{\includegraphics[width=0.35\linewidth, trim=80 240 80 240, clip]{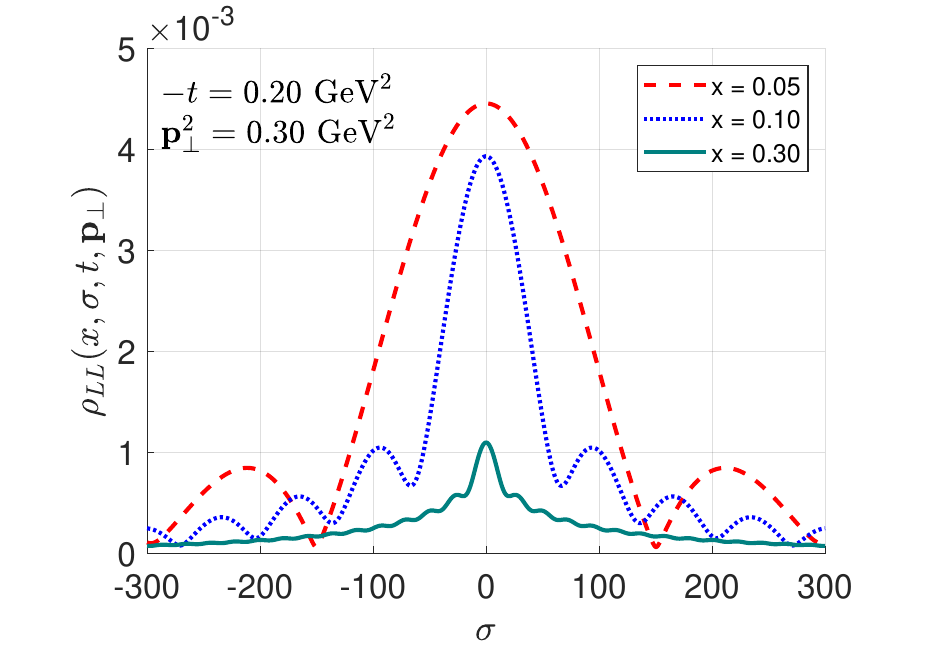}}\hspace{-.5cm}
    \subfigure[]{\includegraphics[width=0.35\linewidth, trim=80 240 80 240, clip]{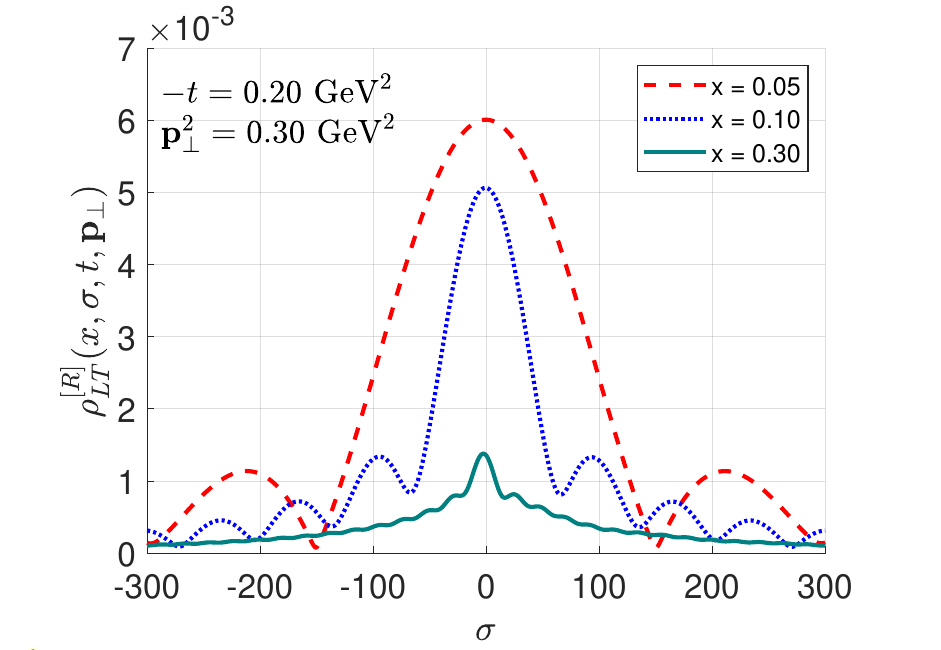}}\hspace{-.5cm}
    \subfigure[]{\includegraphics[width=0.35\linewidth, trim=80 240 80 240, clip]{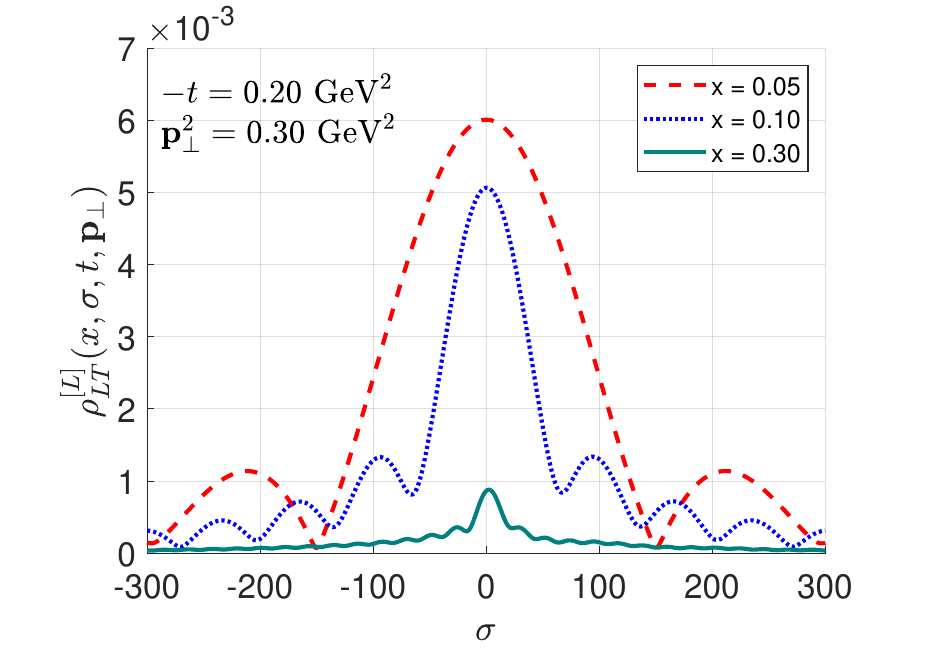}}
    \caption{The longitudinal momentum fraction $(x)$ sensitivity to Wigner distribution $\rho^Z_{LY}$, (Y = U, L, T; Z = R, L) for longitudinally polarized proton in the boost invariant longitudinal position space $\sigma$ at fixed $-t=0.20$ GeV$^2$, $\bfp^2=0.30$ GeV$^2$ and $\bfd\perp\bfp$.}
    \label{fig:Lxvari}
\end{figure}

\begin{figure}[htp]
    \centering
     \subfigure[]{\includegraphics[width=0.35\linewidth, trim=80 240 80 240, clip]{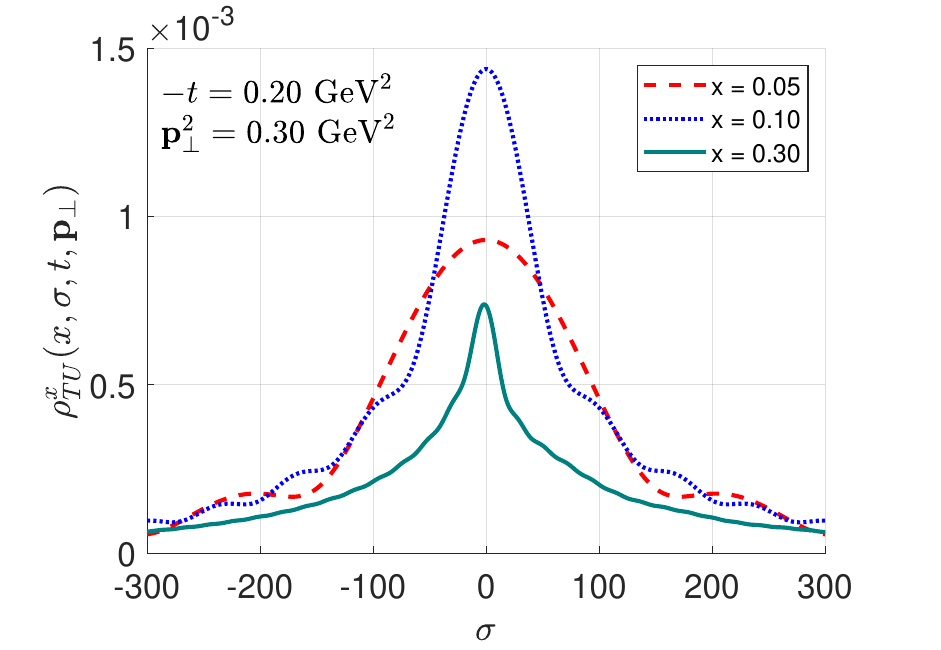}}\hspace{-.5cm}
    \subfigure[]{\includegraphics[width=0.35\linewidth, trim=80 240 80 240, clip]{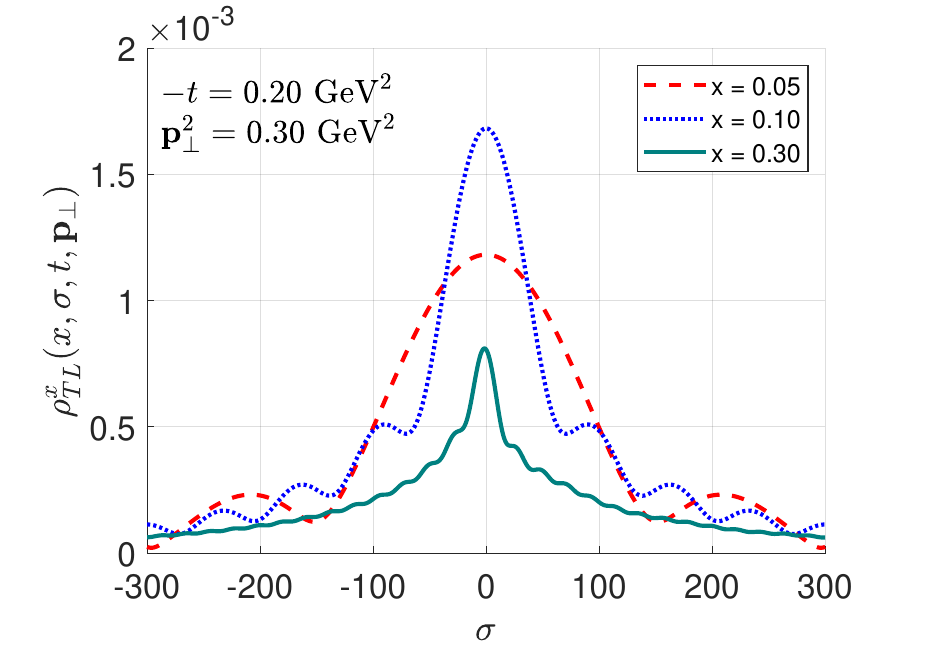}}\hspace{-.5cm}\\
    \subfigure[]{\includegraphics[width=0.35\linewidth, trim=80 240 80 240, clip]{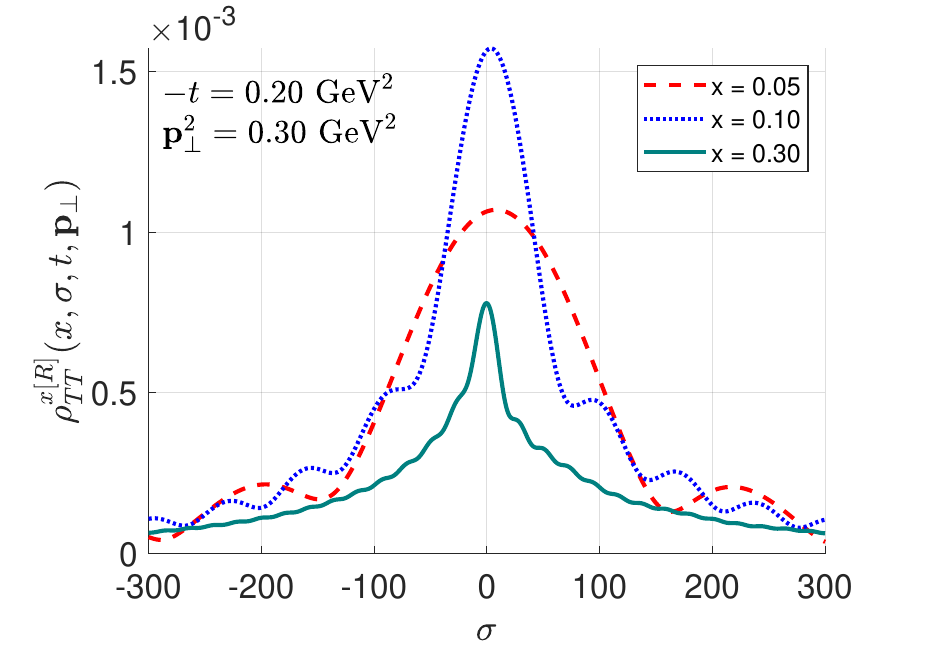}}\hspace{-.5cm}
    \subfigure[]{\includegraphics[width=0.35\linewidth, trim=80 240 80 240, clip]{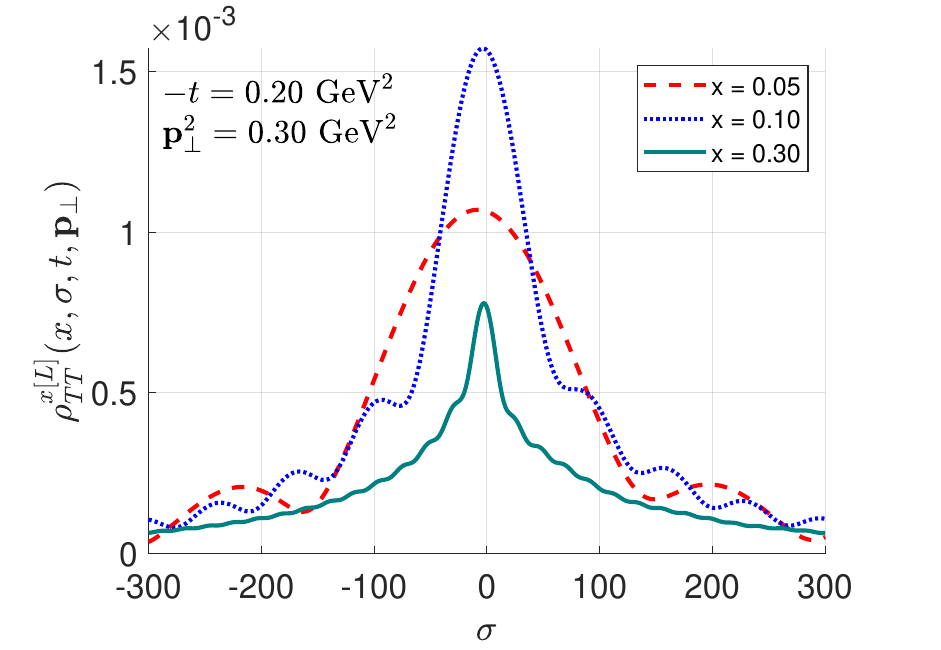}}
    \caption{The longitudinal momentum fraction $(x)$ sensitivity to Wigner distribution $\rho^Z_{TY}$, (Y = U, L, T; Z = R, L) for transversely polarized proton in the boost invariant longitudinal position space $\sigma$ at fixed $-t=0.20$ GeV$^2$, $\bfp^2=0.30$ GeV$^2$ and $\bfd\perp\bfp$.}
    \label{fig:Txvari}
\end{figure}

 All these plots \ref{fig:rho_UU}-\ref{fig:rho_TU} for different $-t$ show very passive sensitivity to the position of the first minima unlike quark WDs \cite{Maji:2022tog}, which leads to the importance of investigating the $x$ dependency in the $\sigma$-space WDs for gluons.
 In Figs.~\ref{fig:Uxvari}--\ref{fig:Txvari}, the three plots in each subfigure correspond to different values of the longitudinal momentum fraction, $x = \{0.05, 0.10, 0.30\}$, represented by red, blue, and teal colors, respectively, at fixed values of energy transfer $-t = 0.2~\text{GeV}^2$ and transverse momentum $\mathbf{p}_\perp^2 = 0.3~\text{GeV}^2$.  
In Fig.~\ref{fig:Uxvari}(a), the distribution of an unpolarized gluon in an unpolarized proton, $\rho_{UU}$, shows a symmetric behavior about $\sigma = 0$ for all values of $x$. it is observed that, the intensity of the central maxima decreases with increasing $x$.  The width of the central maxima varies inversely with increasing $x$ that indicates $x$ can also be treated as effective slit-width. Fig.~\ref{fig:Uxvari}(b) shows the distribution of a longitudinally polarized gluon in an unpolarized proton, $\rho_{UL}$. A similar behavior is observed for lower values of $x$, but for higher $x$, an asymmetry appears in the peak, which shifts toward negative values of $\sigma$.
Furthermore, Figs.~\ref{fig:Uxvari}(c) and (d) show the Wigner distributions of linearly polarized gluons in an unpolarized proton, corresponding to right-handed ($R$), $\rho_{UT}^{R}$, and left-handed ($L$), $\rho_{UT}^{L}$, polarizations, respectively. In both cases, the distributions exhibit a slight asymmetry about $\sigma = 0$ for all considered values of $x$. In the case of $\rho_{UT}^{R}$, the distribution is slightly shifted toward negative values of $\sigma$, while for $\rho_{UT}^{L}$, it is shifted toward positive values of $\sigma$. For both distributions, the peaks at $x = 0.05$ and $x = 0.10$ remain nearly identical.

Fig.~\ref{fig:Lxvari}(a) and (b) show the Wigner distributions of unpolarized and longitudinally polarized gluons inside a longitudinally polarized proton, $\rho_{LU}$ and $\rho_{LL}$, respectively. The overall features of these distributions closely follow those seen in $\rho_{UL}$ and $\rho_{UU}$. The main distinction appears in the magnitude of the distributions, where the central peak in $\rho_{LU}$ and $\rho_{LL}$ is comparatively lower than that of $\rho_{UL}$ and $\rho_{UU}$ for the chosen values of $x$.
Fig.~\ref{fig:rho_LL}(c),(d) show the Wigner distributions of linearly polarized gluons with right-handed ($R$) and left-handed ($L$) polarizations in a longitudinally polarized proton, denoted by $\rho_{LT}^{R}$ and $\rho_{LT}^{L}$, respectively. These distributions follow a similar trend as $\rho_{UT}^{R}$ and $\rho_{UT}^{L}$. This behaviour can be traced to the close similarity among the expressions of $\rho_{UU}$, $\rho_{UL}$, $\rho_{LU}$, and $\rho_{LL}$.

Fig.~\ref{fig:Txvari}(a),(b) shows the Wigner distributions of unpolarized and longitudinally polarized gluons inside a transversely polarized proton, $\rho^x_{TU}$ and $\rho^x_{TL}$, respectively. We observed that the peak of the distribution $\rho^x_{TU}$ first increases with increasing momentum fraction $x$ from $0.05$ to $0.10$, and the corresponding width of the central maxima decreases. 
Whereas, the peak of the distribution decreases significantly when the $x$ value is further increased to $0.30$ with a simultaneous decrease in the width of the distribution, resulting in the first minima shifting close to $\sigma=0$. Additionally, for $x=0.30$, the distribution $\rho^x_{TU}$ is asymmetric and shifted to the left. The distribution $\rho^x_{TL}$ exhibits similar behavior to that of $\rho^x_{TU}$. In Fig.~\ref{fig:Txvari}(c),(d), the WD of linearly polarized gluons (R and L) inside a transversely polarized proton, $\rho^{x[R]}_{TT}$ and $\rho^{x[L]}_{TT}$, display analogous behavior to that of $\rho^x_{TU}$.

\begin{figure}[b]
    \centering
    \includegraphics[width=0.35\linewidth, trim=80 240 80 240, clip]{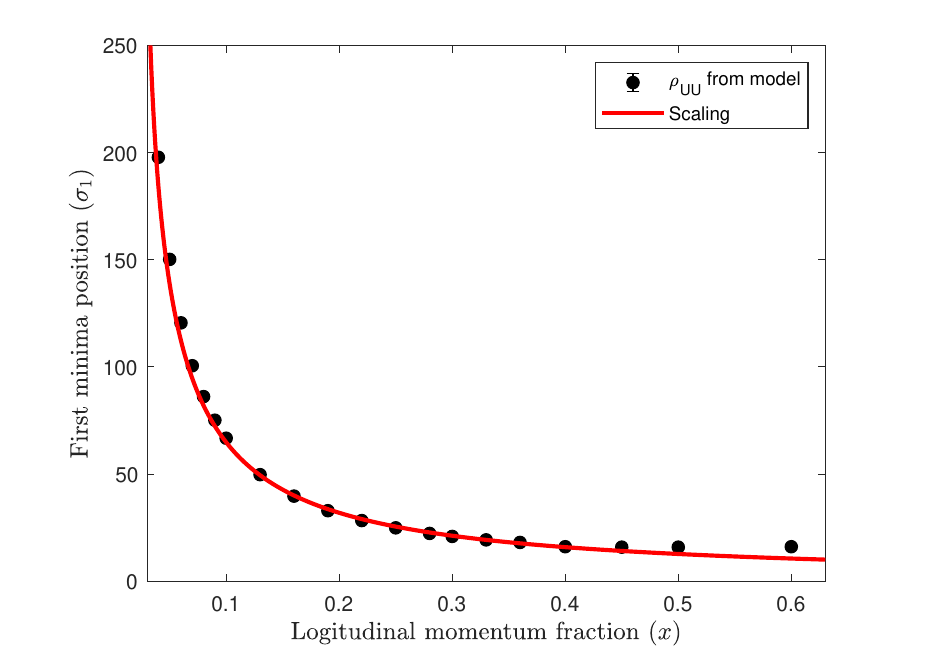}
    \caption{Scaling from Eq.(\ref{eq:xsigvari}) for first minima $\sigma_1$ with respect to longitudinal momentum fraction $x$ of Wigner Distributions $\rho_{UU}$ in $\sigma$-space.}
    \label{fig:xsigvari}
\end{figure}

For all the polarization states of the proton, in Figs.~\ref{fig:Uxvari}-\ref{fig:Txvari}, 
 we observe, as $x$ increases the central maxima of the diffraction-like pattern becomes narrower and the first-minima shifts towards the center significantly unlike $-t$ variation. This inverse power low indicates that $x$ plays a role similar to the effective slit width.
 We also observe that for fixed $x$, the minima appear at integral multiples of the lowest minima of $\sigma_n$ values, as in optical diffraction phenomena. This is consistent with the single-slit diffraction law for the $n^{th}$ minimum,
\begin{align} \label{eq:difrc}
    \sin\theta_n= \frac{\sigma_n}{\sqrt{D^2+\sigma_n^2}} = \frac{n\lambda}{d},
\end{align}
where $\lambda$ is the wavelength and $d$ is the slit-width. The position of the $n^{th}$ minima $\sigma_n$ can be obtained from $\sin\theta_n$ with the separation $(D)$ between the slit and the detector. If the separation $(D)$ is large compare to the $\sigma_n$ one can map the longitudinal momentum fraction $x$ to the dimensionless quantity $d/D$ that leads to relation with $\sigma_n$ as
\begin{align}\label{eq:xsigvari}
    x= \frac{d}{D}=  n\bigg[\frac{A}{\sigma_n}+{B \sigma_n}\bigg],
\end{align}
where the subleading second term can be considered as a correction to the inverse variation between slit-width and the position of the first minima found in single-slit diffraction in Optics. 
The form of the parameters $A$ and $B$ can be extracted as $A=\lambda$ and $B=\lambda/2D^2$. 
These parameters are fitted with variation of $x$ and $\sigma_1$, corresponding to the location of first minima of the Wigner distribution $\rho_{UU}$ for an unpolarized gluon in an unpolarized proton, where $A=6.3552$ GeV$^{-1}$ and $B=3\times10^{-5}$ GeV.  As we can see in Fig.~\ref{fig:xsigvari}, the red line represents the scaling of Eq.(\ref{eq:xsigvari}) that satisfies the first minima values of $\rho_{UU}$ for different $x$. A similar variation is observed in all other Wigner distributions where Eq.(\ref{eq:xsigvari}) is applicable with the the same value of the parameters $A$ and $B$. A similar discussion is included for the DVCS amplitude in boost invariant longitudinal space in \cite{Brodsky:2006ku}.

\section{Wigner distribution in $\bfb$-space \label{b-space}}
The WDs in $\bfb$-space is defined by the Fourier transform of gluon-gluon correlator $W^{[\Gamma^{ij}]}(x,\xi,\bfD,\bfp)$ with respect to $\bfD$, where, the impact parameter $\bfb$ is Fourier conjugate to the variable $\bfD=\bfd/(1-\xi^2)$. 
 For $\xi\ne0$, extensive studies of quark sector in the light-front spectator model and in dressed quark model have been performed in \cite{Maji:2022tog,Ojha:2022fls}. Here, we focus on the gluon sector, and the Wigner distribution for gluons in impact parameter ($\bfb$) space is defined as
\begin{align}
 \rho^{[\Gamma^{ij}]}(x,\xi,\bfb,\bfp)&=\int_{-\infty}^{\infty}\frac{d^2\bfD}{(2\pi)^2}e^{-i\bfD\cdot \bfb}W^{[\Gamma^{ij}]}(x,\xi,\bfD,\bfp).
\end{align}

For unpolarized proton, the Wigner distribution of unpolarized, longitudinally polarized and linearly polarized (R and L) gluons are defined as
\begin{align}
 \rho_{UU}(x,\xi,\bfb,\bfp)&=\int_{-\infty}^{\infty}\frac{d^2\bfD}{(2\pi)^2}e^{-i\bfD\cdot \bfb}\frac{1}{2}\Big[W^{1}_{++}(x,\xi,\bfD,\bfp)+W^{1}_{--}(x,\xi,\bfD,\bfp)\Big],\nn \\
&=\int_{-\infty}^{\infty}\frac{d^2\bfD}{(2\pi)^2}e^{-i\bfD\cdot \bfb}\frac{1}{\sqrt{1-\xi^2}}F^g_{1,1}, \\
\rho_{UL}(x,\xi,\bfb,\bfp)&=\int_{-\infty}^{\infty}\frac{d^2\bfD}{(2\pi)^2}e^{-i\bfD\cdot \bfb}\frac{1}{2}\Big[W^{2}_{++}(x,\xi,\bfD,\bfp)+W^{2}_{--}(x,\xi,\bfD,\bfp)\Big],\nn \\
&=\int_{-\infty}^{\infty}\frac{d^2\bfD}{(2\pi)^2}e^{-i\bfD\cdot \bfb}\Big(-\frac{i}{M^2}\epsilon^{ij}_\perp p^i_\perp D^j_\perp(1-\xi^2)\Big)\frac{G^g_{1,1}}{\sqrt{1-\xi^2}},
 \label{eq:rho_{UL}} \\
\rho^{R}_{UT}(x,\xi,\bfp,\bfb)&=\rho_{UU}(x,\xi,\bfp,\bfb)-\rho_{UL}(x,\xi,\bfp,\bfb),
    \label{eq:rho_{UTR}} \\
\rho^{L}_{UT}(x,\xi,\bfp,\bfb)&=\rho_{UU}(x,\xi,\bfp,\bfb)+\rho_{UL}(x,\xi,\bfp,\bfb).
\label{eq:rho_{UTL}}
\end{align}
In a longitudinally polarized proton, the WDs for the unpolarized, longitudinally polarized, and linearly polarized gluons are defined as
\begin{align}
 \rho_{LL}(x,\xi,\bfb,\bfp)&=\int_{-\infty}^{\infty}\frac{d^2\bfD}{(2\pi)^2}e^{-i\bfD\cdot \bfb}\frac{1}{2}\Big[W^{2}_{++}(x,\xi,\bfD,\bfp)-W^{2}_{--}(x,\xi,\bfD,\bfp)\Big],\nn \\
 &=\int_{-\infty}^{\infty}\frac{d^2\bfD}{(2\pi)^2}e^{-i\bfD\cdot \bfb}\frac{1}{\sqrt{1-\xi^2}}G^g_{1,4}, \\
\rho_{LU}(x,\xi,\bfb,\bfp)&=\int_{-\infty}^{\infty}\frac{d^2\bfD}{(2\pi)^2}e^{-i\bfD\cdot \bfb}\frac{1}{2}\Big[W^{1}_{++}(x,\xi,\bfD,\bfp)-W^{1}_{--}(x,\xi,\bfD,\bfp)\Big],\nn \\
&=\int_{-\infty}^{\infty}\frac{d^2\bfD}{(2\pi)^2}e^{-i\bfD\cdot \bfb}\Big(\frac{i}{M^2}\epsilon^{ij}_\perp \bfp^i \bfD^j(1-\xi^2)\Big)\frac{F^g_{1,4}}{\sqrt{1-\xi^2}}, 
 \label{eq:rho_{LU}} \\
\rho^{R}_{LT}(x,\xi,\bfb,\bfp)&=-\rho_{LL}(x,\xi,\bfb,\bfp)+\rho_{LU}(x,\xi,\bfb,\bfp),
\label{eq:rho_{LTR}} \\
\rho^{L}_{LT}(x,\xi,\bfb,\bfp)&=\rho_{LL}
(x,\xi,\bfb,\bfp)+\rho_{LU}(x,\xi,\bfb,\bfp).
\label{eq:rho_{LTL}} 
\end{align}
For transversely polarized proton, the Wigner distribution for an unpolarized, longitudinally polarized, and linearly polarized (R and L) gluon are given as 
\begin{align}
\rho^x_{TU}(x,\xi,\bfb,\bfp)&=\int_{-\infty}^{\infty}\frac{d^2\bfD}{(2\pi)^2}e^{-i\bfD\cdot \bfb}\frac{1}{2}\Big[W^{1}(x,\xi,\bfD,\bfp,{+\hat{S_x}})-W^{1}(x,\xi,\bfD,\bfp,{-\hat{S_x}})\Big],\\
&=\int_{-\infty}^{\infty}\frac{d^2\bfD}{(2\pi)^2}e^{-i\bfD\cdot \bfb}\frac{2i}{16\pi^3}\Bigg(\bfp^{(2)}\Bigg[\frac{1}{x'}N_2(x'')-\frac{1}{x''}N_1(x')\Bigg] \nn \\
& \hspace{1cm} -\frac{\bfD^{(2)}(1-\xi^2)}{2} \Bigg[\frac{1-x'}{x'}N_2(x'')+\frac{1-x''}{x''}N_1(x')\Bigg]\Bigg)\phi(x',\bfp'^2)\phi(x'',\bfp''^2) , \label{eq:rho_{TU}} 
\end{align}
\begin{align}
\rho^x_{TL}(x,\xi,\bfb,\bfp)&=\int_{-\infty}^{\infty}\frac{d^2\bfD}{(2\pi)^2}e^{-i\bfD\cdot \bfb}\frac{1}{2}\Big[W^{2}(x,\xi,\bfD,\bfp,{+\hat{S_x}})-W^{2}(x,\xi,\bfD,\bfp,{-\hat{S_x}})\Big],\\
&=\int_{-\infty}^{\infty}\frac{d^2\bfD}{(2\pi)^2}e^{-i\bfD\cdot \bfb}\frac{-2}{16\pi^3}\Bigg(\bfp^{(1)} \Bigg[\frac{1}{x''}N_1(x')+\frac{1}{x'}N_2(x'')\Bigg] \nn \\
&\hspace{1cm} -\frac{\bfD^{(1)}(1-\xi^2)}{2}\Bigg[\frac{1-x'}{x'}N_2(x'')-\frac{1-x''}{x''}N_1(x')\Bigg]\Bigg) \phi(x',\bfp'^2)\phi(x'',\bfp''^2), \label{eq:rho_{TL}} \\
\rho^{x[R]}_{TT}(x,\xi,\bfb,\bfp)&=\rho^x_{TU}(x,\xi,\bfb,\bfp)-\rho^x_{TL}(x,\xi,\bfb,\bfp),
\label{eq:rho_{TTR}} \\
\rho^{x[L]}_{TT}(x,\xi,\bfb,\bfp)&=\rho_{TU}(x,\xi,\bfb,\bfp)+\rho_{TL}(x,\xi,\bfb,\bfp).
    \label{eq:rho_{TTL}}
\end{align}
\begin{figure}[h]
    \centering
     \includegraphics[width=0.30\linewidth, trim=80 240 100 240, clip]{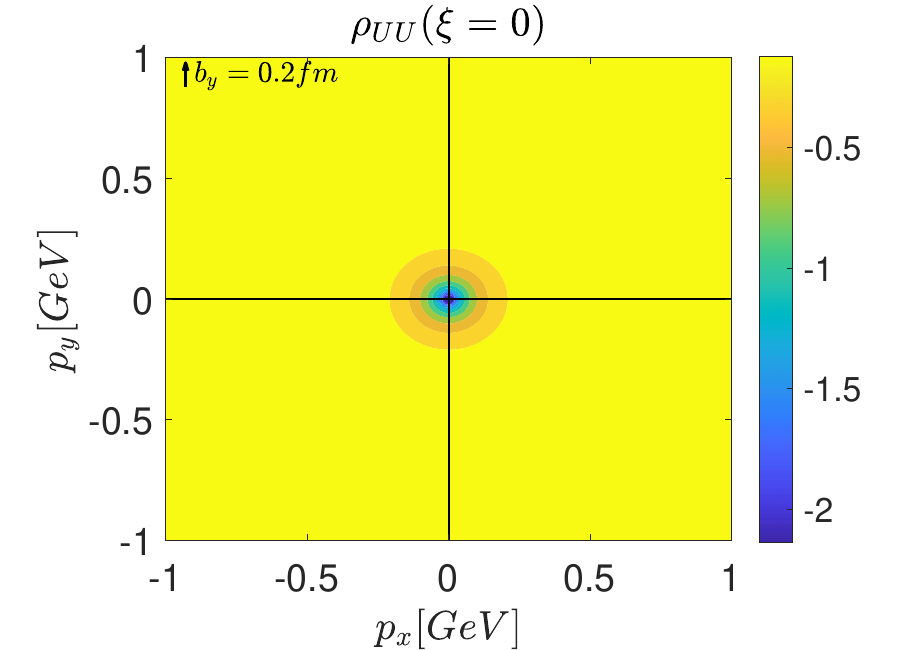}
     \includegraphics[width=0.30\linewidth, trim=80 240 100 240, clip]{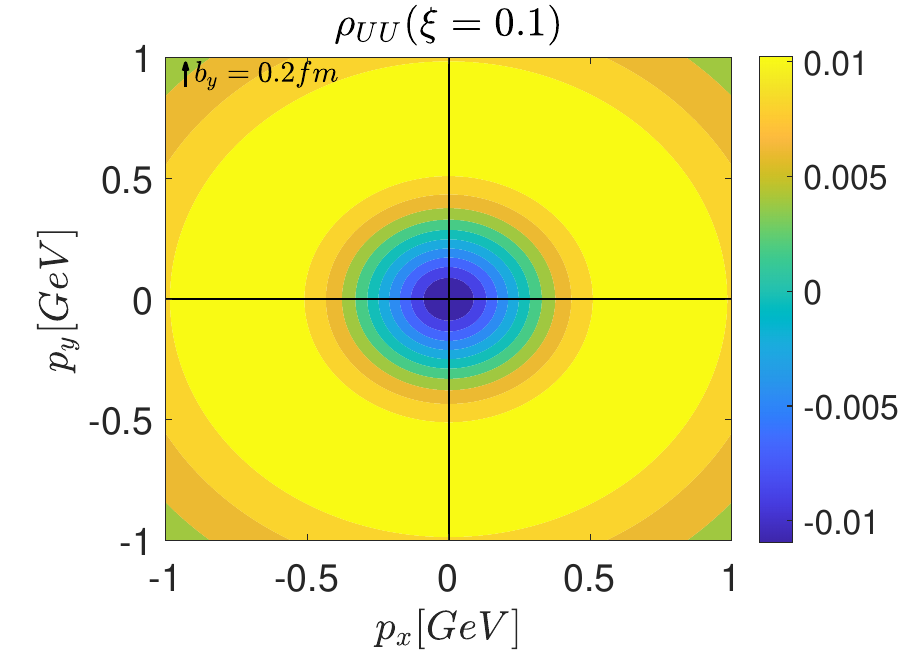}
    \includegraphics[width=0.30\linewidth, trim=80 240 100 240, clip]{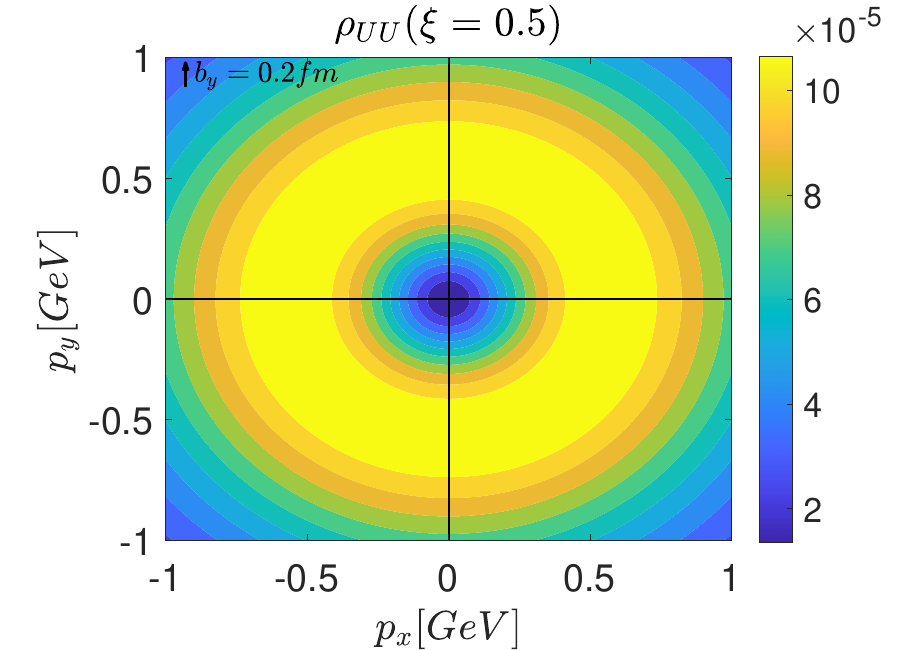}\\ \vspace{0cm}
      \includegraphics[width=0.30\linewidth, trim=80 240 100 240, clip]{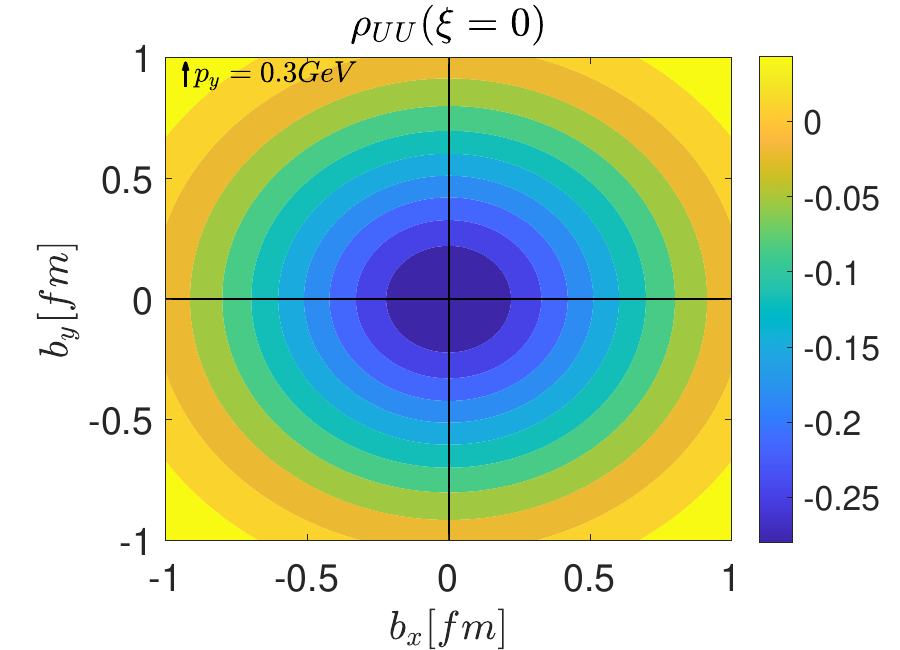}
     \includegraphics[width=0.30\linewidth, trim=80 240 100 240, clip]{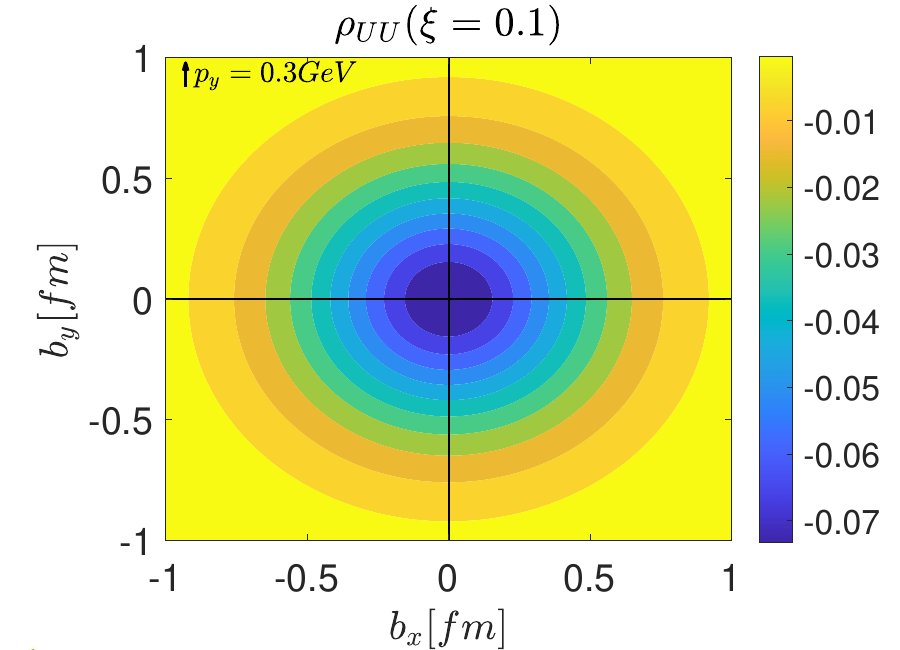}
     \includegraphics[width=0.30\linewidth, trim=80 240 100 240, clip]{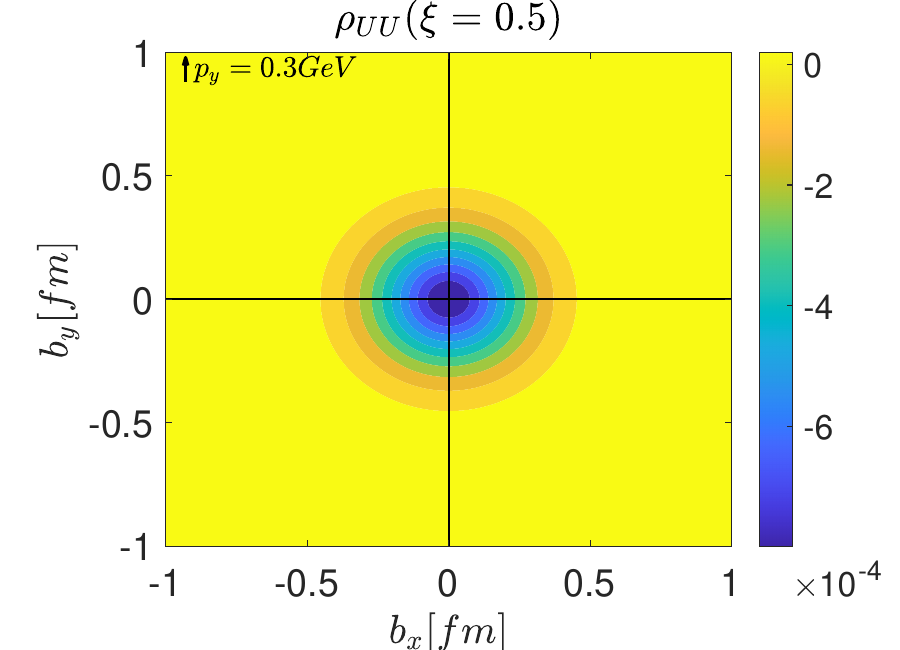}
    \caption{The first Mellin moment of gluon Wigner distribution $\rho_{UU}$ for different values of skewness parameter ($\xi= 0, 0.1,0.5$) in transverse momentum space (upper panel) and impact parameter space (lower panel) for fixed $\bfb=0.2$ fm $\hat{y}$ and $\bfp=0.3$ GeV $\hat{y}$, respectively, with the condition $\bfp\perp\bfd$.}
    \label{fig:b_rhoUU}
\end{figure}
All the Wigner correlators are written in terms of $\bfD$, which is Fourier conjugate to impact parameter $(\bfb)$ for $\xi \neq 0$ using the relation $\bfD=\bfd/(1-\xi^2)$.
We plot the first Mellin moment of Wigner distribution, $\rho^{[Z]}_{XY}(x,\xi,\bfb,\bfp)$ $(X,Y = U, L, T; Z=R, L)$ for unpolarized, longitudinally polarized, and linearly polarized (R and L) gluons for different polarization states of proton; unpolarized, longitudinally polarized, and transversely polarized. The first Mellin moment of WDs characterizes the transverse phase-space distribution of gluons in a proton. We present the variation of WDs with $\xi$. Fig.~\ref{fig:b_rhoUU} show the projection of the Mellin moments of unpolarized gluon inside unpolarized proton $\rho_{UU}(\bfb,\bfp)$ on transverse momentum plane (upper row) with fixed $\bfb=0.2$ fm along $\hat{y}$ and transverse impact parameter plane with $\bfp=0.3$ GeV along $\hat{y}$. Three columns are for three different values of $\xi=0, 0.1, 0.5$ respectively, and $\bfp$ is considered to be perpendicular to $\bfd$. 
 We observe that for all $\xi$ values, the distribution is circularly symmetric with a negative maxima at the center $(\{p_x,p_y\}=\{0,0\})$, and $(\{b_x,b_y\}=\{0,0\})$ for the choice of $\bfp \perp \bfb$, which is expected as the average quadrupole distortions $Q^{ij}_b(\bfp)$ and $Q^{ij}_p(\bfb)$ for $\rho_{UU}$, defined as~\cite{Lorce:2011kd}
\begin{align} 
Q^{ij}_b(\bfp)= \frac{\int d^2\bfb (2b^i_\perp b^j_\perp-\delta^{ij}\bfb^2)\rho_{UU}(\bfb,\bfp)}{\int d^2\bfb \bfb^2 \rho_{UU}(\bfb,\bfp)},\\
Q^{ij}_p(\bfb)= \frac{\int d^2\bfp (2p^i_\perp p^j_\perp-\delta^{ij}\bfp^2)\rho_{UU}(\bfb,\bfp)}{\int d^2\bfp \bfp^2 \rho_{UU}(\bfb,\bfp)},\label{Eq_Qdis}
\end{align}
vanish for the Gaussian type form (over $\bfp$ and $\bfb$ ) of wave functions of soft-wall AdS/QCD model. However, for any other choice, the distribution would exhibit axial symmetry.  While, for quarks, distributions are found with positive maxima \cite{Chakrabarti:2017teq}. However for gluon, the polarity of the distribution in this model shows qualitatively similar behavior as presented in \cite{Tan:2023vvi} for both the transverse planes and dressed quark model results shows opposite polarity with distorted circular peaks \cite{More:2017zqp}. In the transverse momentum space, the distribution decreases sharply around the center for $\xi=0$, and in comparison, for $\xi=0.1, 0.5$ the distributions are oscillatory with a positive secondary peak around $p_x=0.5$ GeV, and gradually increase for large $p_x$. One can also observe that as the skewness variable $\xi$ increases from $0$ to $0.1$, the intensity of the peak is decreased by approximately $10$-times, and for larger momentum transferred in the longitudinal direction $\xi=0.5$, the peak value decreases by a large amount. In the impact parameter plane, as the skewness variable $\xi$ is increased from $0$ to $0.5$, the distributions become more concentrated at the center. 
   \begin{figure}[h]
    \centering
     \includegraphics[width=0.30\linewidth, trim=80 240 100 240, clip]{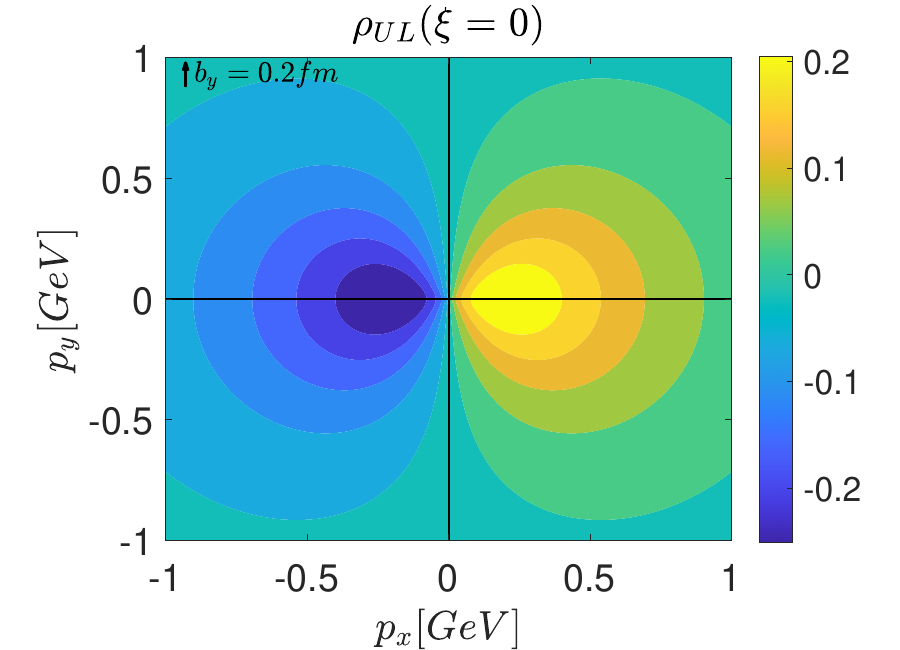}
     \includegraphics[width=0.30\linewidth, trim=80 240 100 240, clip]{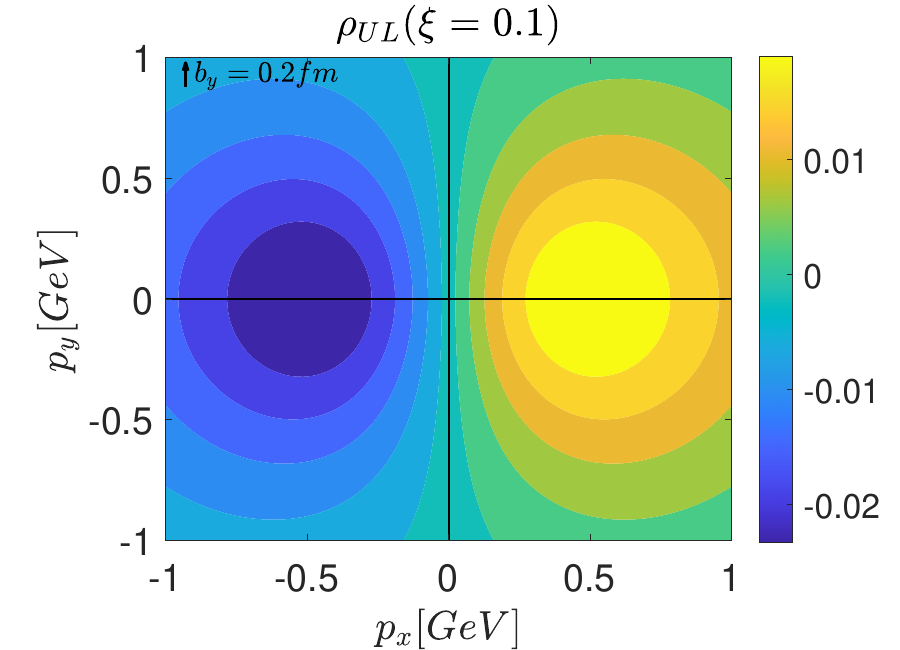}
     \includegraphics[width=0.30\linewidth, trim=80 240 100 240, clip]{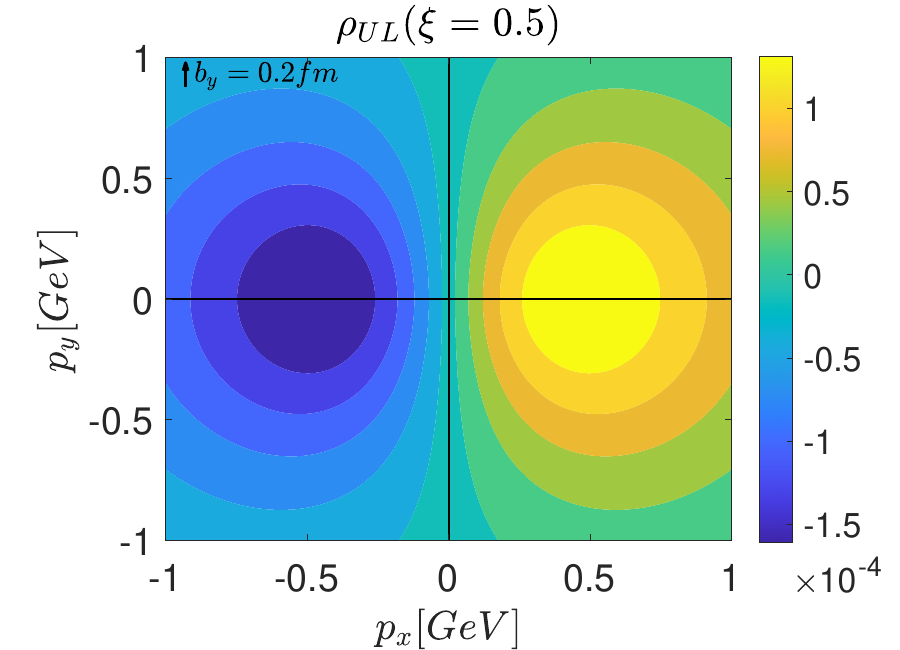}\\ \vspace{-0.0cm}
     \includegraphics[width=0.30\linewidth, trim=80 240 100 240, clip]{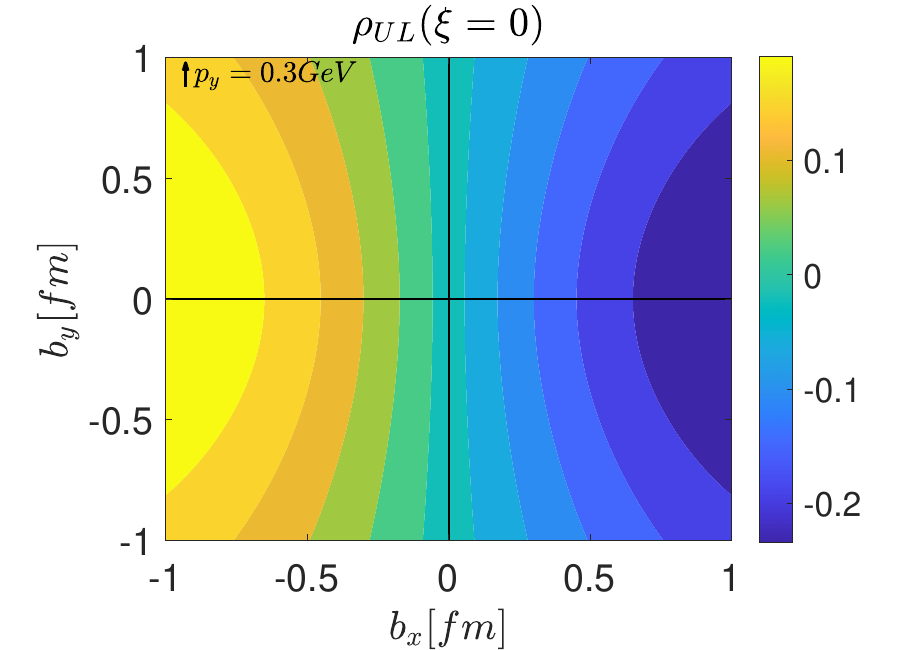}
     \includegraphics[width=0.30\linewidth, trim=80 240 100 240, clip]{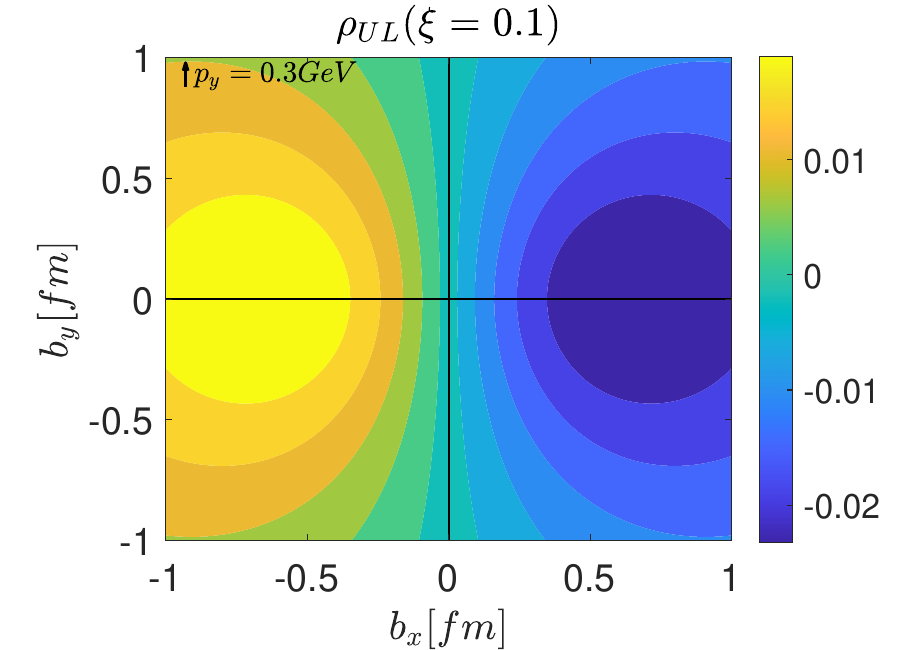}
     \includegraphics[width=0.30\linewidth, trim=80 240 100 240, clip]{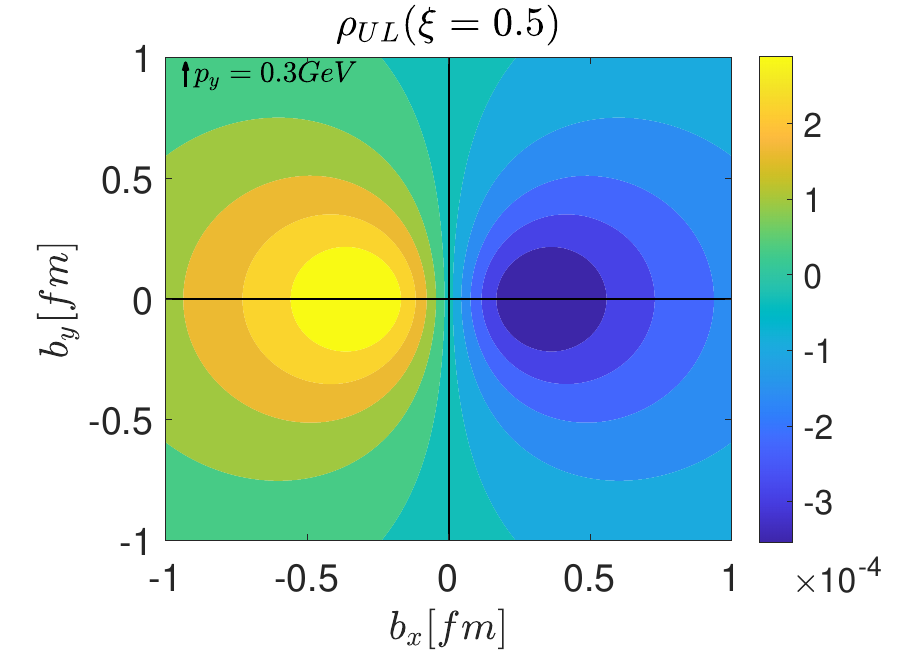}
    \caption{The first Mellin moment of gluon Wigner distribution $\rho_{UL}$ for different values of skewness parameter ($\xi= 0, 0.1,0.5$) in transverse momentum space (upper panel) and impact parameter space (lower panel) for fixed $\bfb=0.2$ fm $\hat{y}$ and $\bfp=0.3$ GeV $\hat{y}$, respectively, with the condition $\bfp\perp\bfd$.}
    \label{fig:b_rhoUL}
\end{figure}

In Fig.~\ref{fig:b_rhoUL}, we present the Wigner distribution $\rho_{UL}(\bfb,\bfp)$, which describes the transverse phase-space distribution of longitudinally polarized gluons inside an unpolarized proton. The Wigner distribution $\rho_{UL}(\boldsymbol{b}_\perp,\boldsymbol{p}_\perp)$ exhibits a dipolar structure in both $\bfp$ and $\bfb$ planes due to the presence of the term $\bfp \times \bfb$ in Eq.~(\ref{eq:rho_{UL}}). Notably, the polarity of the distribution in momentum space is opposite to that observed in impact-parameter space. A similar qualitative behavior is observed in other models for gluons \cite{Tan:2023vvi, More:2017zqp} and also for quarks \cite{Chakrabarti:2017teq}.  Moreover, \cite{More:2017zqp} reports a distorted dipolar structure in $\bfb-$space. As $\xi$ value increases from $\xi=0$ to $\xi=0.1$, the peak magnitudes of the distributions decrease by about $10$-times in both spaces, while for $\xi=0.5$, the peak magnitude decreases by a larger amount.
Furthermore, in transverse momentum space, the dipolar separation between the maxima and minima peaks shows oscillatory behavior over a change in $\xi$. For $\xi = 0$, the distribution attains its maximum at $p_x \approx 0.19$ GeV and with increasing skewness, the peak shifts toward larger values of $p_x$, reaching $p_x \approx 0.49$ GeV for $\xi = 0.1$, which corresponds to an increase of about $157\%$ whereas, slightly decreases towards center to $p_x \approx 0.47$ GeV for $\xi = 0.5$. In contrast, in impact parameter space, the position of the maximum moves closer to the center with increasing $\xi$. 
In the case of $\bfb$-plane, the dipolar separation gradually decreases with increasing $\xi$. The distribution attains its maxima at $b_x \approx 1.42$ fm for $\xi = 0$. As the skewness increases, the peak position shifts toward smaller values of $b_x$, reaching $b_x \approx 0.65$ fm at $\xi = 0.1$, corresponding to a decrease of approximately $54\%$, and further for $\xi = 0.5$, the separation is reduced by half to $b_x \approx 0.31$ fm. One can also observe that the magnitude of peaks is approximately of the similar order in both the transverse momentum and impact parameter planes corresponding to the respective three $\xi$ values. 
\begin{figure}[h]
    \centering
     \includegraphics[width=0.30\linewidth, trim=80 240 100 240, clip]{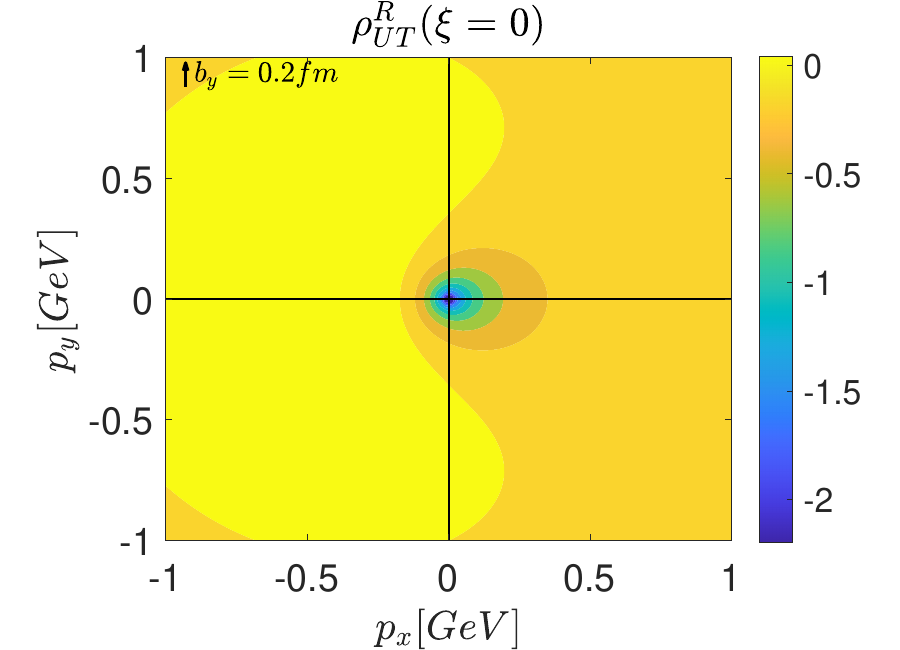}
     \includegraphics[width=0.30\linewidth, trim=80 240 100 240, clip]{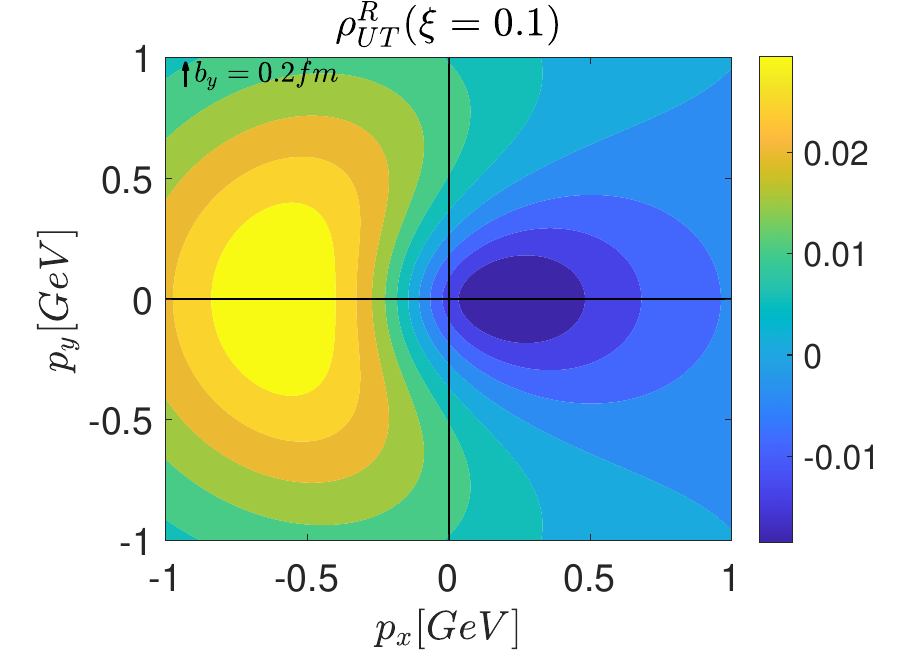}
     \includegraphics[width=0.30\linewidth, trim=80 240 100 240, clip]{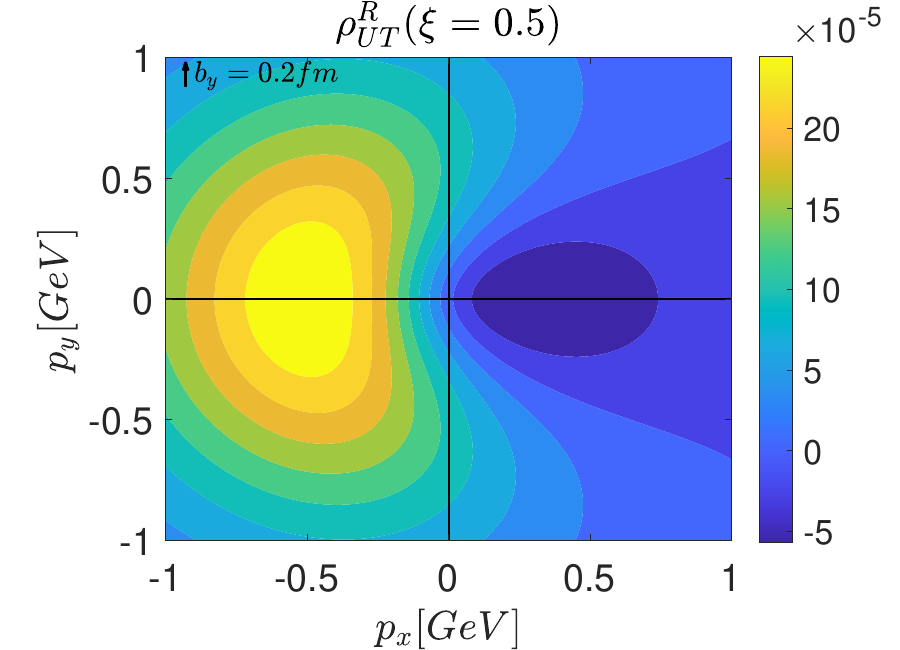}\\ \vspace{0cm}
     \includegraphics[width=0.30\linewidth, trim=80 240 100 240, clip, trim=80 240 100 240, clip]{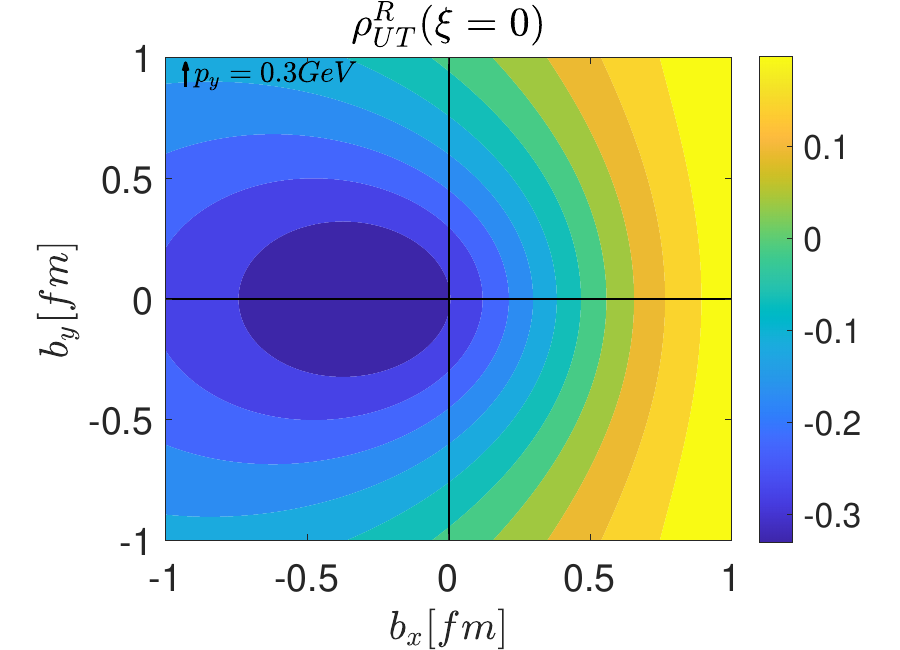}
     \includegraphics[width=0.30\linewidth, trim=80 240 100 240, clip]{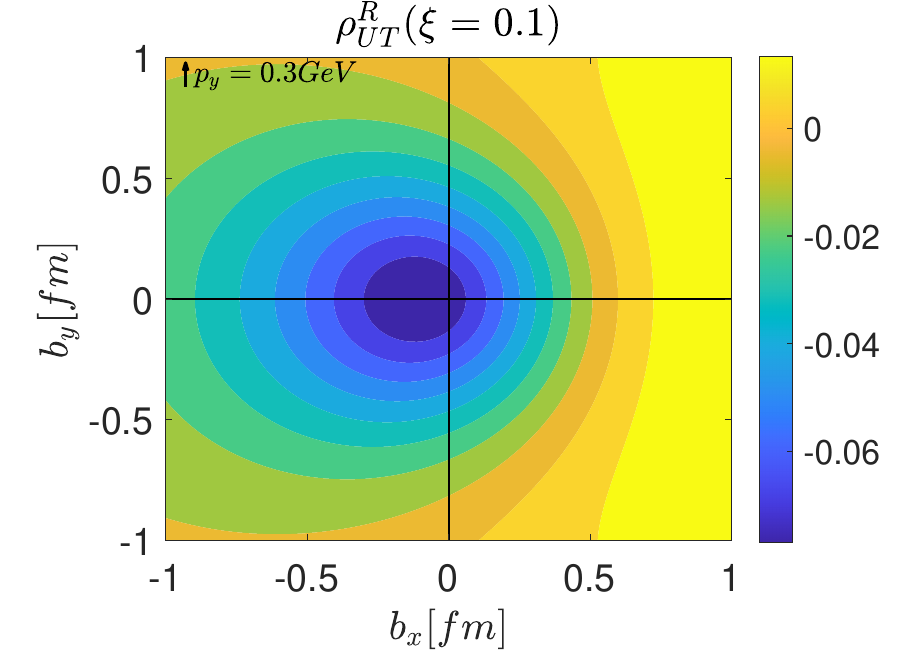}
     \includegraphics[width=0.30\linewidth, trim=80 240 100 240, clip]{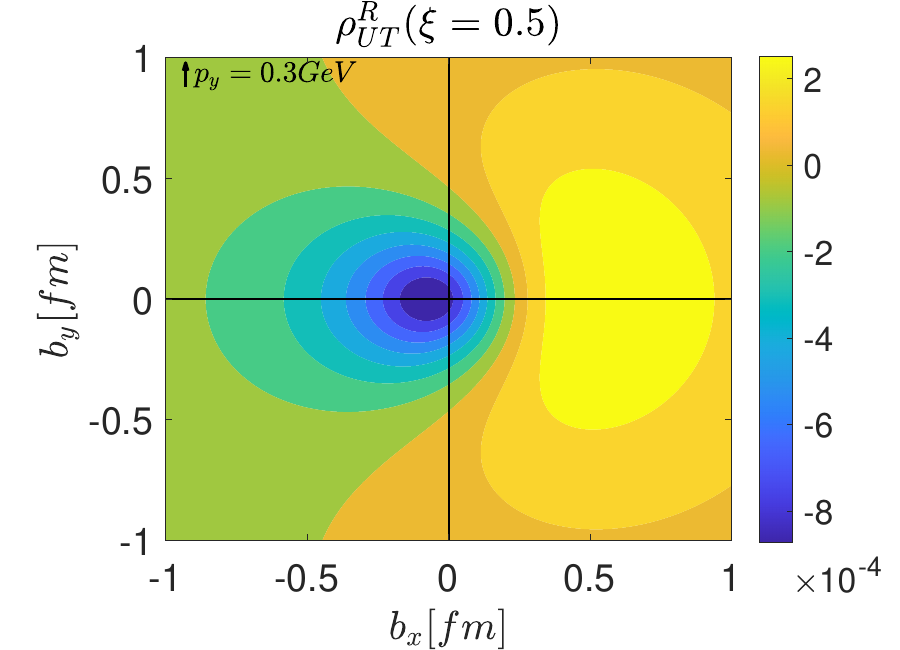}
    \caption{The first Mellin moment of gluon Wigner distribution $\rho^{R}_{UT}$ for different values of skewness parameter ($\xi= 0, 0.1,0.5$) in transverse momentum space (upper panel) and impact parameter space (lower panel) for fixed $\bfb=0.2$ fm $\hat{y}$ and $\bfp=0.3$ GeV $\hat{y}$, respectively, with the condition $\bfp\perp\bfd$.}
    \label{fig:b_rhoUTR}
\end{figure}
\begin{figure}[h]
    \centering
     \includegraphics[width=0.30\linewidth, trim=80 240 100 240, clip]{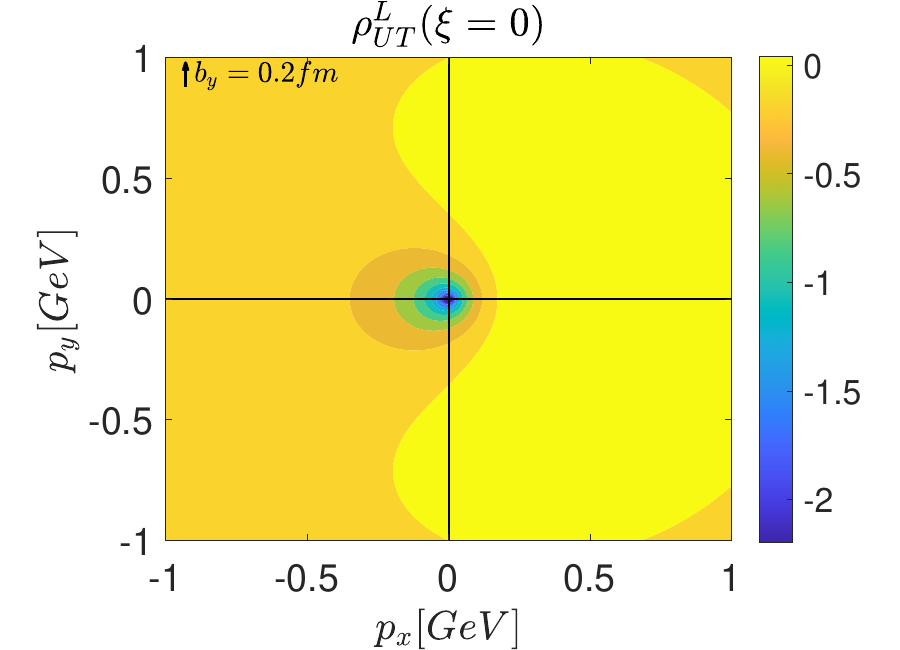}
     \includegraphics[width=0.30\linewidth, trim=80 240 100 240, clip]{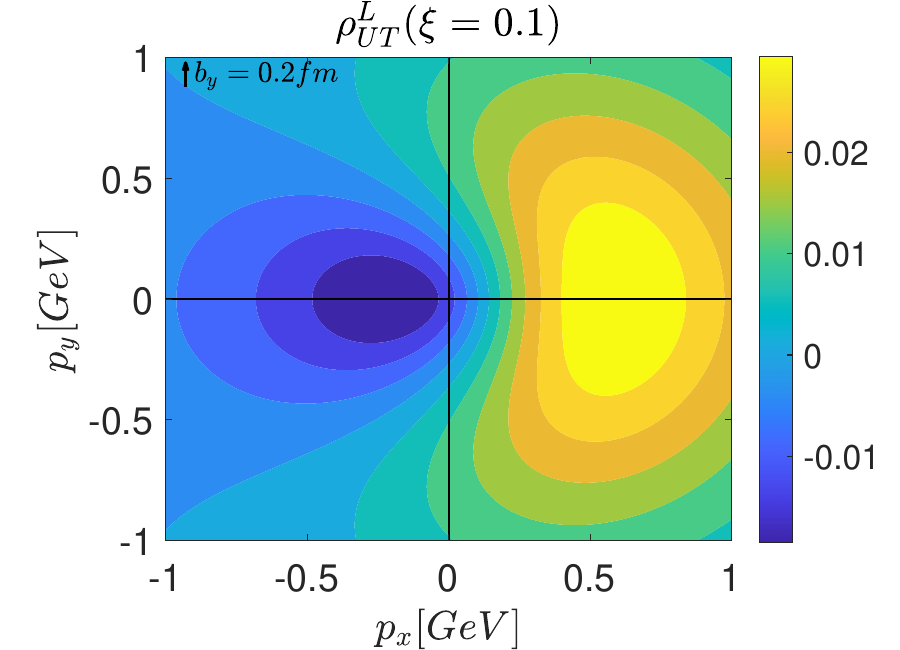}
     \includegraphics[width=0.30\linewidth, trim=80 240 100 240, clip]{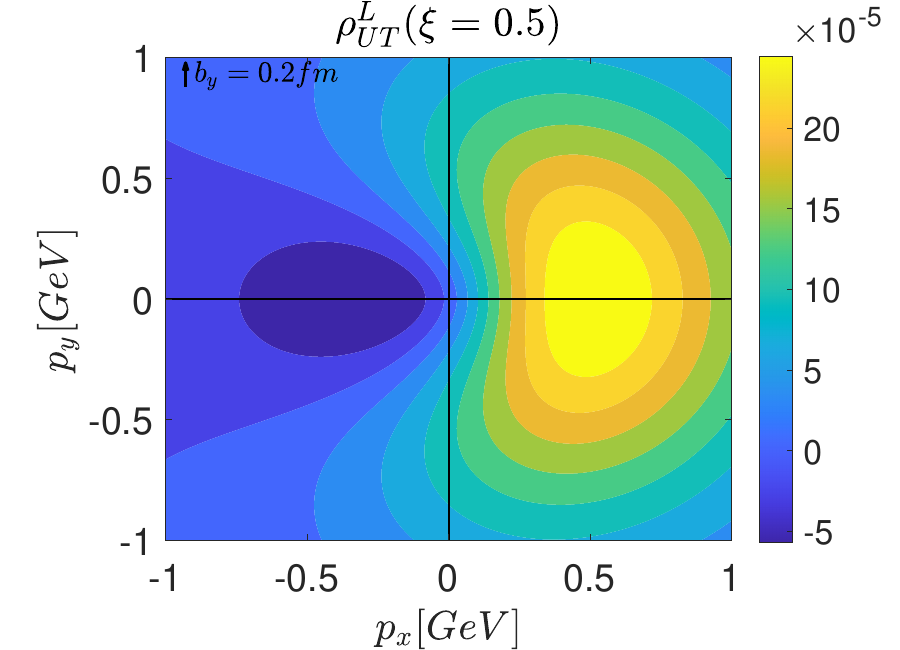}\\ \vspace{0cm}
      \includegraphics[width=0.30\linewidth, trim=80 240 100 240, clip]{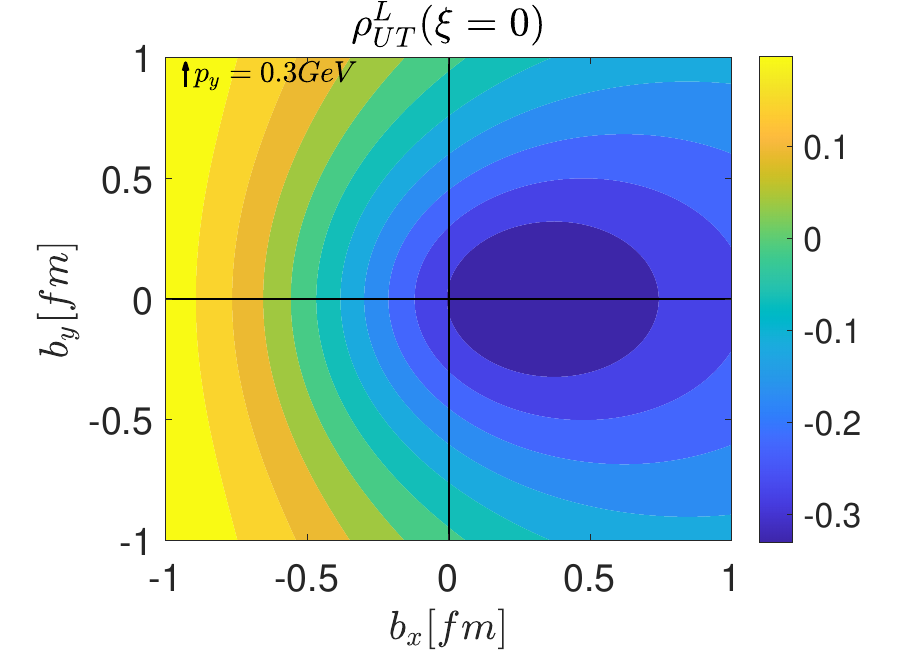}
     \includegraphics[width=0.30\linewidth, trim=80 240 100 240, clip]{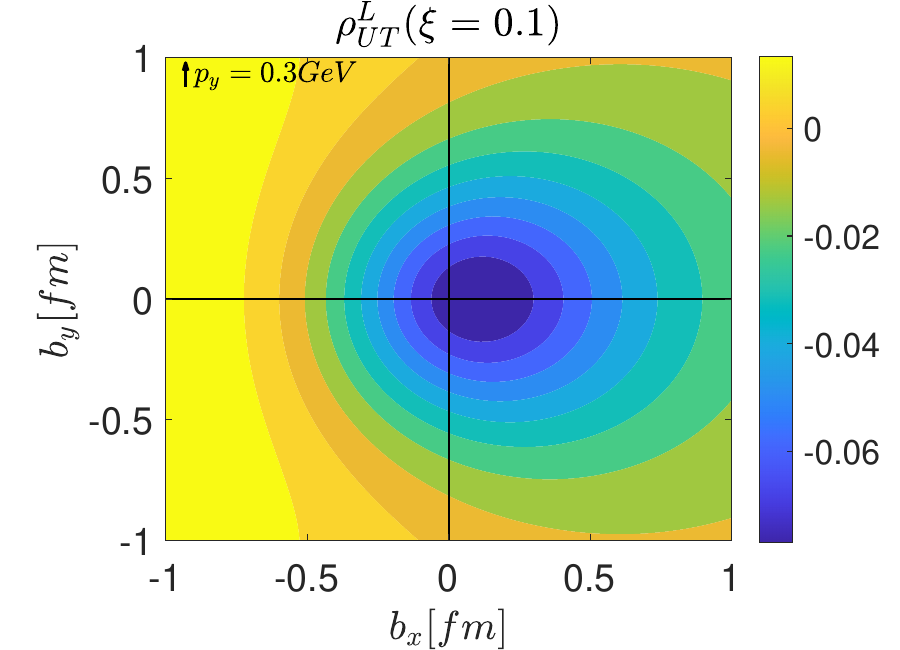}
     \includegraphics[width=0.30\linewidth, trim=80 240 100 240, clip]{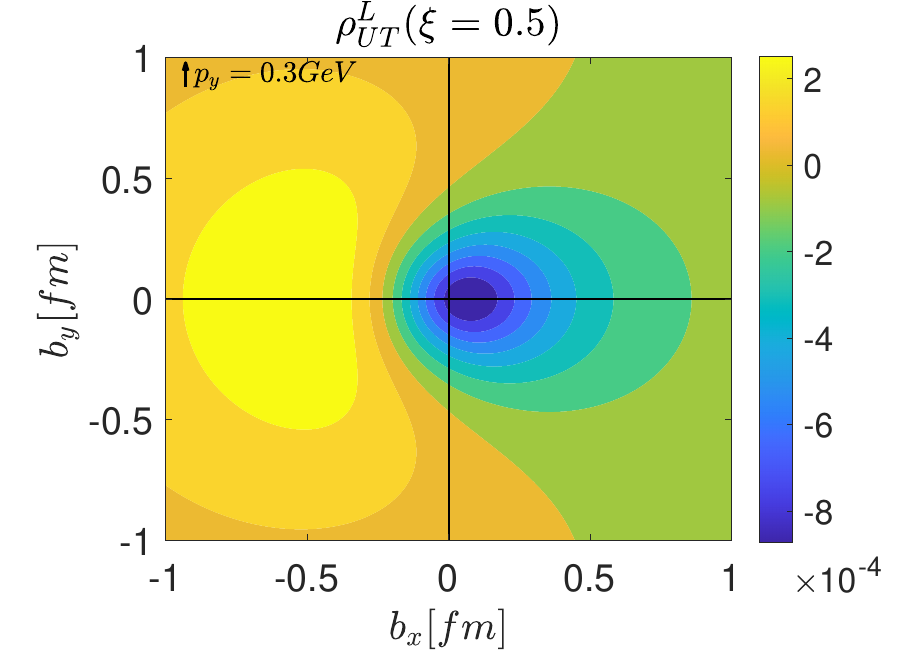}
    \caption{The first Mellin moment of gluon Wigner distribution $\rho^{L}_{UT}$ for different values of skewness parameter ($\xi= 0, 0.1,0.5$) in transverse momentum space (upper panel) and impact parameter space (lower panel) for fixed $\bfb=0.2$ fm $\hat{y}$ and $\bfp=0.3$ GeV $\hat{y}$, respectively, with the condition $\bfp\perp\bfd$.}
    \label{fig:b_rhoUTL}
\end{figure}
\begin{figure}[h]
    \centering
     \includegraphics[width=0.30\linewidth, trim=80 240 100 240, clip]{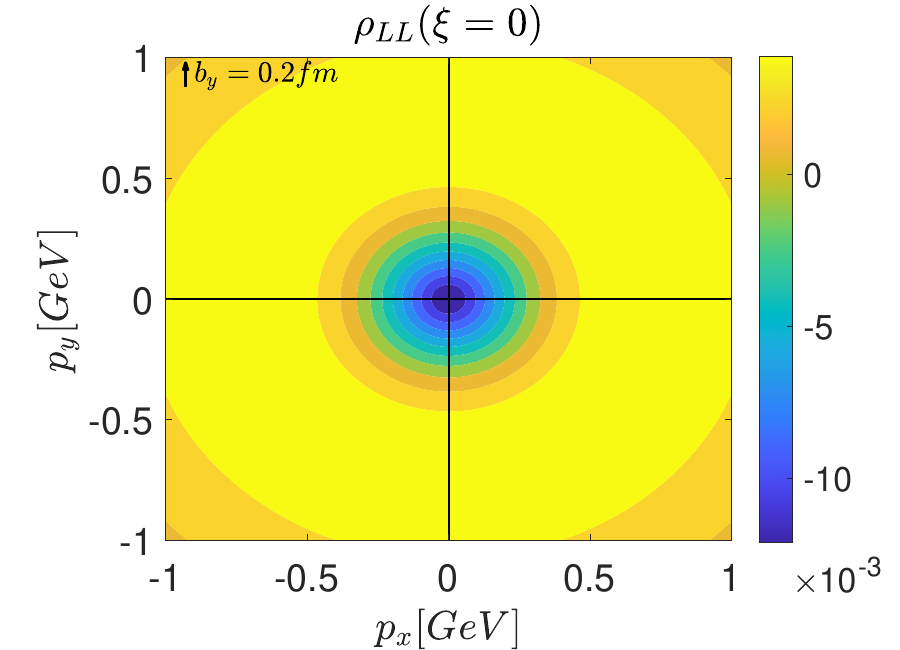}
     \includegraphics[width=0.30\linewidth, trim=80 240 100 240, clip]{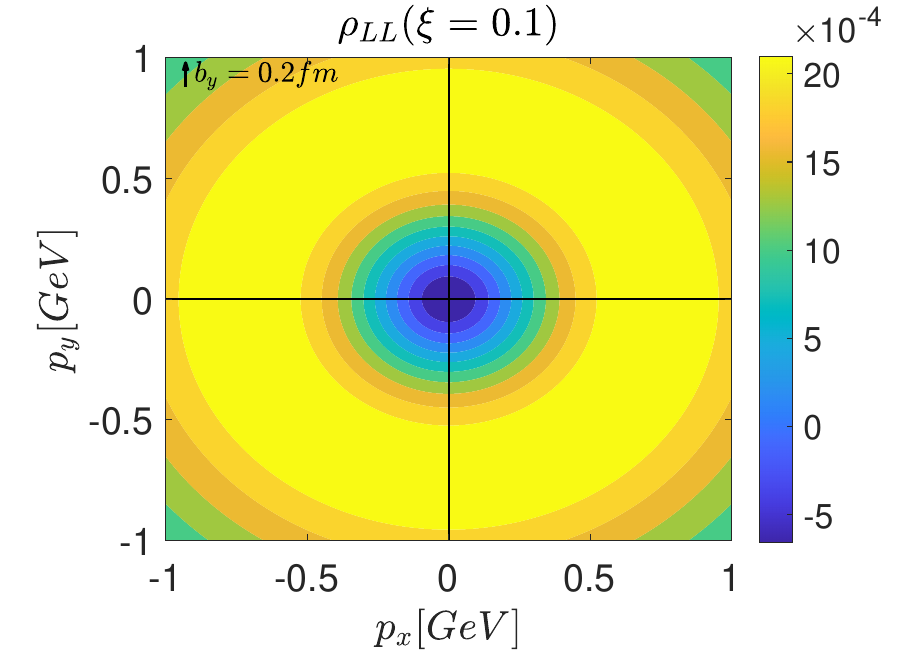}
     \includegraphics[width=0.30\linewidth, trim=80 240 100 240, clip]{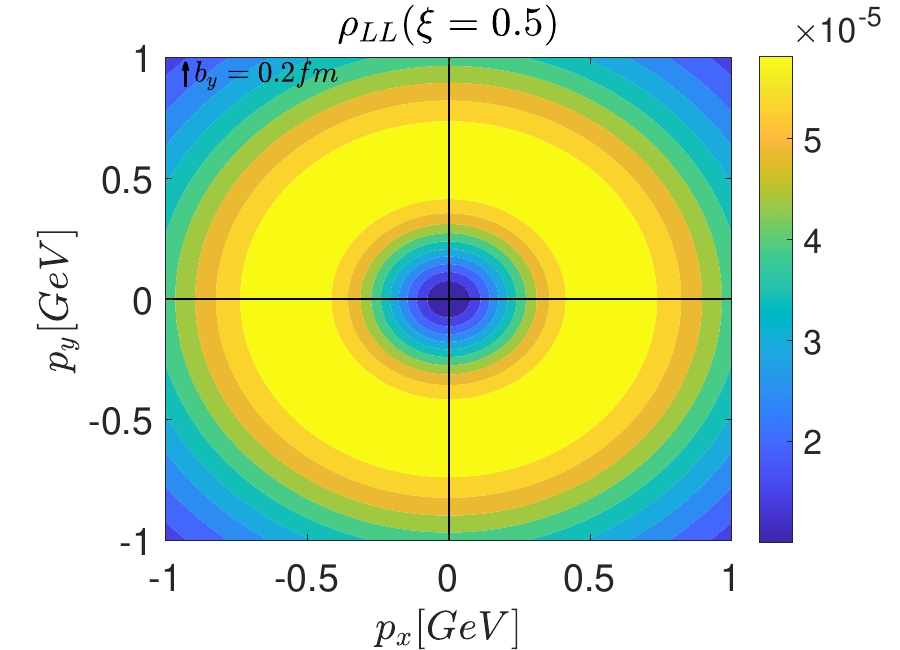}\\ \vspace{0cm}
      \includegraphics[width=0.30\linewidth, trim=80 240 100 240, clip]{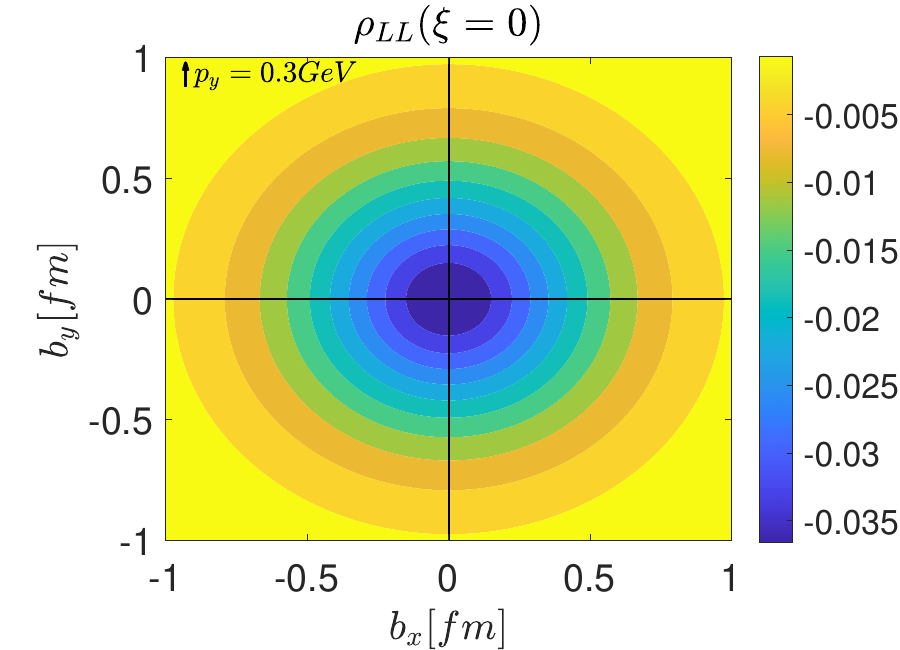}
     \includegraphics[width=0.30\linewidth, trim=80 240 100 240, clip]{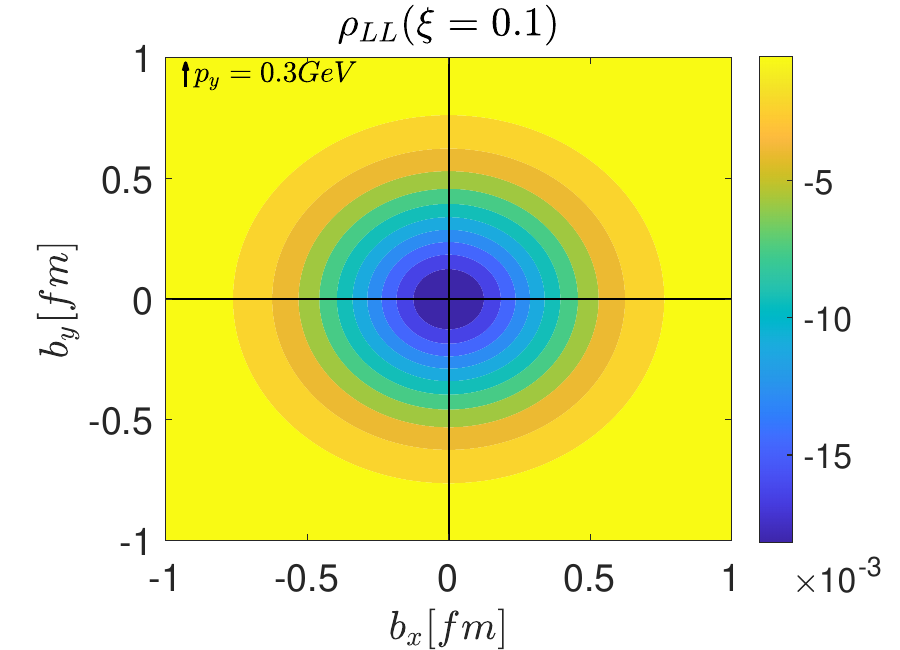}
     \includegraphics[width=0.30\linewidth, trim=80 240 100 240, clip]{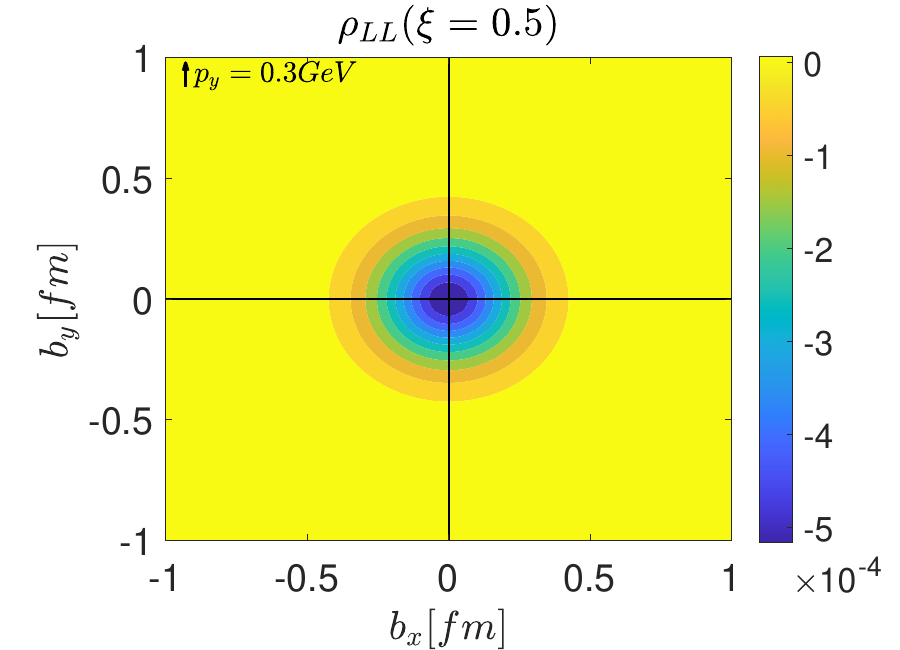}
    \caption{The first Mellin moment of gluon Wigner distribution $\rho_{LL}$ for different values of skewness parameter ($\xi= 0, 0.1,0.5$) in transverse momentum space (upper panel) and impact parameter space (lower panel) for fixed $\bfb=0.2$ fm $\hat{y}$ and $\bfp=0.3$ GeV $\hat{y}$, respectively, with the condition $\bfp\perp\bfd$.}
    \label{fig:b_rhoLL}
\end{figure} 
The distribution $\rho_{UL}(\bfb,\bfp)$ is related to gluon spin-orbit correlation, $C^g_z$, defined as
$$
C^g_z=\int dxd^2\bfp d^2\bfb (\bfb\times\bfp) \rho_{UL}(x,0,\bfb,\bfp).
$$ 
within the forward limit $\bfd=0$ and $\xi=0$. The qualitative behavior of the spin-OAM correlation of the gluon can be analyzed by calculating $C^g_z$- the positive(negative) value of $C^g_z$ indicates a parallel(anti-parallel) configuration of the gluon spin to its OAM.
In this model, we obtain $C^g_z = -15.6$, which is a bit larger in magnitude than the value $C^g_z \simeq -10.05$ reported in \cite{Tan:2023vvi}, however, consistent in the negative sign that indicates the gluon spin and gluon OAM are anti-aligned to each other. 
In this model, the strengths of the correlation for gluon spin-OAM 
exhibit a higher value than the other model prediction, that inherent model dependence may arise because of its sensitivity to transverse-momentum correlations or higher-twist operators that involve the quark-gluon interactions. The light-front quark-diquark(LFQDM) model shows quark spin and OAM are mutually anti-parallel in a proton \cite{Chakrabarti:2017teq}; however, this correlation strength for gluon is found to be significantly dominating over gluon. 

In Figs.~\ref{fig:b_rhoUTR}, \ref{fig:b_rhoUTL}, we show Wigner distributions of linearly polarized gluons (R and L) inside an unpolarized proton represented by $\rho^{R}_{UT}(\bfb,\bfp)$ and $\rho^{L}_{UT}(\bfb,\bfp)$, respectively. For $\rho^{R}_{UT}$, the distribution exhibits a distorted dipolar structure in both the transverse momentum and impact parameter space for all $\xi$ values. 
For $\xi = 0$, a pronounced asymmetry in the dipole structure having a narrow higher negative peak near origin can be understood from Eq.~(\ref{eq:rho_{UTR}}) that contains the dominant term $\rho_{UU}$ compare to $\rho_{UL}$ as shown in Fig.~\ref{fig:b_rhoUU} (first upper subfigure) and Fig.~\ref{fig:b_rhoUL} (first upper subfigure). 
As the skewness increases, the second term $\rho_{UL}$ of Eq.~(\ref{eq:rho_{UTR}}) dominates and the prominent dipole structures are observed for $\xi = 0.1$ and $\xi = 0.5$. However, the magnitude of peaks decreases rapidly, and the separation between poles is not significantly sensitive for larger $\xi$ region. 
In the impact parameter space, with the increase of the skewness, the distribution becomes a distorted dipolar type with suppressed peak intensity and the negative pole approaches towards the origin. 
\begin{figure}[h]
    \centering
     \includegraphics[width=0.30\linewidth, trim=80 240 100 240, clip]{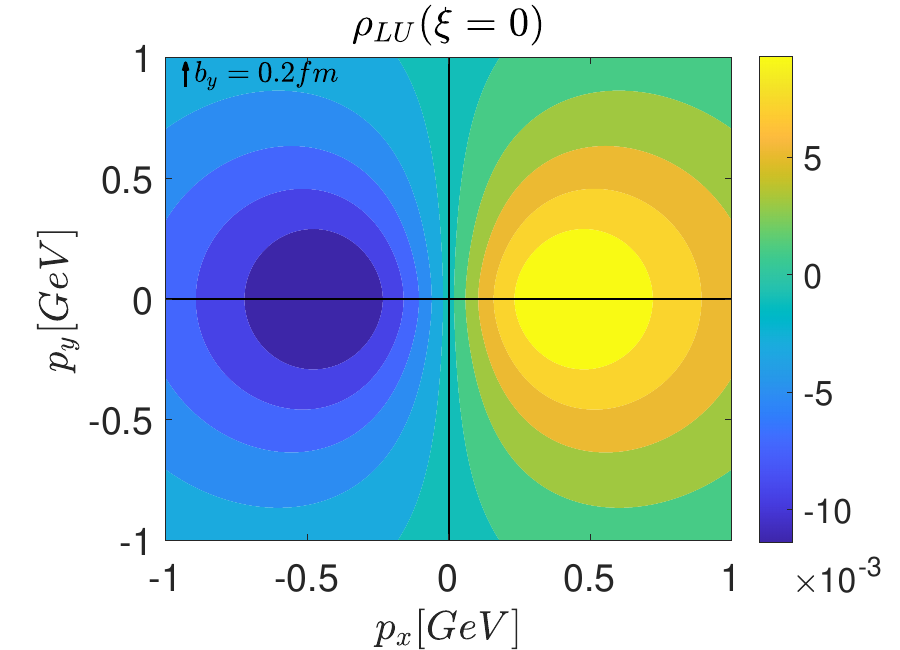}
     \includegraphics[width=0.30\linewidth, trim=80 240 100 240, clip]{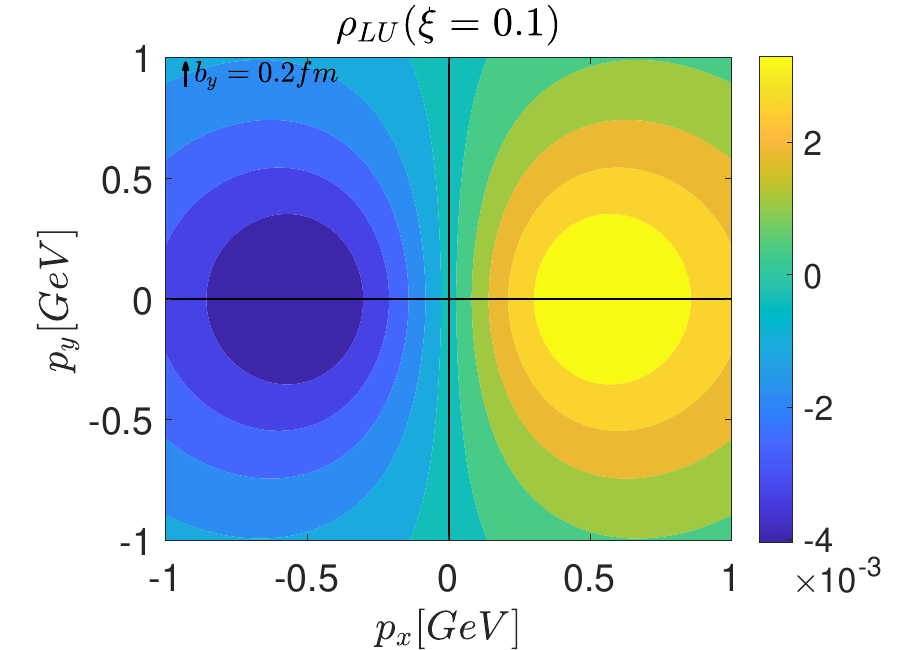}
     \includegraphics[width=0.30\linewidth, trim=80 240 100 240, clip]{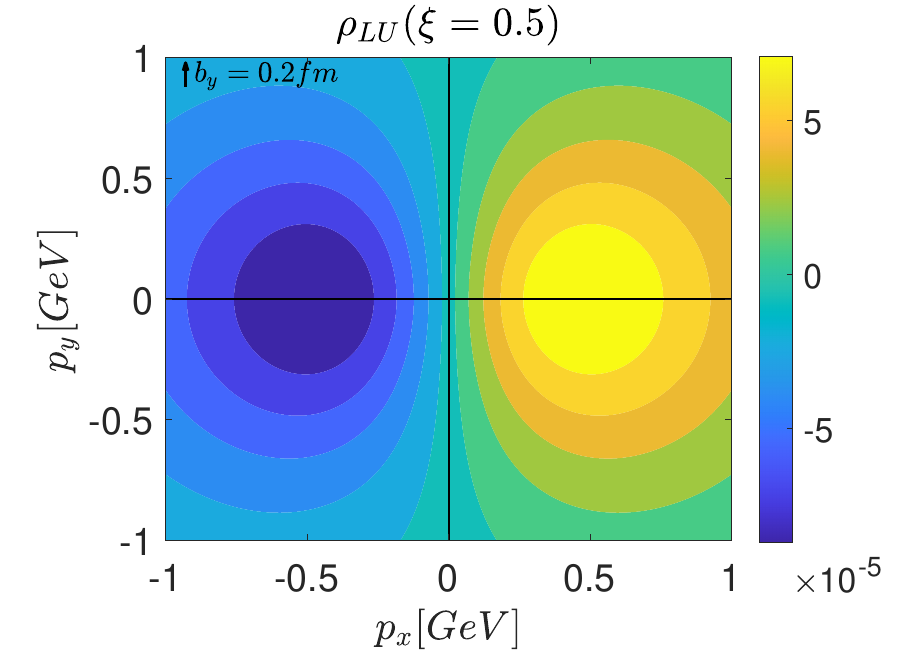}\\ \vspace{0cm}
      \includegraphics[width=0.30\linewidth, trim=80 240 100 240, clip]{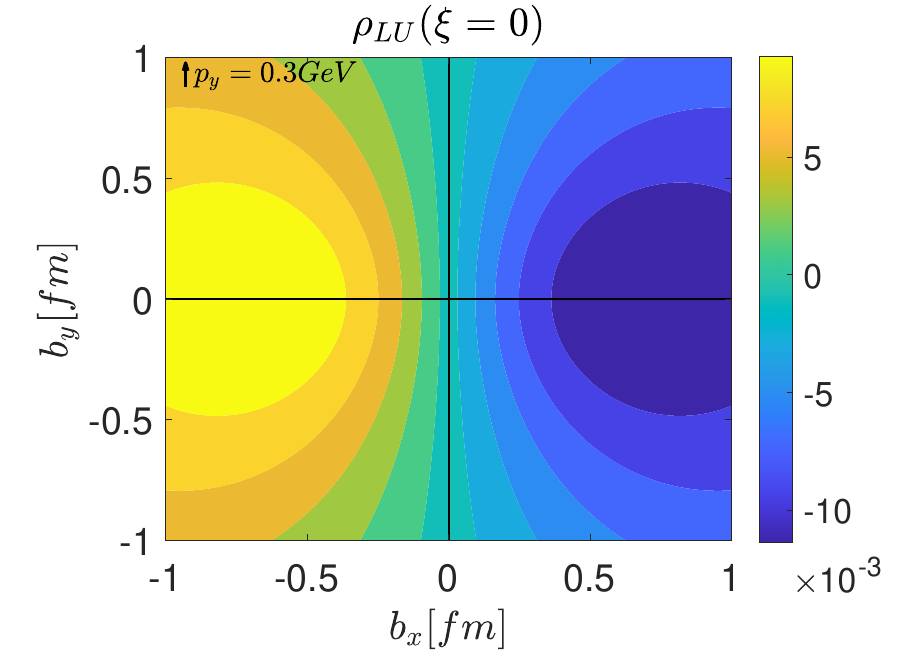}
     \includegraphics[width=0.30\linewidth, trim=80 240 100 240, clip]{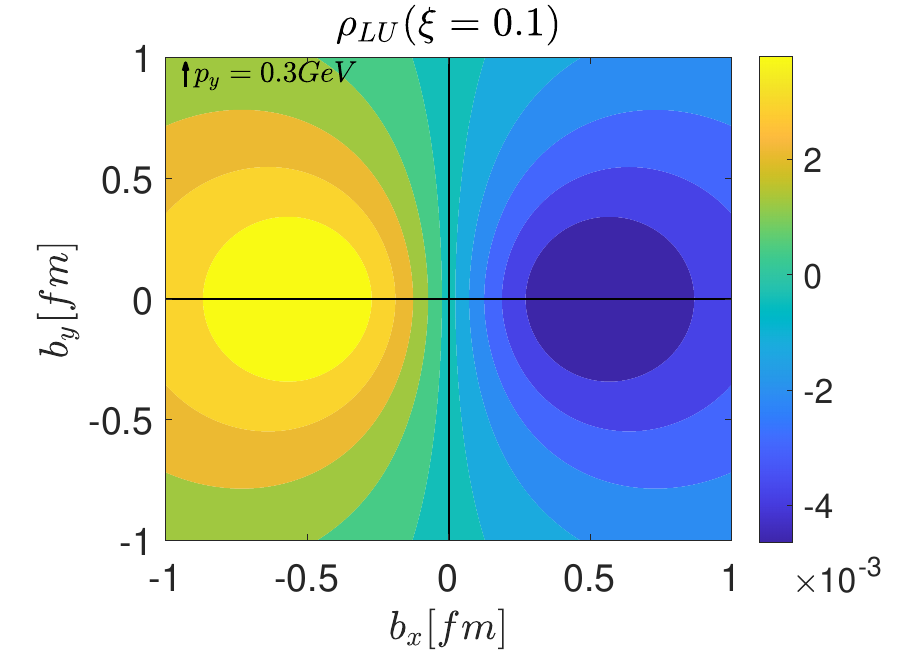}
     \includegraphics[width=0.30\linewidth, trim=80 240 100 240, clip]{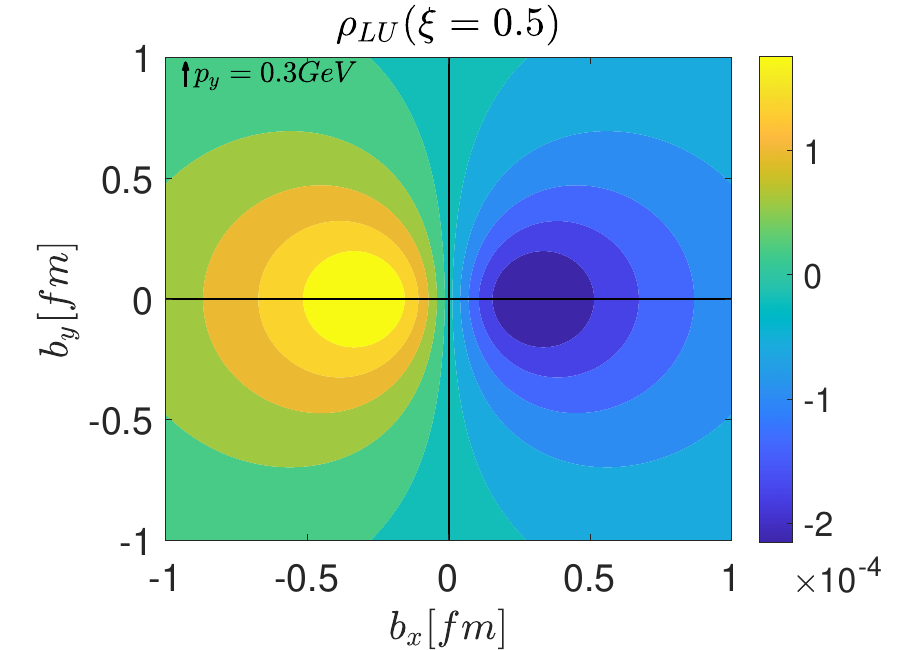}
    \caption{The first Mellin moment of gluon Wigner distribution $\rho_{LU}$ for different values of skewness parameter ($\xi= 0, 0.1,0.5$) in transverse momentum space (upper panel) and impact parameter space (lower panel) for fixed $\bfb=0.2$ fm $\hat{y}$ and $\bfp=0.3$ GeV $\hat{y}$, respectively, with the condition $\bfp\perp\bfd$.}
    \label{fig:b_rhoLU}
\end{figure}
\begin{figure}[h]
    \centering
     \includegraphics[width=0.30\linewidth, trim=80 240 100 240, clip]{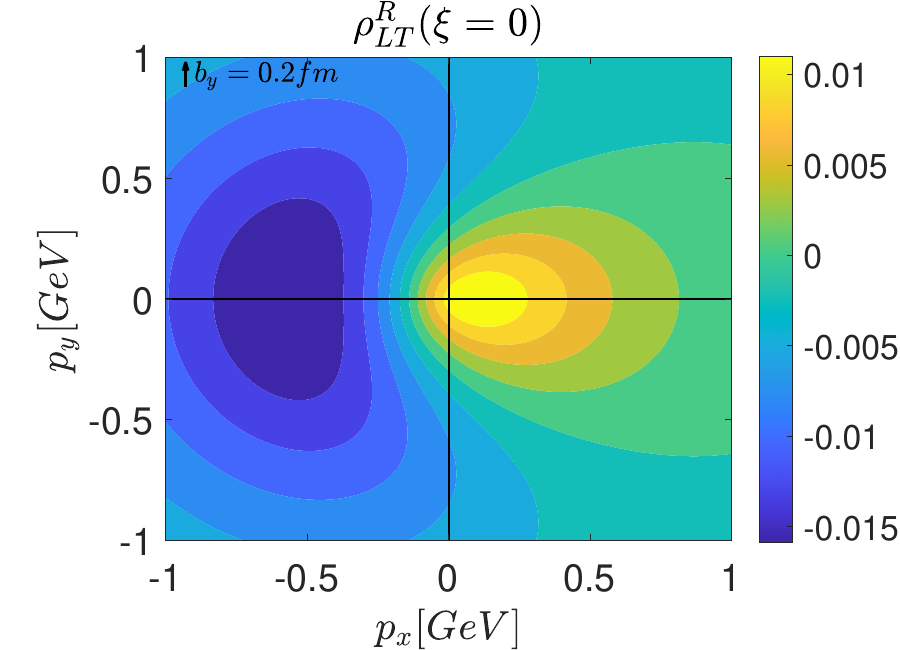}
     \includegraphics[width=0.30\linewidth, trim=80 240 100 240, clip]{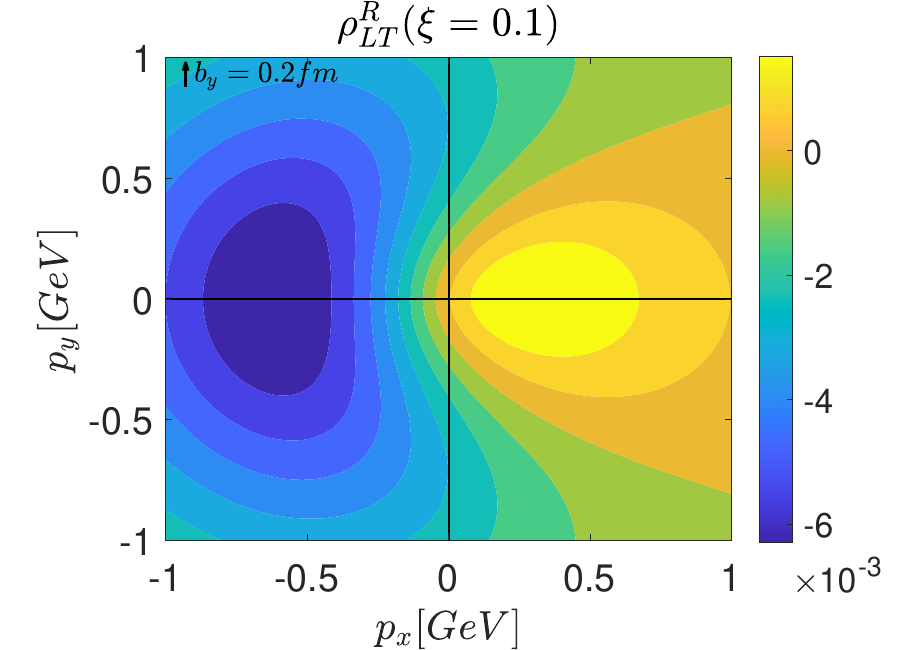}
     \includegraphics[width=0.30\linewidth, trim=80 240 100 240, clip]{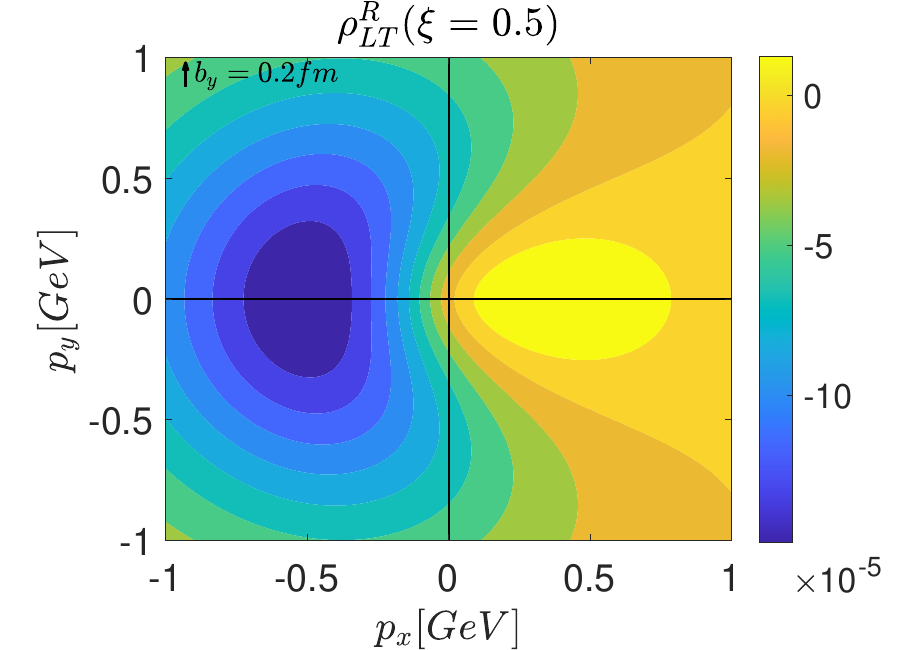}\\ \vspace{0cm}
     \includegraphics[width=0.30\linewidth, trim=80 240 100 240, clip]{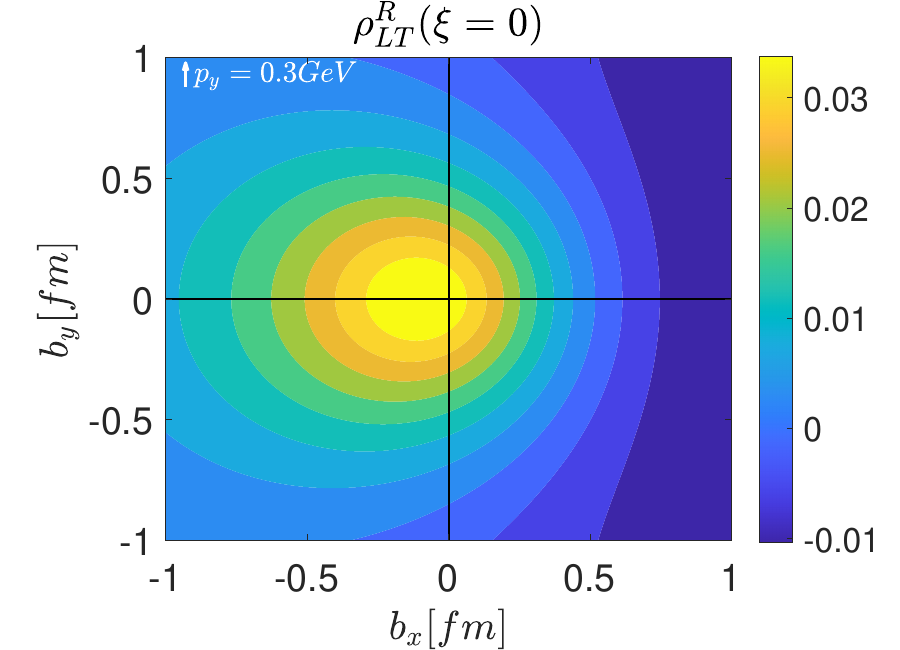}
     \includegraphics[width=0.30\linewidth, trim=80 240 100 240, clip]{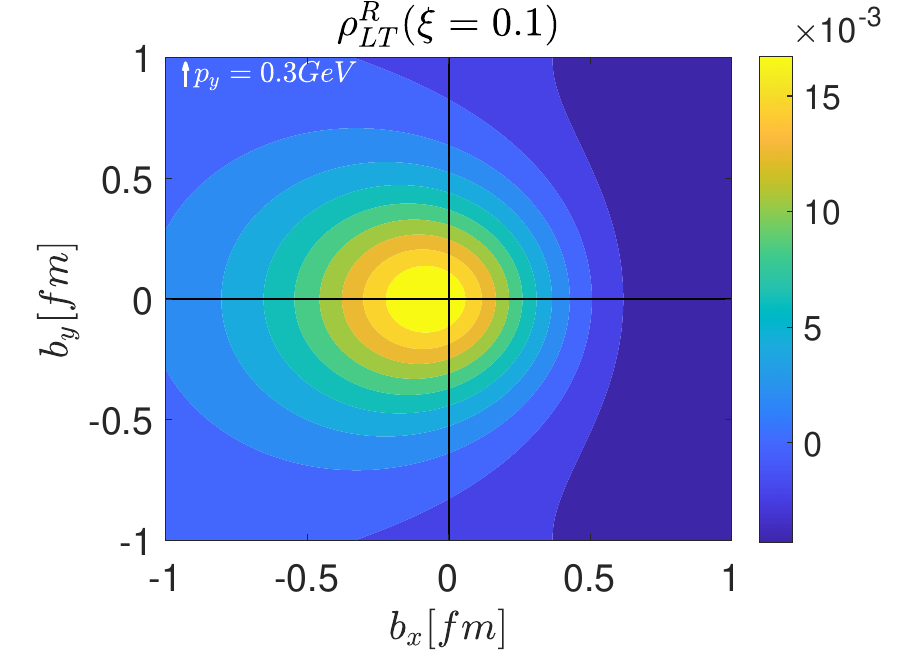}
     \includegraphics[width=0.30\linewidth, trim=80 240 100 240, clip]{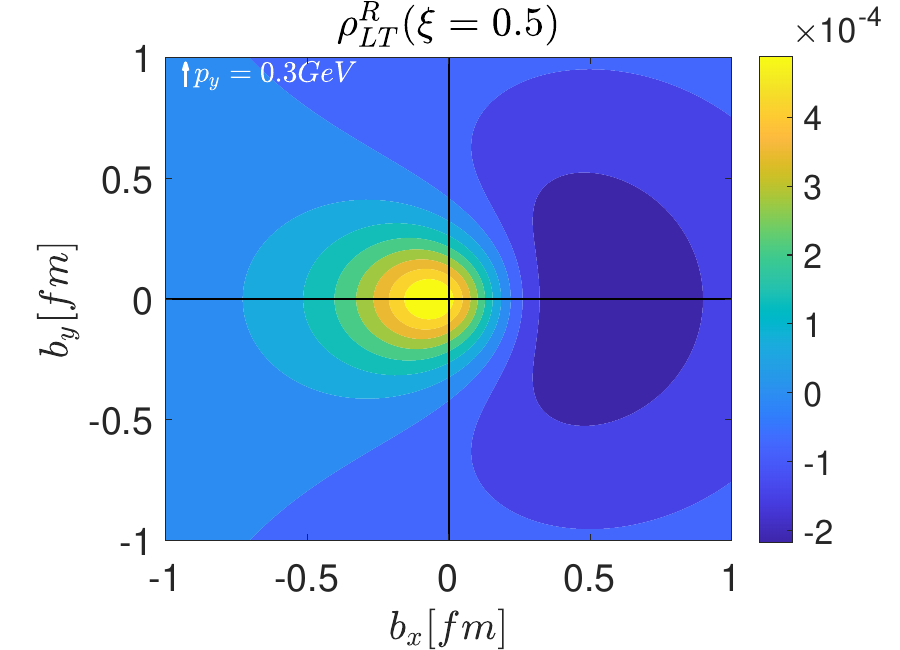}
    \caption{The first Mellin moment of gluon Wigner distribution $\rho^R_{LT}$ for different values of skewness parameter ($\xi= 0, 0.1,0.5$) in transverse momentum space (upper panel) and impact parameter space (lower panel) for fixed $\bfb=0.2$ fm $\hat{y}$ and $\bfp=0.3$ GeV $\hat{y}$, respectively, with the condition $\bfp\perp\bfd$.}
    \label{fig:b_rhoLTR}
\end{figure}
\begin{figure}[h]
    \centering
     \includegraphics[width=0.30\linewidth, trim=80 240 100 240, clip]{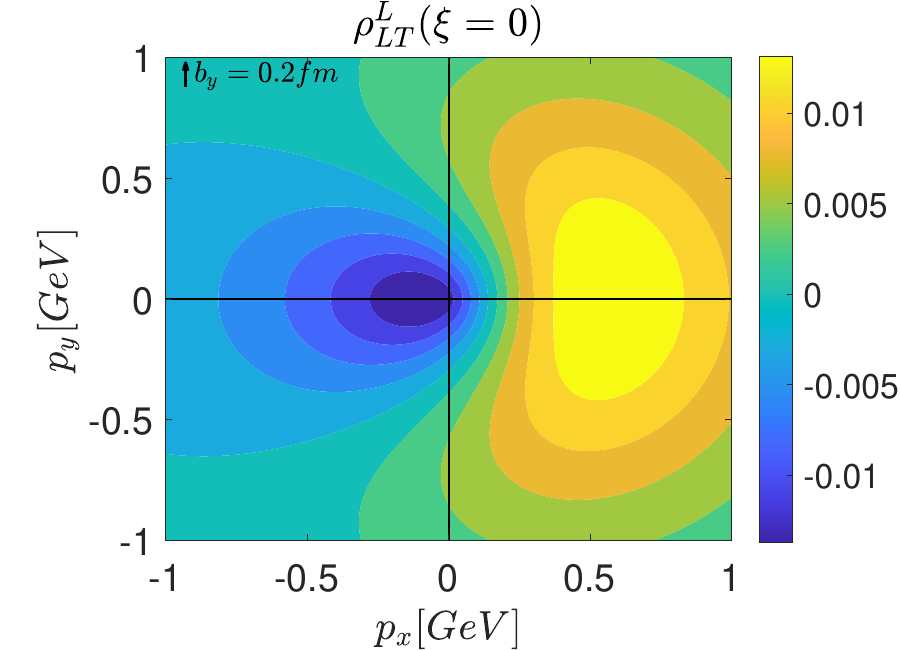}
     \includegraphics[width=0.30\linewidth, trim=80 240 100 240, clip]{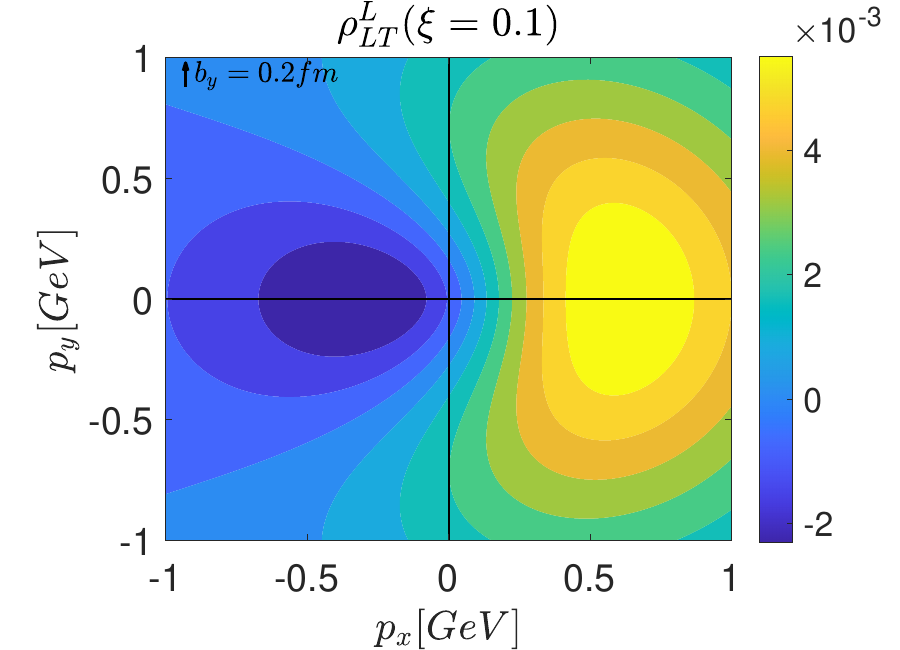}
     \includegraphics[width=0.30\linewidth, trim=80 240 100 240, clip]{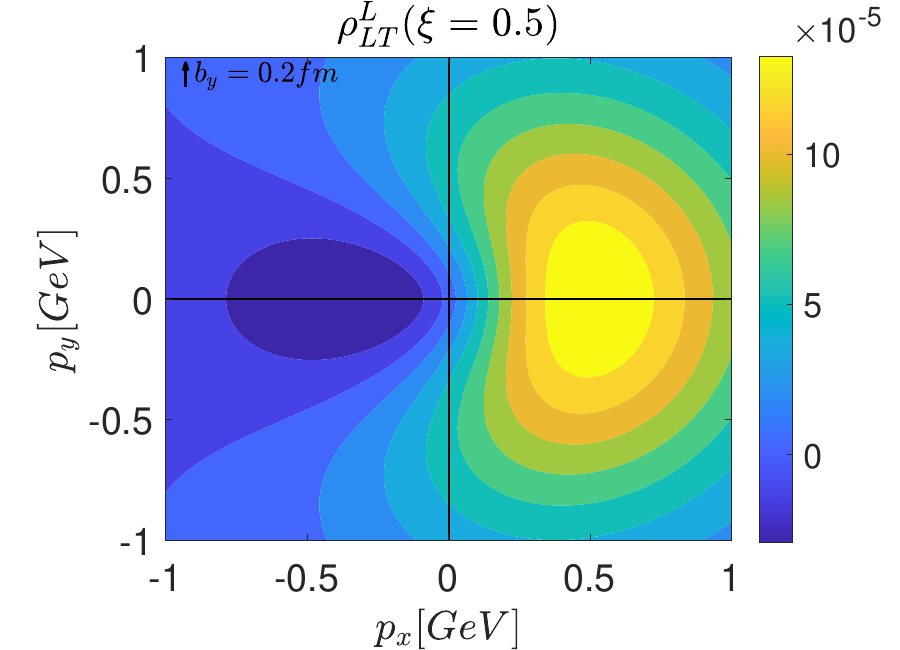}\\ \vspace{0cm}
      \includegraphics[width=0.30\linewidth, trim=80 240 100 240, clip]{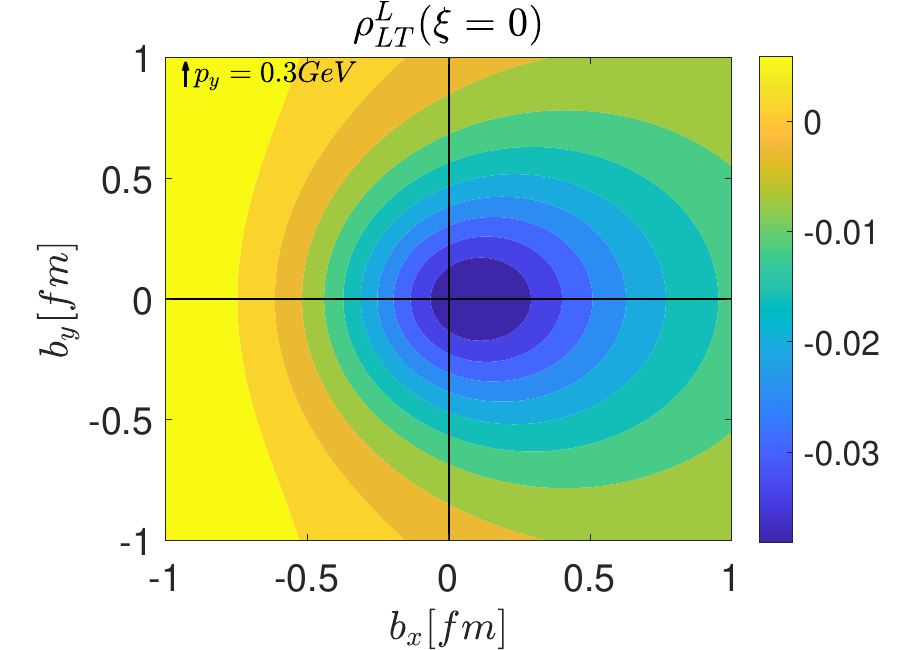}
     \includegraphics[width=0.30\linewidth, trim=80 240 100 240, clip]{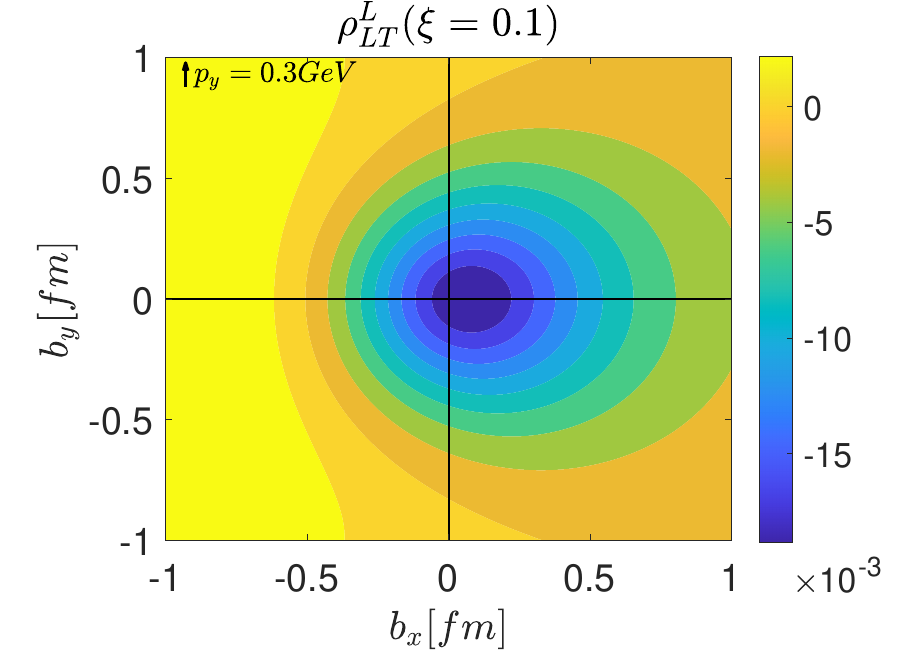}
     \includegraphics[width=0.30\linewidth, trim=80 240 100 240, clip]{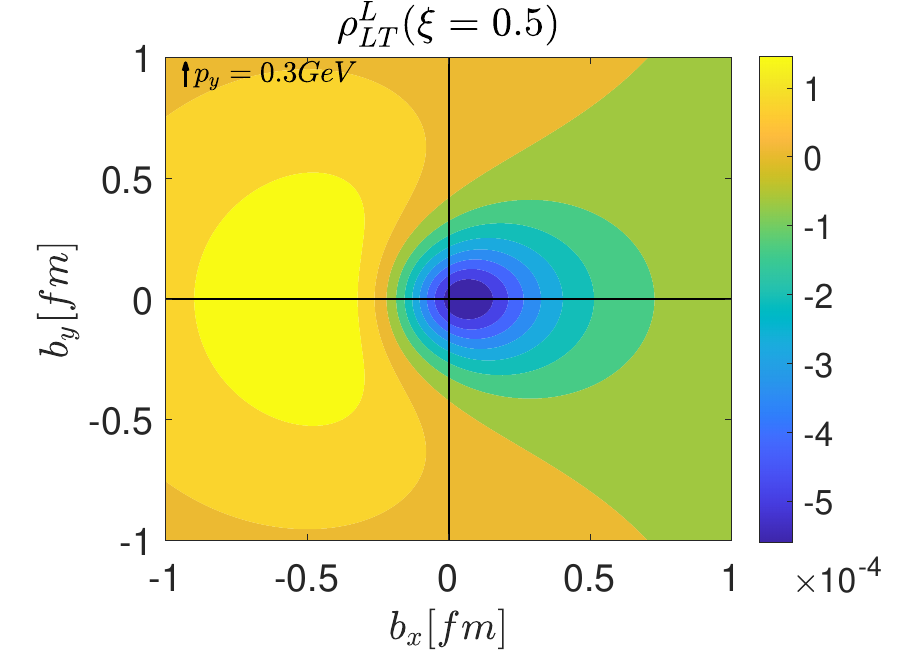}
    \caption{The first Mellin moment of gluon Wigner distribution $\rho^L_{LT}$ for different values of skewness parameter ($\xi= 0, 0.1,0.5$) in transverse momentum space (upper panel) and impact parameter space (lower panel) for fixed $\bfb=0.2$ fm $\hat{y}$ and $\bfp=0.3$ GeV $\hat{y}$, respectively, with the condition $\bfp\perp\bfd$.}
    \label{fig:b_rhoLTL}
\end{figure}
In Fig.~\ref{fig:b_rhoUTL}, the left-handed linearly polarized gluon distribution $\rho^{L}_{UT}$ exhibits the mirror image orientation of dipoles than the right-handed distribution $\rho_{UT}^{R}$ which can be justified from the opposite helicity structure of the right- and left-handed linearly polarized gluon states. An analogous behavior is observed in \cite{More:2017zqp} with opposite polarity of dipoles.
Fig.~\ref{fig:b_rhoLL} represents the WDs $\rho_{LL}$, where both proton and constituent gluons are longitudinally polarized. The distribution is axially symmetric in both $\bfp$-plane and $\bfb$-plane and shows opposite trend in the change of the peak-width with increasing $\xi$. Our model results exhibits similar polarity as reported in \cite{Tan:2023vvi}, while opposite polarity is observed in \cite{More:2017zqp}. This model result shows that, $\rho_{UU} > \rho_{LL}$.
WDs $\rho_{LL}$ has contribution from GTMD $G_{1,4}$ which encodes the spin contribution of gluons to the proton defined as
\be
s^g= \frac{1}{2}\int dx~ d^2p_\perp ~ G^g_{1,4}(x, 0, \bfp^2, 0,0),
\ee
at the forward limit $\bfd=0$ and $\xi=0$. In this model,  $s^g=0.215 $ for the whole range of the integration limit over $x$. It is observed that, most of the gluon contribution comes from the lower $x$ region. For the $x-$range $\{0.05 \to 0.3;~ 0.05 \to 0.2;~ 0.05 \to 1\}$, our model results $s^g=\{0.126,0.103,0.139 \}$ significantly consistent to the other phenomenological results $s^g=\{0.126,0.103,0.139 \}$ \cite{PHENIX:2008swq,deFlorian:2009vb,deFlorian:2014yva} respectively. A reasonably high gluon contribution to the proton spin is reported in \cite{Kaur:2019kpe,Joo:2019bzr}. 
The latest lattice prediction to the gluon total angular momentum contribution is $J_g=0.187$ \cite{Alexandrou:2020sml}.

In Fig.~\ref{fig:b_rhoLU}, we show the longitudinal-unpolarized Wigner distributions $\rho_{LU}(\bfb,\bfp)$ which describe the unpolarized gluon phase-space distributions in a longitudinal polarized proton, which exhibits dipolar behavior qualitatively similar to that of $\rho_{UL}$ because of the term associated to $\bfp\times\bfb$ in Eq. (\ref{eq:rho_{LU}}). A similar behavior is observed in other models \cite{More:2017zqp,Tan:2023vvi} for gluon.  While for quarks, an overall opposite polarity is observed in \cite{Chakrabarti:2017teq} for both the planes. 
In both planes, the peak magnitudes and the dipolar-separation are less sensitive to the change in $\xi=0, 0.1,$ and $0.5$, unlike $\rho_{UL}$. The model result shows that, $\rho_{UL} > \rho_{LU}$. 
The distribution $\rho_{LU}$ encodes information on the correlation between  canonical OAM of gluon and proton spin.
The gluon canonical orbital angular momentum (OAM), $l^g_z$, in the forward limit, $\bfd = 0$ and at skewness $\xi = 0$ is given by 
\be 
l^g_z=\int dxd^2\bfp d^2\bfb (\bfb\times\bfp) \rho_{LU}(x,0,\bfb,\bfp).
\ee 
The positive(negative) value of $l^g_z$ indicates a parallel(anti-parallel) configuration of the gluon OAM to the proton spin.  
In this model, we obtain a numerical value of $l^g_z = -0.375$, which is in good agreement with the result $l^g_z \simeq -0.33$ reported in \cite{Tan:2023vvi}. The negative sign indicates that the gluon OAM is found to be anti-parallel to the proton spin in this model. Whereas, in the case of quark, the quark OAM is reported parallel to the proton spin \cite{Chakrabarti:2017teq}. That can be justified from the sign flip polarity of the GTMDs $F_{1,4}$ to the $G_{1,1}$ for gluon (Eqs.~\ref{eq:F14}, \ref{eq:G11}).

Figs.~\ref{fig:b_rhoLTR}, \ref{fig:b_rhoLTL} illustrate the Wigner phase-space distributions corresponding to linearly polarized right-handed and left-handed gluons in a longitudinally polarized proton, $\rho^{R}_{LT}(\bfb,\bfp)$ and $\rho^{L}_{LT}(\bfb,\bfp)$, respectively. In transverse momentum space, the two distributions exhibit distorted dipolar structures because of the dominance of the first term in the Eqs. (\ref{eq:rho_{LTR}}) and (\ref{eq:rho_{LTL}}). With increasing $\xi$, the position of maxima is shifting away from the origin by a larger amount compared to the negative peak, and the separation between the poles gradually increases. 
In $\bfb$-space, the distributions are circularly asymmetric due to the dominance of the second term in  Eqs. (\ref{eq:rho_{LTR}}) and (\ref{eq:rho_{LTL}}). With increasing skewness parameter $\xi$, the distribution shows dipolar behavior and the position of the maxima becomes progressively localized closer to the origin, accompanied by a simultaneous reduction in both intensity and width.
The Wigner distribution $\rho_{LT}^{L}$ exhibits magnitudes and structural features identical to those of $\rho_{LT}^{R}$, differing only by an interchange of relative sign with a mirror symmetry along $y$-axis in both planes. In the dressed quark model \cite{More:2017zqp}, a dipolar behavior is reported with polarity flip for gluon. 

\begin{figure}[h]
    \centering
     \includegraphics[width=0.30\linewidth, trim=80 240 100 240, clip]{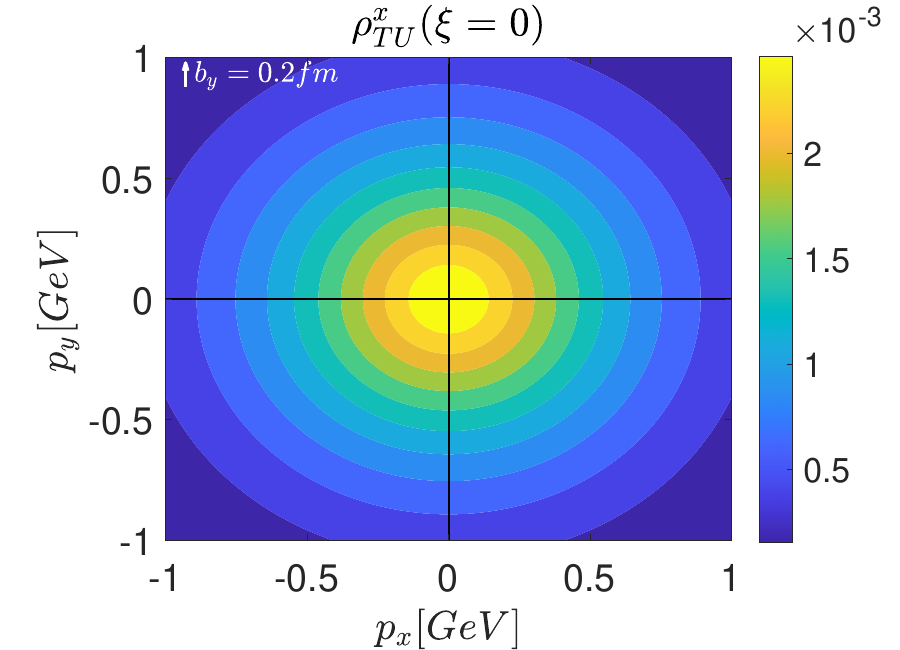}
     \includegraphics[width=0.30\linewidth, trim=80 240 100 240, clip]{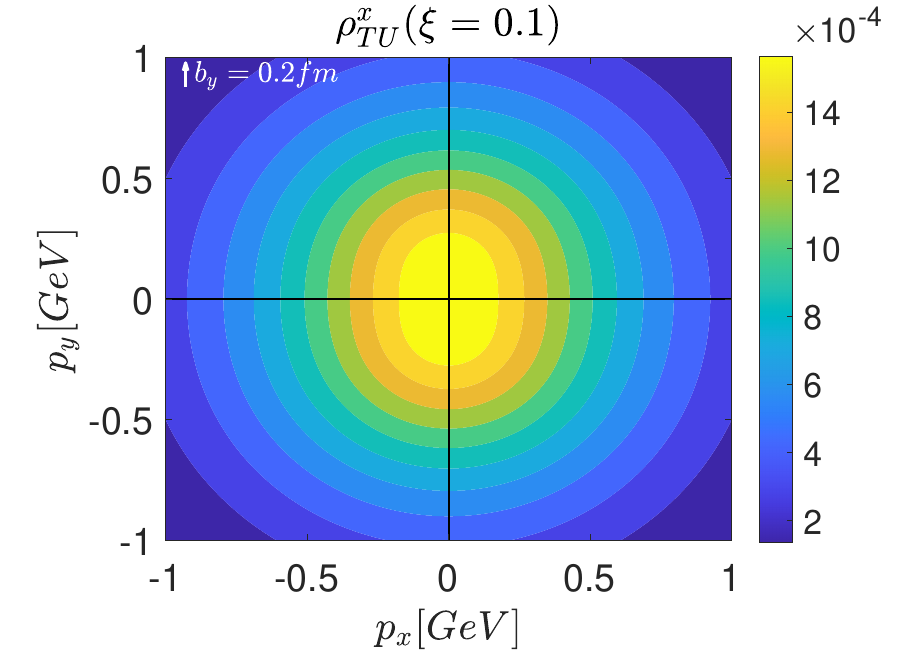}
     \includegraphics[width=0.30\linewidth, trim=80 240 100 240, clip]{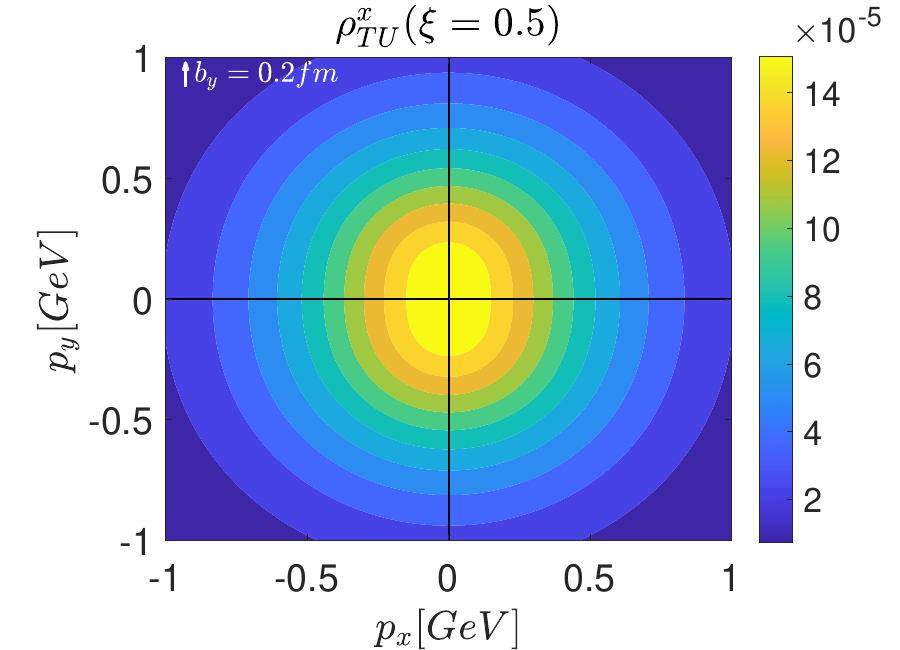}\\ \vspace{0cm}
      \includegraphics[width=0.30\linewidth, trim=80 240 100 240, clip]{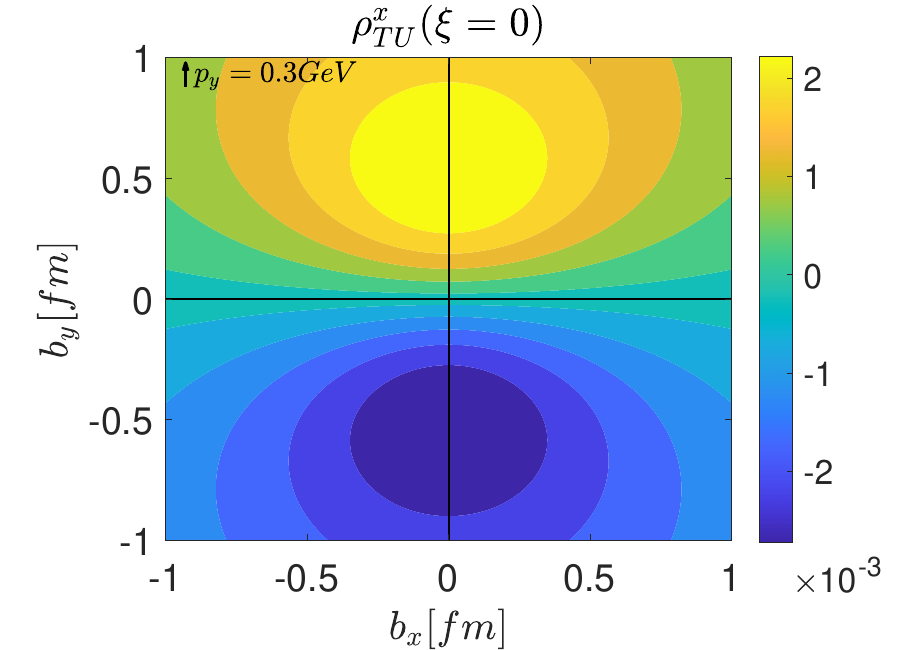}
     \includegraphics[width=0.30\linewidth, trim=80 240 100 240, clip]{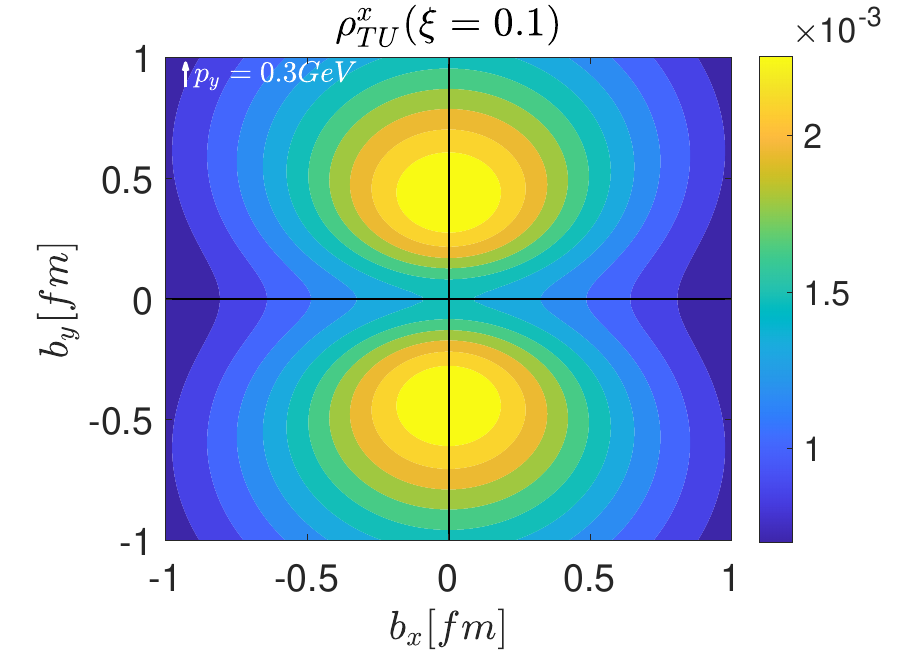}
     \includegraphics[width=0.30\linewidth, trim=80 240 100 240, clip]{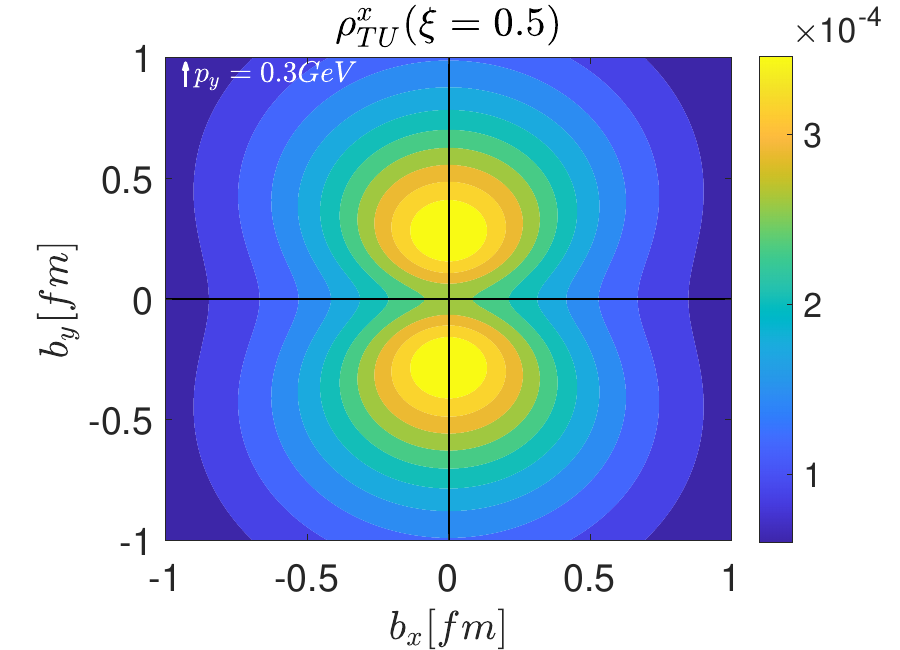}
    \caption{The first Mellin moment of gluon Wigner distribution $\rho^x_{TU}$ for different values of skewness parameter ($\xi= 0, 0.1,0.5$) in transverse momentum space (upper panel) and impact parameter space (lower panel) for fixed $\bfb=0.2$ fm $\hat{y}$ and $\bfp=0.3$ GeV $\hat{y}$, respectively, with the condition $\bfp\perp\bfd$.}
    \label{fig:b_rhoTU}
\end{figure}
\begin{figure}[h]
    \centering
     \includegraphics[width=0.30\linewidth, trim=80 240 100 240, clip]{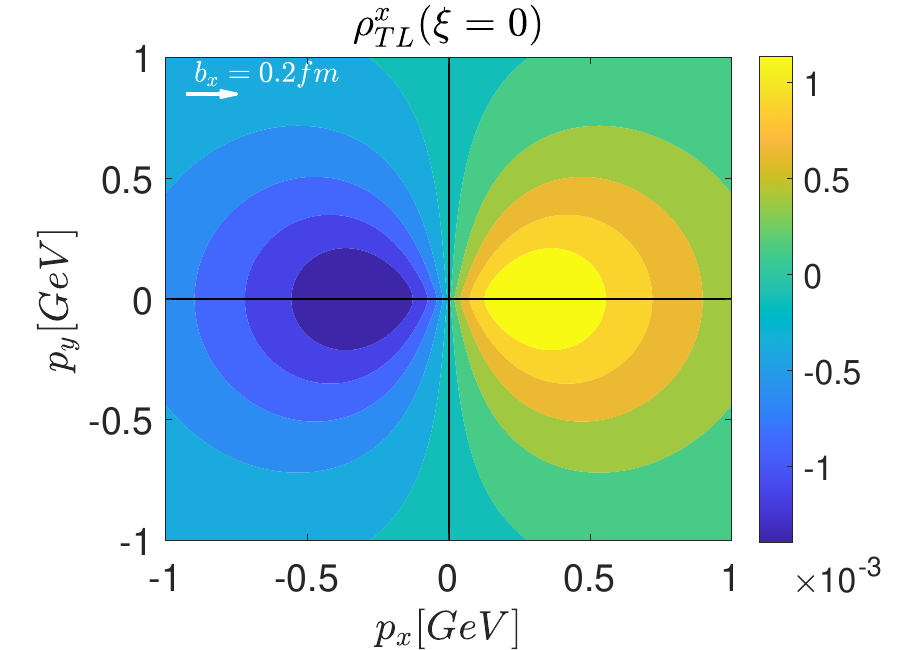}
     \includegraphics[width=0.30\linewidth, trim=80 240 100 240, clip]{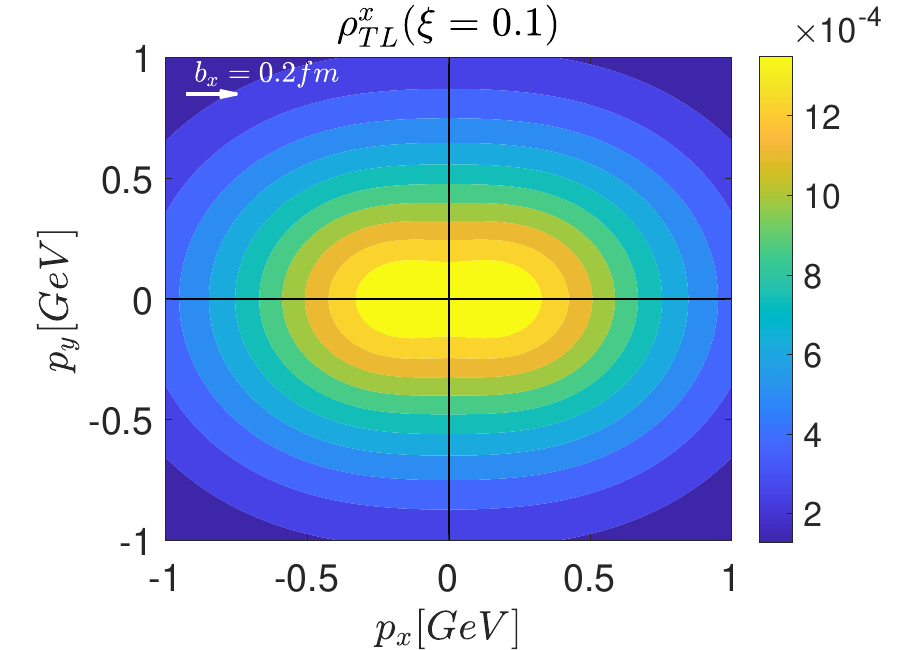}
     \includegraphics[width=0.30\linewidth, trim=80 240 100 240, clip]{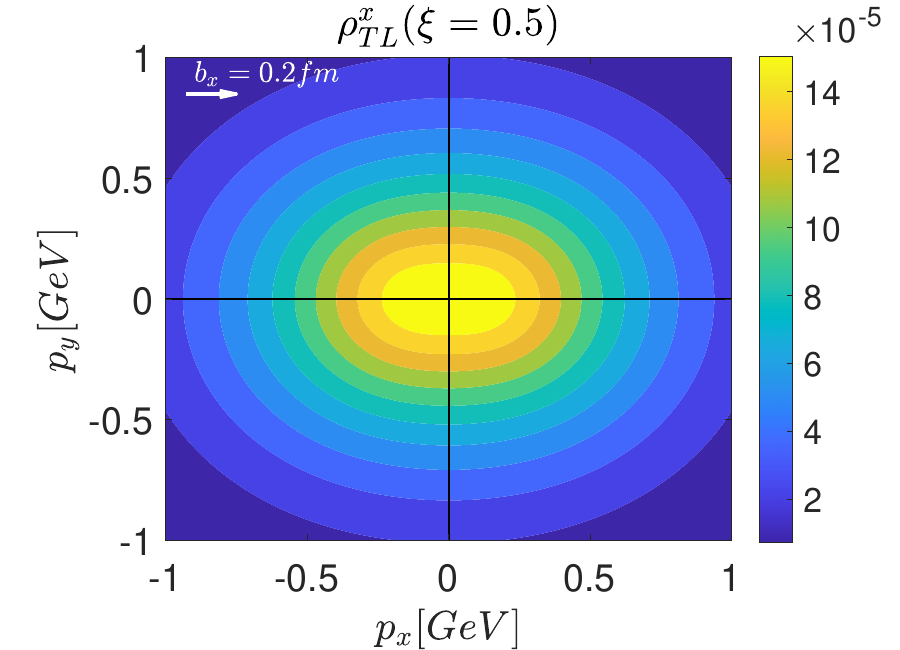}\\ \vspace{0cm}
      \includegraphics[width=0.30\linewidth, trim=80 240 100 240, clip]{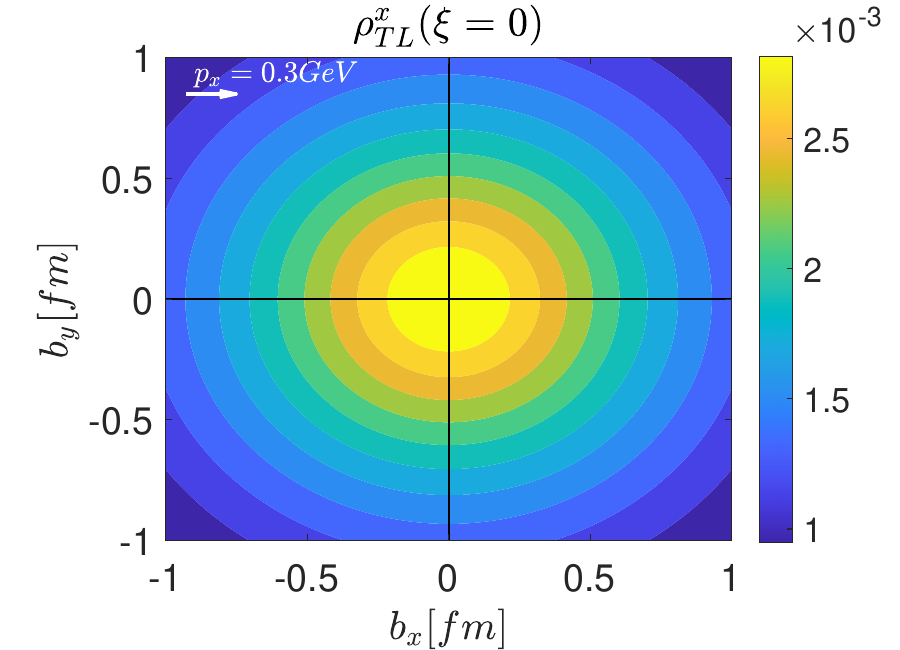}
     \includegraphics[width=0.30\linewidth, trim=80 240 100 240, clip]{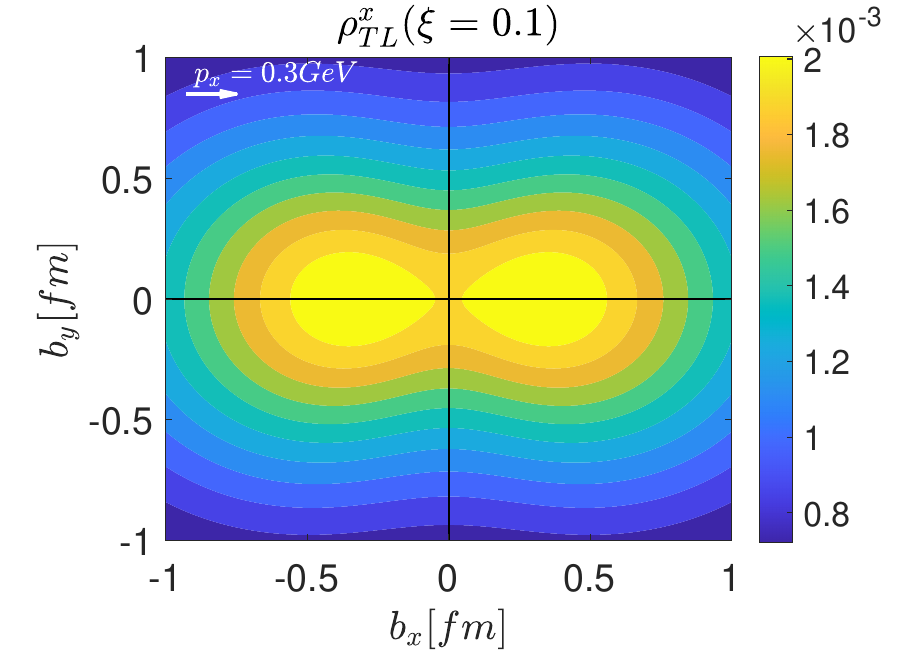}
     \includegraphics[width=0.30\linewidth, trim=80 240 100 240, clip]{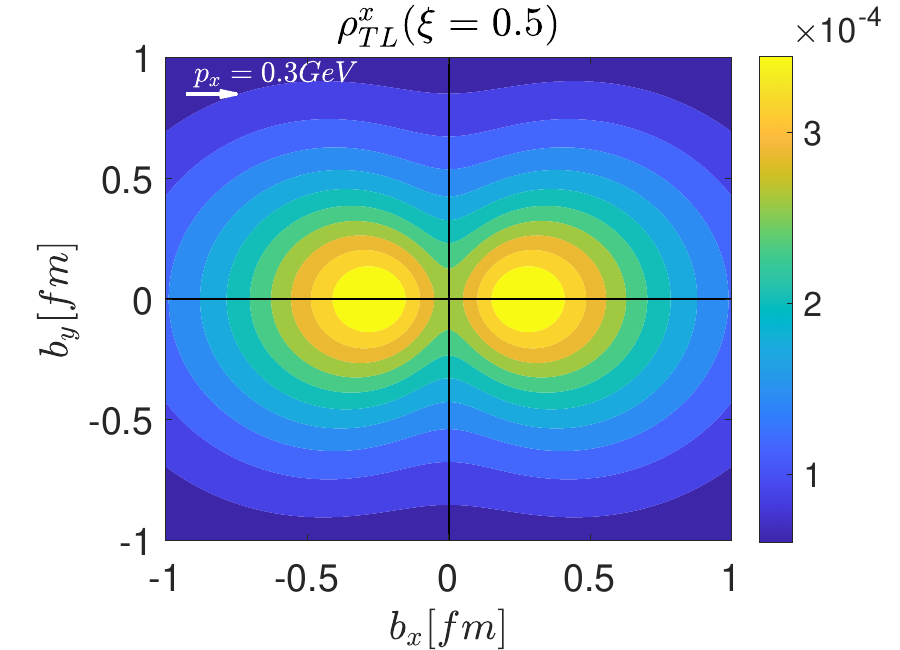}
    \caption{The first Mellin moment of gluon Wigner distribution $\rho^x_{TL}$ for different values of skewness parameter ($\xi= 0, 0.1,0.5$) in transverse momentum space (upper panel) and impact parameter space (lower panel) for fixed $\bfb=0.2$ fm $\hat{y}$ and $\bfp=0.3$ GeV $\hat{y}$, respectively, with the condition $\bfp\perp\bfd$.}
    \label{fig:b_rhoTL}
\end{figure}

\begin{figure}[h]
    \centering
     \includegraphics[width=0.30\linewidth, trim=80 240 100 240, clip]{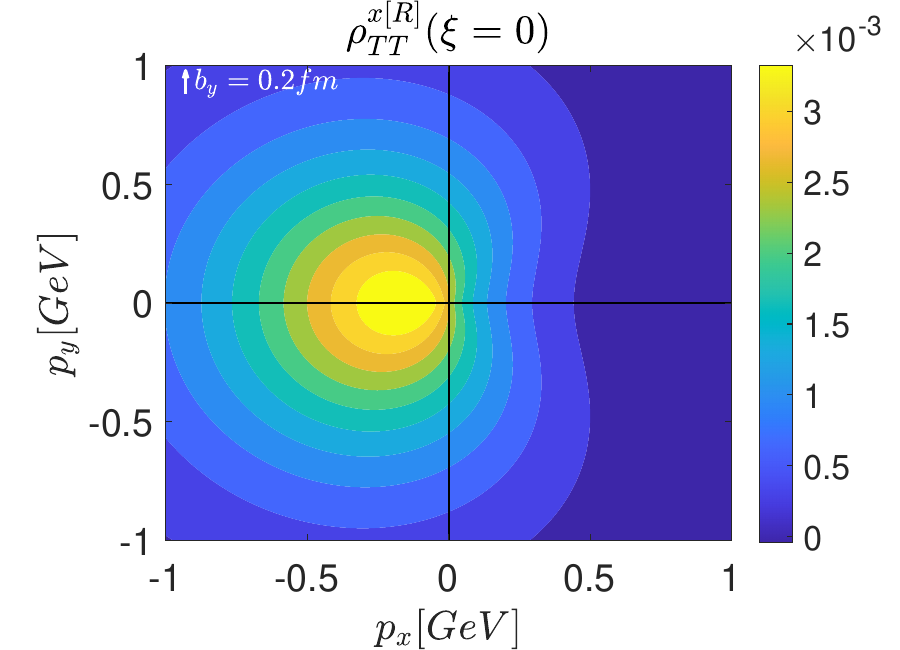}
     \includegraphics[width=0.30\linewidth, trim=80 240 100 240, clip]{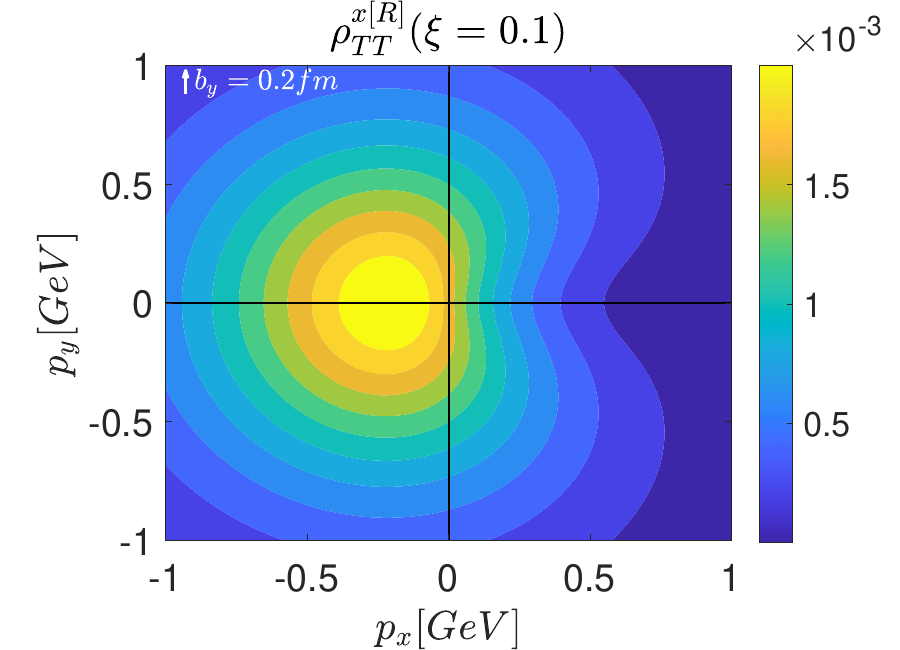}
     \includegraphics[width=0.30\linewidth, trim=80 240 100 240, clip]{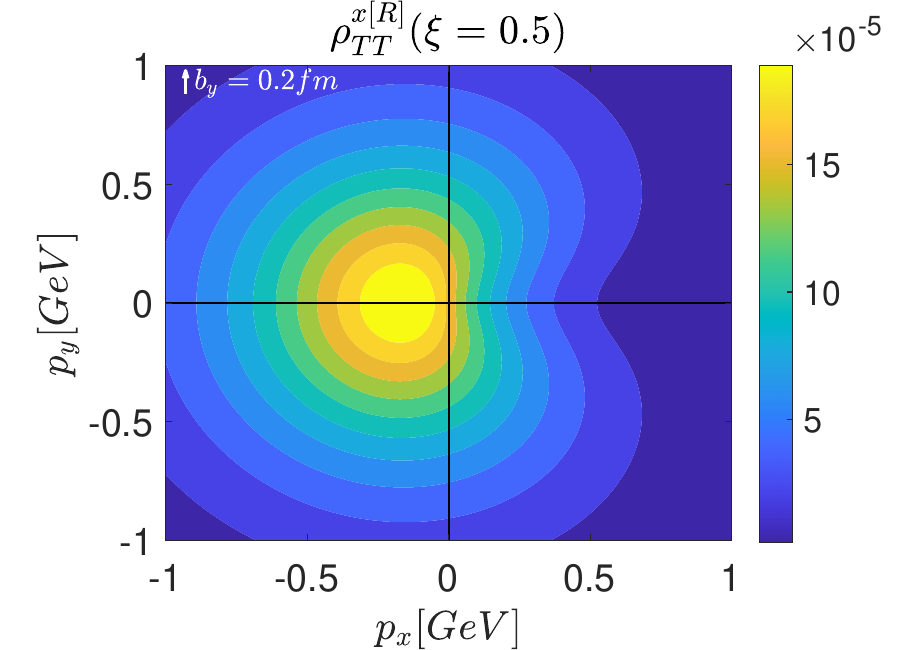}\\ \vspace{0cm}
      \includegraphics[width=0.30\linewidth, trim=80 240 100 240, clip]{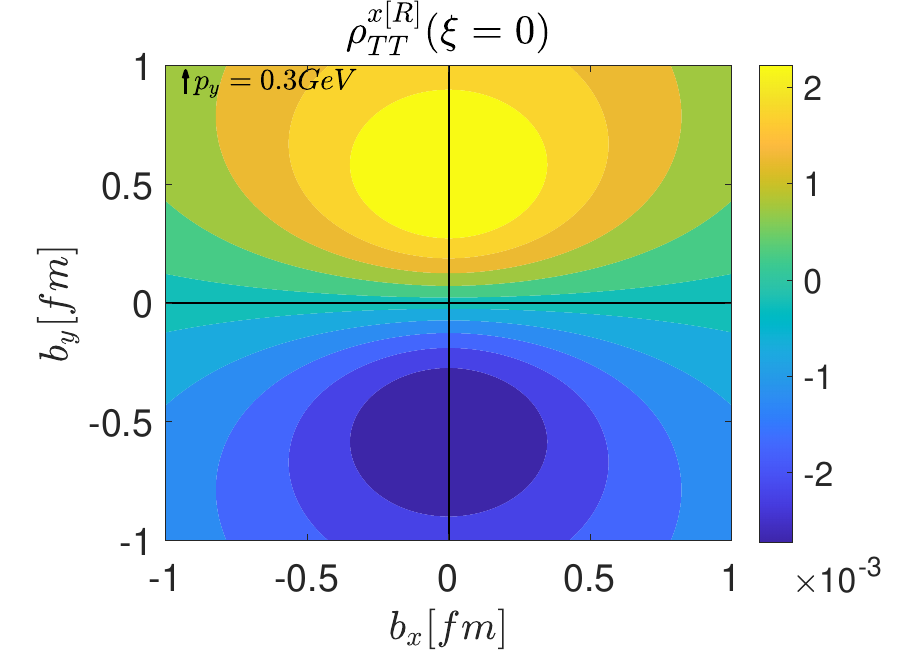}
     \includegraphics[width=0.30\linewidth, trim=80 240 100 240, clip]{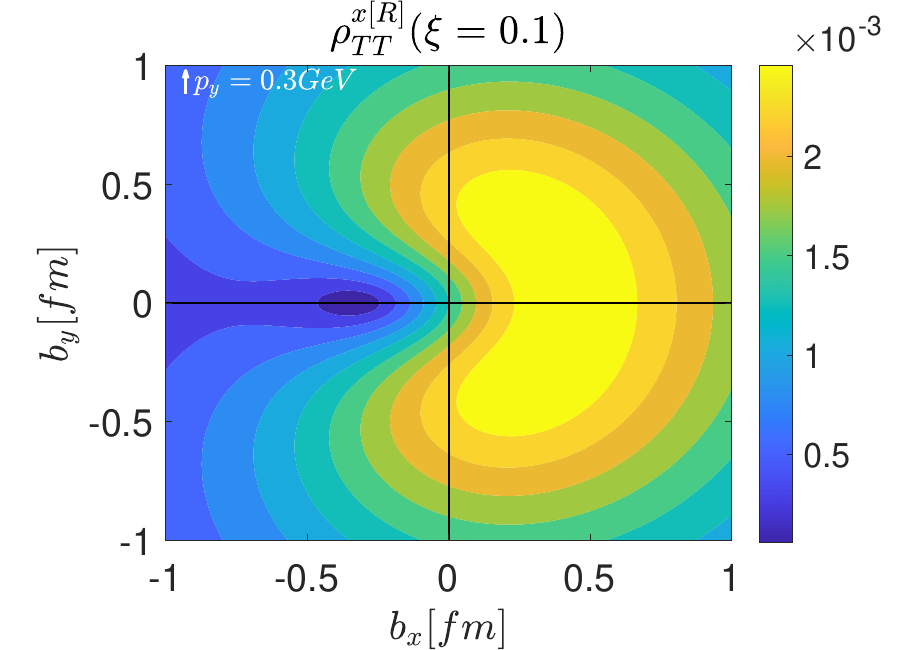}
     \includegraphics[width=0.30\linewidth, trim=80 240 100 240, clip]{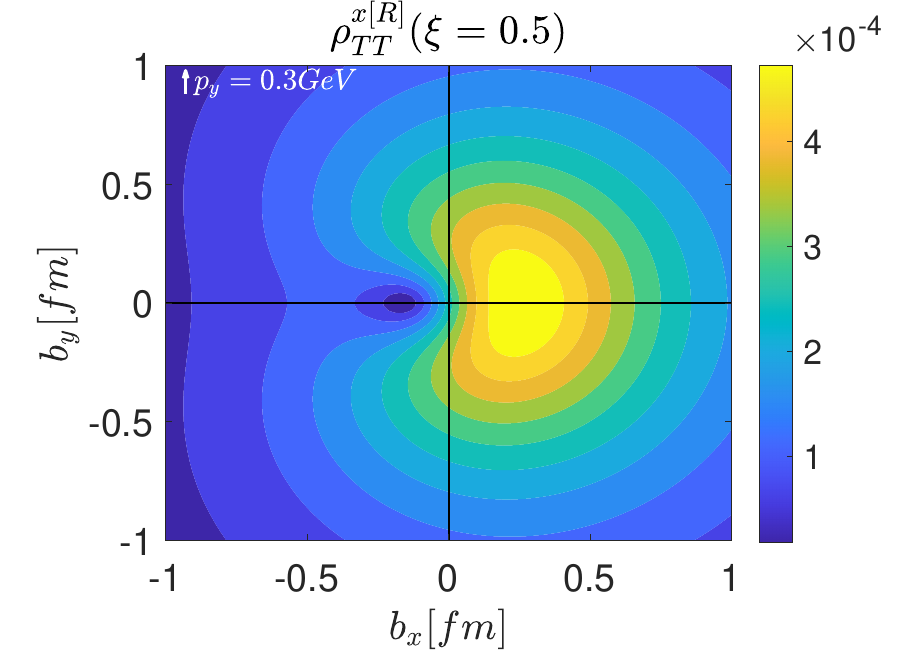}
    \caption{The first Mellin moment of gluon Wigner distribution $\rho^{x[R]}_{TT}$ for different values of skewness parameter ($\xi= 0, 0.1,0.5$) in transverse momentum space (upper panel) and impact parameter space (lower panel) for fixed $\bfb=0.2$ fm $\hat{y}$ and $\bfp=0.3$ GeV $\hat{y}$, respectively, with the condition $\bfp\perp\bfd$.}
    \label{fig:b_rhoTTR}
\end{figure}
\begin{figure}[h]
    \centering
     \includegraphics[width=0.30\linewidth, trim=80 240 100 240, clip]{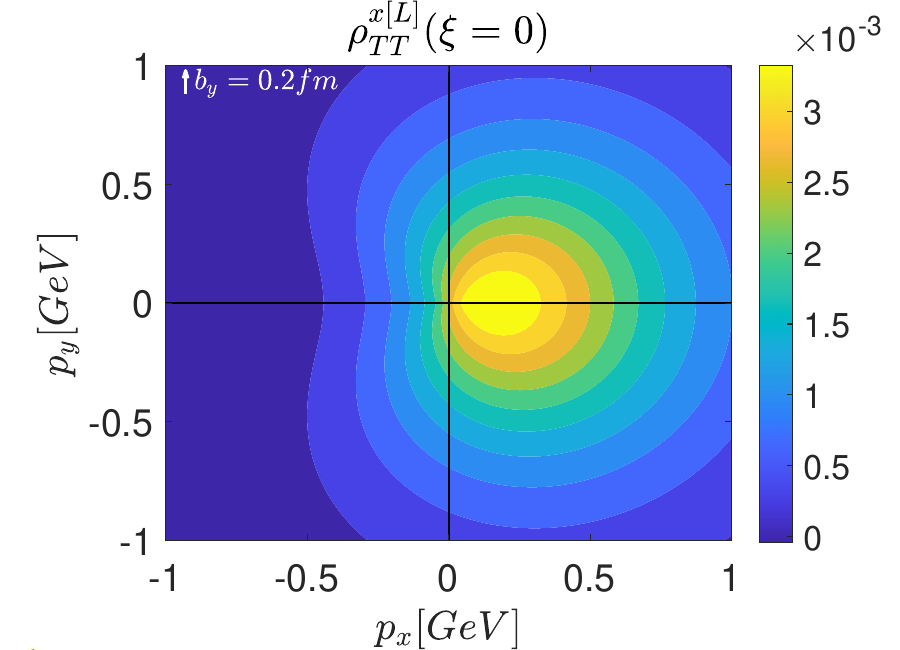}
     \includegraphics[width=0.30\linewidth, trim=80 240 100 240, clip]{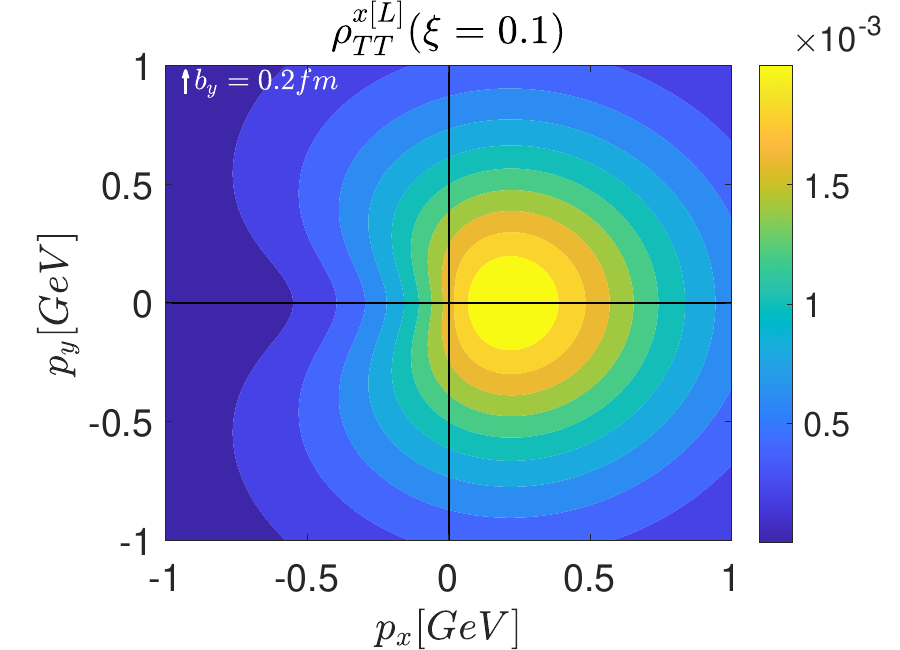}
     \includegraphics[width=0.30\linewidth, trim=80 240 100 240, clip]{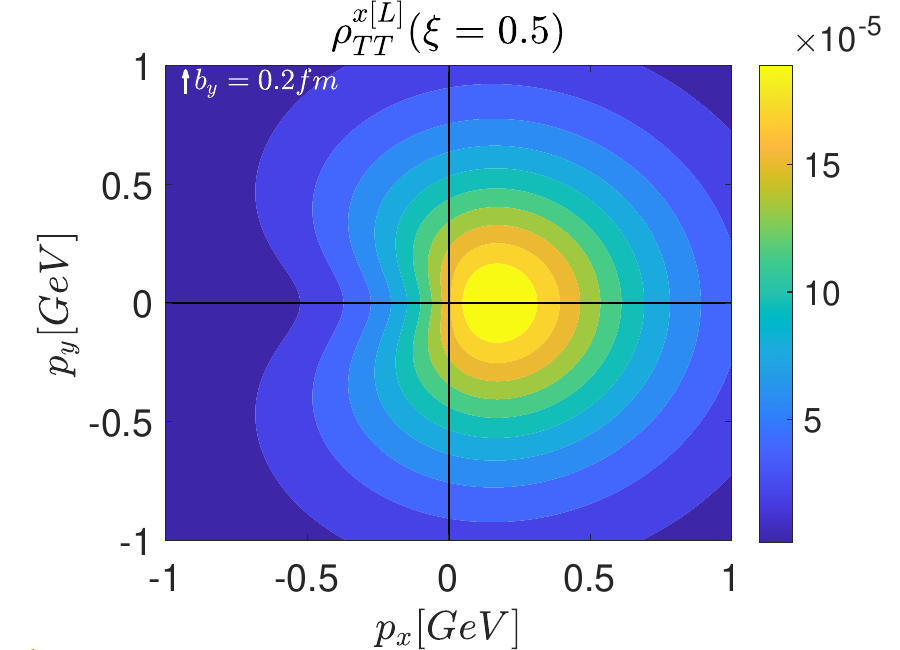}\\ \vspace{0cm}
      \includegraphics[width=0.30\linewidth, trim=80 240 100 240, clip]{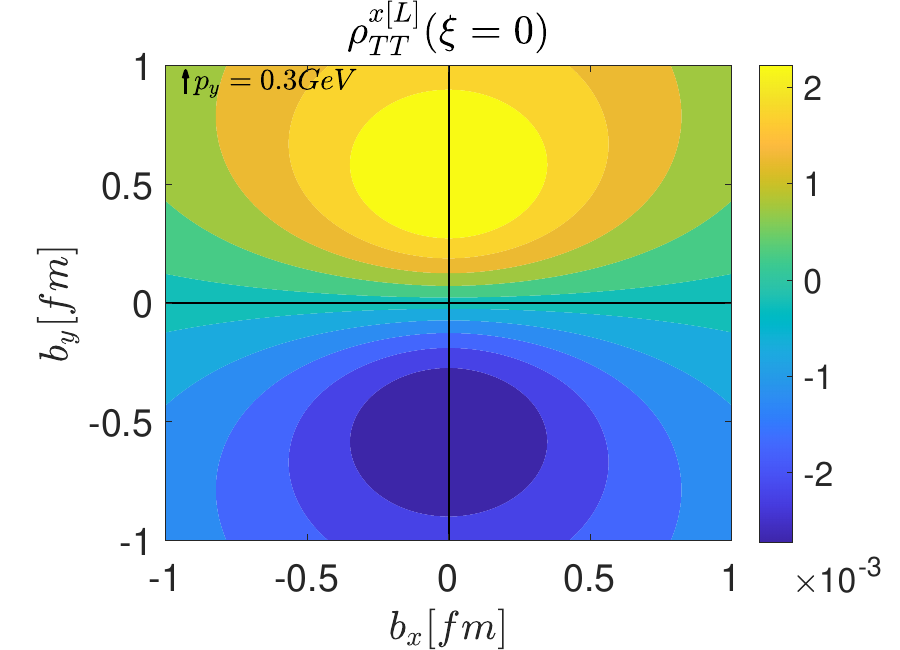}
     \includegraphics[width=0.30\linewidth, trim=80 240 100 240, clip]{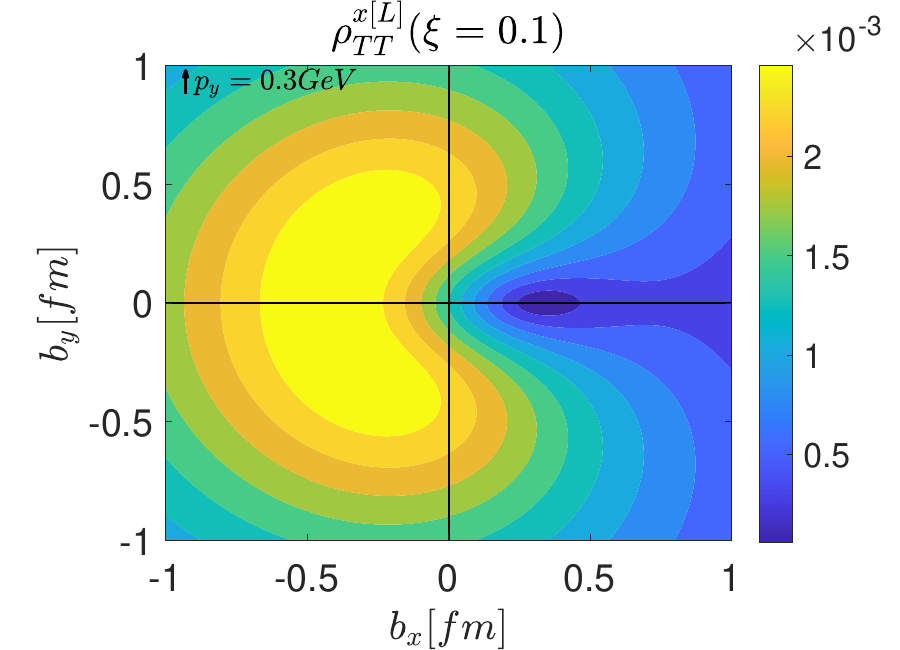}
     \includegraphics[width=0.30\linewidth, trim=80 240 100 240, clip]{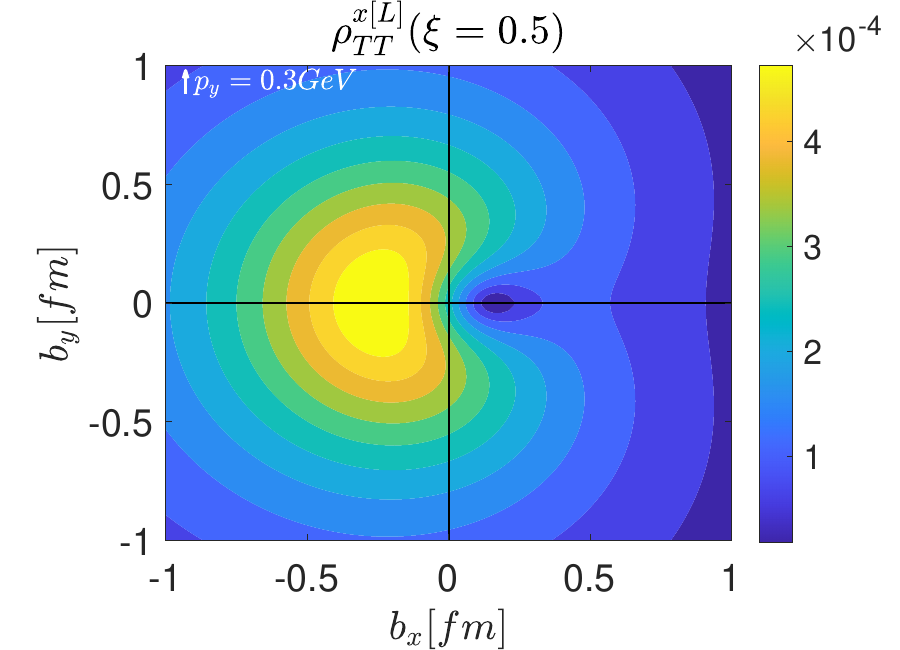}
    \caption{The first Mellin moment of gluon Wigner distribution $\rho^{x[L]}_{TT}$ for different values of skewness parameter ($\xi= 0, 0.1,0.5$) in transverse momentum space (upper panel) and impact parameter space (lower panel) for fixed $\bfb=0.2$ fm $\hat{y}$ and $\bfp=0.3$ GeV $\hat{y}$, respectively, with the condition $\bfp\perp\bfd$.}
    \label{fig:b_rhoTTL}
\end{figure}

For transversely polarized proton, Fig.~\ref{fig:b_rhoTU} displays the distribution of unpolarized gluons $\rho^x_{TU}(\bfp,\bfb)$ in both the $\bfp$ and $\bfb-$planes.
We observe that in $\bfp-$ space, for $\xi=0$, the distribution is circularly symmetric, having positive maxima localized at the origin. While the dressed quark model shows a quadrupole behavior \cite{More:2017zqp}. For non-zero skewness, the absolute value of $\rho^x_{TU}(p_x,p_y)$ shows elliptically symmetric behavior which becomes more concentrated near the origin with an increasing skewness.
In $\bfb-$ space, for $\xi=0$, the distribution exhibits dipolar structure that is symmetric about $y$-axis caused by the presence of $\bfb^{(2)}$-term in Eq.~(\ref{eq:rho_{TU}}). A similar qualitative behavior is reported in the dressed quark model \cite{More:2017zqp} with a flip in axis. 
 The positive poles move closer to each other towards the origin for the increasing skewness. 
In Fig.~\ref{fig:b_rhoTL}, we present the distribution of longitudinally polarized gluons inside a transversely polarized proton $\rho^x_{TL}(\bfp,\bfb)$ in both the $\bfp$ and $\bfb-$planes with fixed impact parameter $\bfb=0.2$ fm along $\hat{x}$ and with fixed transverse momentum $\bfp=0.3$ GeV along $\hat{x}$, respectively. 
In transverse momentum space, for $\xi=0$, the distribution $\rho^x_{TL}$ have dipolar configuration due to the presence of the term $\bfp^{(1)}$ in Eq.~(\ref{eq:rho_{TL}}), while in impact parameter space, the distribution is circularly symmetric exhibit positive maxima at the center. An analogous behavior is observed in the dressed quark model for both the planes \cite{More:2017zqp}.
For non-zero skewness, the absolute value of the distribution $\rho^x_{TL}(p_x,p_y)$ becomes elliptical and the width of the central maxima decreases with increasing skewness.
In Figs.~\ref{fig:b_rhoTTR} and \ref{fig:b_rhoTTL}, the linearly polarized gluon distributions $\rho^{x[R]}_{TT}$ and $\rho^{x[L]}_{TT}$ exhibits a distorted circularly distribution in $\bfp$-space and a dipolar distribution in $\bfb$-space at $\xi = 0$. However, \cite{More:2017zqp} reported distorted dipolar distribution in $\bfp-$space and a distorted circular behavior in $\bfb-$space.
For nonzero skewness, the absolute value of $\rho^{x[R]}_{TT}$ reveals a distorted dipolar pattern that becomes progressively more localized with increasing $\xi$. The distribution $\rho_{TT}^{x[L]}$ exhibits the same qualitative features as $\rho_{TT}^{x[R]}$, but with an opposite dipole orientation. This demonstrates that $\rho_{TT}^{x[R]}$ and $\rho_{TT}^{x[L]}$ are related by a reflection symmetry in both transverse momentum and impact parameter spaces, arising from the opposite helicity structures of the corresponding gluon polarization states.

\section{Conclusions}\label{con}
We have investigated the gluon Wigner distributions for unpolarized, longitudinally polarized, and linearly polarized gluon within a proton, in the boost-invariant longitudinal space $\sigma$ for the first time. The variable $\sigma=\frac{1}{2}b^- P^+$ provides information on the change in the longitudinal light-front coordinate of the constituent gluon in a proton. We use the light-front gluon spectator model for nucleon inspired by soft-wall ADS/QCD, where the momentum transfer is considered in both the transverse and the longitudinal directions. For the numerical calculation, the finite limit of the Fourier integration over skewness is governed by the DGLAP region $\xi < x < 1$.
The gluon Wigner distributions in $\sigma$ space are analyzed by varying the total energy transferred $-t$ at fixed $x$, as well as by varying the longitudinal momentum fraction $x$ for fixed $-t$. The distributions exhibit a characteristic oscillatory pattern, analogous to single-slit optical diffraction. The width of the central maxima varies inversely with $x$ and $-t$, those can be considered as the effective slit-width of the diffraction pattern.  We observe that the position of the first-minima of gluon WDs shifts by large amount towards the center with increasing $x$, whereas the width of the central maxima is less sensitive to $-t$, unlike the quark WDs in a proton.
A detailed investigation of the skewness sensitivity to gluon WDs in the impact parameter $\bfb-$space is also presented in both the transverse momentum and impact parameter planes which would be useful for future experimental measurements.  The model result shows gluon canonical OAM-spin correlation factor $l_z <0 $, which indicates OAM of the gluon is found to be anti-parallel to the proton spin which is opposite for the constituent quark. Whereas $C_z <0 $, which implies gluon spin and the gluon OAM are mutually anti-parallel. The model results inspired by AdS/QCD wave functions shows significantly dominating spin-OAM correlation for gluon over quark. These model dependent results seek for further experimental verification.

\begin{acknowledgments}
This work is supported
by the Anusandhan National Research Foundation (ANRF), Department of Science and Technology, Government of India and Science and Engineering Research Board (SERB) through the SRG (Start-up Research Grant) of File No.
SRG/2023/001093.

\end{acknowledgments}

\appendix 

\section{Bilinear Decomposition of the Gluon-Gluon Correlator And GTMDs}\label{AppA}
The bilinear decomposition of the gluon-gluon correlator Eq.(\ref{Wdef}) is related to the leading twist GTMDs as~\cite{Stephan}
\begin{align}
	W^{[\delta_\perp^{ij}]}_{\lambda^\prime,\lambda^{\prime\prime}}
	=&\frac{1}{2M}\bar{u}(\textbf{P}^{\prime \prime},\lambda^{\prime \prime})\bigg[F_{1,1}^g+\frac{i\sigma^{i+}\bfp^i}{P^+}F^g_{1,2} +\frac{i\sigma^{i+}\bfd^i}{P^+}F^g_{1,3}+\frac{i\sigma^{ij}\bfp^i\Delta^j_\perp}{M^2}F^g_{1,4}\bigg]u(\textbf{P}^\prime,\lambda^\prime) \nonumber\\
	=&\frac{1}{M\sqrt{1-\xi^2}} \bigg\{\bigg[M\delta_{\lambda^\prime,\lambda^{\prime \prime}}-\frac{1}{2}(\lambda^\prime\bfd^{(1)}+i\bfd^{(2)})\delta_{\lambda^\prime,-\lambda^{\prime \prime}}\bigg]F^g_{1,1} +(1-\xi^2)(\lambda^\prime \bfp^{(1)}+i\bfp^{(2)})\delta_{\lambda^\prime,-\lambda^{\prime\prime}}F_{1,2}^g \nonumber \\
	&+(1-\xi^2)(\lambda^\prime\bfd^{(1)}+i\bfd^{(2)})\delta_{\lambda^\prime,-\lambda^{\prime\prime}}F^g_{1,3} +\frac{i\epsilon_\perp^{ij}\bfp^i\bfd^j}{M^2} \bigg[\lambda^\prime M \delta_{\lambda^\prime,\lambda^{\prime\prime}}-\frac{\xi}{2}(\bfd^{(1)}+i\lambda^\prime\bfd^{(2)})\delta_{\lambda^\prime,-\lambda^{\prime\prime}}\bigg]F_{1,4}^g \bigg\}
	\label{eq:F-GTMDs}
\end{align}
and
\begin{align}
 W^{[-i\epsilon_\perp^{ij}]}_{\lambda^\prime,\lambda^{\prime \prime}}
	=&\frac{1}{2M}\bar{u}(\textbf{P}^{\prime \prime},\lambda^{\prime \prime})\bigg[-\frac{i\epsilon_\perp^{ij}\bfp^i\bfd^j}{M^2} G_{1,1}^g+\frac{i\sigma^{i+}\gamma_5 \bfp^i}{P^+}G^g_{1,2} +\frac{i\sigma^{i+}\gamma_5 \bfd^i}{P^+}G^g_{1,3}+i\sigma^{+-}\gamma_5 G^g_{1,4}\bigg]u(\textbf{P}^\prime,\lambda^\prime) \nonumber\\
	=&\frac{1}{M\sqrt{1-\xi^2}} \bigg\{-\frac{i\epsilon^{ij}_\perp \bfp^i\bfd^j}{M^2} \bigg[M\delta_{\lambda^\prime,\lambda^{\prime \prime}}-\frac{1}{2}(\lambda^\prime\bfd^{(1)}+i\bfd^{(2)}\delta_{\lambda^\prime,-\lambda^{\prime \prime}}\bigg]G^g_{1,1} +(1-\xi^2)(\bfp^{(1)}+i\lambda^\prime \bfp^{(2)})\delta_{\lambda^\prime,-\lambda^{\prime \prime}}G_{1,2}^g \nonumber \\
	&+(1-\xi^2)(\bfd^{(1)}+i\lambda^\prime\bfd^{(2)})\delta_{\lambda^\prime,-\lambda^{\prime \prime}}G^g_{1,3} +\bigg[\lambda^\prime M \delta_{\lambda^\prime,\lambda^{\prime \prime}}-\frac{\xi}{2}(\bfd^{(1)}+i\lambda^\prime\bfd^{(2)})\delta_{\lambda^\prime,-\lambda^{\prime \prime}}\bigg]G_{1,4}^g \bigg\}
	\label{eq:G-GTMDs},
\end{align}
The spinors $u(k,\lambda) $ with the momentum $k$ and the helicity $\lambda\,(=\pm )$ are given by
\begin{center}
$u(k, +)=\frac{1}{\sqrt{2 k^+}}\left( \begin{matrix} 
  k^+ + m_F\\
  k^1 + i k^2\\
  k^+ - m_F\\
  k^1 + i k^2\\
\end{matrix} \right) $, \hspace{1cm}
$u(k, -)=\frac{1}{\sqrt{2 k^+}}\left( \begin{matrix} 
    - k^1 + i k^2\\
    k^+ + m_F\\
  k^1- i k^2\\
  - k^+ + m_F\\
\end{matrix} \right) $
\end{center}
with $m_F$ being the mass of the fermion.
Using the kinematics given in  Eqs.~(\ref{Pp}), (\ref{Ppp}), one can find out the spinors $u(P^\prime,\lambda^\prime) $  and $u(P^{\prime \prime},\lambda^{\prime \prime}) $ and compute the matrix elements of $\bar{u}(P^{\prime \prime},\lambda^{\prime \prime}) \Gamma u(P^\prime,\lambda^\prime)$, where $\Gamma$ represents the Dirac matrix structure.
\bibliography{WD_arXiv_v3}
\end{document}